\pdfoutput=1
\RequirePackage[T1]{fontenc}
\documentclass[12pt]{article}

\usepackage[height=8.85in,width=6.25in]{geometry}

\usepackage[utf8]{inputenc}
\usepackage{amsmath}
\usepackage{amssymb}
\usepackage{mathtools}
\numberwithin{equation}{section}
\usepackage{slashed}
\usepackage{braket}
\usepackage[svgnames,dvipsnames]{xcolor}
\usepackage[colorlinks,citecolor=DarkGreen,linkcolor=FireBrick,urlcolor=FireBrick,linktocpage,breaklinks=true]{hyperref}
\usepackage{cite}
\usepackage{graphicx}
\usepackage{tikz}
\usepackage{times}
\usepackage{courier}
\usepackage{bm}
\usepackage{subfig}
\usepackage{mathabx}
\let\upeta\eta

\usepackage{tikz-cd}

\newcommand{\diff}{\breve}     
\DeclareMathOperator{\ph}{ph}
\DeclareMathOperator{\tr}{tr}
\protected\def\[#1\]{\begin{equation}\begin{split}#1\end{split}\end{equation}}

\usetikzlibrary{calc}

\def\bZ{\mathbb{Z}}
\def\bR{\mathbb{R}}
\def\bC{\mathbb{C}}

\def\ch{\mathrm{ch}}

\def\sS{\mathsf{S}}
\def\tors{\text{tors}}
\def\Tors{\mathop{\mathrm{Tors}}}
\def\Ker{\mathop{\mathrm{Ker}}}
\def\ph{\mathop{\mathrm{ph}}}

\def\Hom{\mathop{\mathrm{Hom}}}

\def\Im{\mathop{\mathrm{Im}}}

\let\check\widecheck
\let\diff\widecheck
\let\breve\widecheck
\def\cl{\text{cl}}

\def\PD{\mathrm{PD}}

\begin{document}

\begin{titlepage}

\begin{flushright}

\end{flushright}

\vskip 3cm

\begin{center}

{\Large \bfseries Type I anomaly cancellation revisited}

\vskip 1cm
Saghar S. Hosseini,
Yuji Tachikawa
and
Hao Y. Zhang
\vskip 1cm

\begin{tabular}{ll}
 & Kavli Institute for the Physics and Mathematics of the Universe (WPI), \\
& University of Tokyo,  Kashiwa, Chiba 277-8583, Japan
\end{tabular}

\vskip 2cm

\end{center}

\noindent 

We revisit the issue of how the perturbative and global fermion anomaly of Type I string theory in ten dimensions is cancelled by the Green-Schwarz mechanism using the RR fields.
This will be done by realising the RR fields as boundary modes of an eleven-dimensional bulk theory described 
in terms of a quadratic refinement of the differential KO-theory pairing. 
We will then generalise this analysis to Sugimoto's $usp(32)$ string and Sagnotti's $u(32)$ string.

We also discuss in a more general setting the procedures which need to be followed when we try to cancel fermion anomalies in terms of $p$-form fields based on differential K-theory classes.
This we illustrate by performing an analysis of the mod-2 anomaly cancellation in nine dimensions arising from the $S^1$ compactification of the Type I theory.

\end{titlepage}

\setcounter{tocdepth}{2}
\tableofcontents

\section{Introduction and summary}
\label{sec:intro}

\subsection{Some historical background}

In Type I string theory, the perturbative part of the spacetime fermion anomaly is famously cancelled 
by the Green-Schwarz mechanism \cite{Green:1984sg,Green:1984ed,Green:1984qs}, 
which uses the following couplings of the Ramond-Ramond (RR) fields: \begin{align}
dG_3 &=X_4^{GS}=\frac{p_1(R)}2 -\frac{p_1(F)}2,\label{eq:introX4}\\
dG_7 &= Y_8^{GS}=\frac{3p_1(R)^2-4p_2(R)}{192}-\frac1{12}\frac{p_1(R)}2\frac{p_1(F)}2+\frac{p_1(F)^2-2p_2(F)}{12}\label{eq:introY8}
\end{align} where $G_3$ is the gauge-invariant 3-form field strength, $G_7$ is its Hodge dual, and
$X_4^{GS}$, $Y_8^{GS}$ are characteristic classes constructed from the spacetime Pontryagin classes $p_i(R)$ and the $so(32)$ gauge Pontryagin classes $p_i(F)$.
But this fact alone does not guarantee that no subtle global anomaly remains. 
Although the absence of the global anomaly in the analogous heterotic $E_8\times E_8$ case was already settled by Witten in  \cite{Witten:1985bt}
using a bordism analysis performed by a homotopy theorist \cite{StongAppendix},
the case of the Type I $so(32)$ theory was not settled for a long time.

A detailed outline to resolve this vexing issue was put forward by Freed in \cite{Freed:2000ta} almost a quarter century ago,
which was an extension built upon a previous work by Freed and Hopkins  \cite{Freed:2000tt} on anomaly cancellations in Type II theory with D-branes, using differential K theory. While \cite{Freed:2000ta} has provided deep insight and has guided much of the thinking in this area, it was formulated prior to the development of the modern notion of anomaly field theory, and certain aspects remain to be understood that were not fully addressed in the available literature.
Subsequent efforts by Distler, Freed and Moore around 2009
aimed to extend \cite{Freed:2000ta} to completely general orientifold spacetimes. However, this work
did not result in a published full paper accessible to the broader community.\footnote{%
It is to be noted that an announcement of the results was made in \cite{Distler:2009ri}
and that the computer files for multiple talks and lectures given by D. Freed and G. W. Moore on this topic
 can be found on their  webpages \url{https://people.math.harvard.edu/~dafr/} and  
\url{https://www.physics.rutgers.edu/~gmoore/}.
In addition, the authors of this paper  thank J. Distler, D. Freed and G. W. Moore for making the unpublished draft version of 
their paper available to them, which significantly helped the preparation of this paper.}
The main aim of this paper is to partially fill this gap in the literature, by
providing a self-contained exposition of the perturbative and global anomaly cancellation of the Type~I theory in ten dimensions,
using the differential KO-theoretic formulation of the Type I RR fields. \footnote{See \cite{Grady:2018suz,Gomi:2021bhy,berwick2023families} for more mathematical physics literature that discuss differential KO theory.}

\subsection{Precise formulation of RR fields and the cancellation}

To be slightly more concrete, the Type I RR fields $G_3$, $G_7$ in ten dimensions will be realised as edge modes of an eleven-dimensional bulk theory, whose action for topologically trivial fields is given by \begin{equation}
S_{11} = 2\pi i \int_{M_{11}}  C_3 d C_7 + \cdots.\label{eq:naive-action}
\end{equation} 
To discuss the global anomaly cancellation, 
we need to specify how the fluxes $F_4$ and $F_8$ of the bulk form fields $C_3$ and $C_7$ are 
quantised.
Naively one might try to say that their fluxes are quantised by classes in $H^4(M_{11};\bZ)$ and $H^8(M_{11};\bZ)$, respectively.
If this were the case, we would have proceeded to use differential cohomology classes 
in $\check H^4(M_{11};\bZ)$ and $\check H^8(M_{11};\bZ)$
to describe the bulk form fields.
Here, differential cohomology is a framework which allows us to treat $p$-form fields 
whose flux is quantised in terms of integer cohomology classes in a mathematically precise manner,
 the use of which is becoming increasingly common in the theoretical physics literature.

But this does not suffice, because the characteristic class $Y_8$ \eqref{eq:introY8} above satisfies no easy integrality properties.
Rather, the K theory interpretation of the RR fields  \cite{Witten:1998cd,Moore:1999gb} dictates 
that the combination of the fluxes of $F_4$ and $F_8$ is quantised at the same time 
by a single class in $KO^0(M_{11})$,
whose differential version contains both $C_3$ and $C_7$.
A proper setting for such a form field is to use the differential KO theory classes
valued in $\check{KO}{}^0(M_{11})$, whose properties will be detailed later.

Still using a schematic notation, the action \eqref{eq:naive-action} becomes \begin{equation}
S_{11} = 2\pi i \cdot \frac12 \int_{M_{11}} CdC + \cdots \label{eq:better-action}
\end{equation} 
in terms of a single combination $C=(C_3,C_7)$.
This makes our bulk action to behave more as an Abelian Chern-Simons theory
rather than as a BF theory, as the action \eqref{eq:naive-action} might have suggested.
Also, the $\int CdC$ term is only defined modulo $\bZ$,
and therefore taking one half of it as in \eqref{eq:better-action} requires an additional machinery,
called quadratic refinements.

The edge modes of Abelian Chern-Simons theories based on differential K theory were already studied in \cite{Belov:2006jd,Belov:2006xj,Hsieh:2020jpj}, 
whose analyses we follow closely.
The end result is that the bulk Abelian Chern-Simons theory with this action is an invertible field theory,
which we will show to exactly cancel the invertible theory for the fermion anomalies.
Then the boundary RR fields $G_3$, $G_7$ are considered 
as chiral edge modes of this Abelian Chern-Simons theory,
even though no self-duality is naively involved at the level of  differential forms.

After explaining our approach in the case of Type I theory in detail, which uses 
a quadratic refinement of the pairing of differential KO theory,
we treat the global anomaly cancellation in two other 10-dimensional string theories,
namely Sugimoto's $usp(32)$ theory \cite{Sugimoto:1999tx}
and Sagnotti's $u(32)$ theory \cite{Sagnotti:1995ga}.
In Sugimoto's case, we use a quadratic refinement of the pairing of differential KSp theory,
while in Sagnotti's case, we use an alternative quadratic refinement of the pairing of differential K theory,
which is different from the quadratic refinement used in the case of Type IIB theory.

We then consider the $S^1$ compactification of Type I theory, by taking $M_{11}=M_{10} \times S^1$.
On this manifold, depending on whether the differential form has a component along $S^1$ or not,
the field $C=(C_3,C_7)$ splits into $A=(A_3,A_7)$ and $B=(B_2,B_6)$.
The action \eqref{eq:better-action} then becomes \begin{equation}
S_{10}=2\pi i \int_{M_{10}} AdB +\cdots 
= 2\pi i\int_{M_{10}} (A_3 d B_6 + A_7 d B_2) + \cdots,
\end{equation}
meaning that the theory now behaves more as a BF theory even in terms of the differential KO classes,
and the boundary RR fields no longer involve the self-duality.
Studying the anomaly cancellation in this simpler setup, without the extra complication due to self-duality,
gives us some additional insight into what is involved in trying to cancel fermion anomalies 
by form fields described by differential K or KO theory.
This also gives an example where mod-2 anomalies are cancelled in terms of continuous form fields.

It is true that it requires a significant amount of efforts to learn 
differential K-theory and KO-theory, but the readers will see that, 
with this formalism,  it is almost automatic that the form-fields
based on them can  cancel both perturbative and global
parts of gauge and mixed gauge-gravity anomalies of fermions.
What is still non-trivial is that the pure gravitational part of the anomalies is cancelled.
We will find that the anomaly of the chiral RR fields
based on differential KO theory is uniquely fixed by the formalism.
In ten dimensions, we will find that the resulting anomaly equals 
that of the anomaly of the dilatino and the gravitino,
guaranteeing the full cancellation of anomalies of Type I theory.
Not only that, although the formalism itself allows the gauge fields to be described by 
an arbitrary real bundle $V$, we will see that 
solving the bulk 11d equation of motion
we automatically have $\dim V=32$, $w_1(V)=0$ and $w_2(V)=0$,\footnote{%
Due to the authors' lack of ability, so far it can only be proved that $\int_{M_2} w_2(V)=0$ 
on \emph{orientable} two-dimensional submanifolds $M_2$ of the ten-dimensional spacetime
in the approach of this paper, although the authors believe that it can be generalised to 
non-orientable submanifolds.
The authors would hope to come back to fix this deficiency in a future paper.
\label{foot:disclaimer}
}
so that the gauge group is $Spin(32)$.

\subsection{Comparison to the cancellation in the heterotic dual}

Before proceeding, we would like to mention a comparison to the anomaly cancellation in the S-dual heterotic $so(32)$ theory.
The global anomaly cancellation on the heterotic side in general is performed 
by including the $H_3$ field, the NS-NS field S-dual to our $G_3$, as part of the spacetime structure.
The condition $dH_3=X_4^{GS}=p_1(R)/2 - p_1(F)/2$ is then upgraded to the integral cohomology level,
promoting the spin structure of the spacetime to a structure known in the mathematical literature as the twisted string structure.
One then performs the bordism-type analysis of the anomalies which became commonplace lately. 

In fact, the very early analysis in \cite{Witten:1985bt,StongAppendix}  on the cancellation of the global anomaly of the $E_8\times E_8$ theory was  performed exactly as outlined above.
No direct analysis of the ten-dimensional $so(32)$ case was performed to the authors' knowledge,\footnote{%
An indirect result about the general anomaly cancellation in heterotic theory in \cite{Tachikawa:2021mby,Yonekura:2022reu} is also applicable to this case, 
but it relies on a conjecture relating the worldsheet superconformal theory to the mathematical theory of topological modular forms.
}
most probably due to the difficulty in the corresponding bordism computation.
Indeed, the necessary bordism group $\Omega^\text{string}_{11}(BSpin(32)/\bZ_2)$ was determined only very recently in \cite{Kneissl:2024zox}.

Structurally, our analysis on the Type I side is totally different:
\begin{itemize}
\item From a physics perspective, on the heterotic side, only the 3-form field $H_3$ and its Hodge dual are introduced to the system.
In contrast, on the Type I side, not only the 3-form field $G_3$ and its Hodge dual but also some discrete $\bZ_2$-valued fields are introduced to the system, as they are part of a single differential KO class.
\item From a mathematical perspective, 
on the heterotic side, the spin structure of the spacetime needs to be upgraded to a twisted string structure as outlined above.
In contrast, on the Type I side, we do not change the spacetime structure from the spin structure,
and we simply analyse the anomaly of the RR fields as the anomaly of a spin theory.
\item More technically, the modified Bianchi on the heterotic side was fixed to be $dH_3=p_1(R)/2-p_1(F)/2$ from the consideration of the worldsheet anomaly cancellation.
In contrast, on the Type I side, our formalism guarantees $dH_3=cp_1(R)/2 - p_1(F)/2$ for some coefficient $c$,
but the fact $c=1$ emerges after a long computation,
and indeed $c\neq 1$ if we apply the KO-theory formulation of the RR fields in other dimension of the form $8k+2$. \footnote{Via inflow, this question is related to the question of consistently identifying the global symmetry of a 6D (1,0) SCFT that admit both a type I construction and a heterotic construction, extending the analysis in \cite{Zhang:2024oas}.
This should also be related to the anomaly cancellation on torsion D(-1)- and D0-branes of Type I theory. }
\end{itemize}
These were somewhat surprising to the authors,
since the heterotic $so(32)$ theory and the Type I theory share the same set of massless fields,
and the perturbative anomaly cancellation works in a parallel fashion.
The two differences mentioned above make it quite unclear 
how the anomaly cancellation on both sides are related.
It would be desirable to have a better understanding of these issues, but it needs to await another paper. 

\subsection{Organisation of the paper}

The rest of the paper is organised as follows. 
\begin{itemize}
\item
In Section \ref{sec:BF}, we present a general framework writing the BF theory as bulk invertible phase for non-chiral field using generalised differential cohomology, where we treat $E = H, K, KO$ uniformly. 
\item
Then, in Section \ref{sec:CS}, we extend our framework to cover the bulk invertible phase for chiral $p$-form fields,
assuming the existence of a quadratic refinements. 
We will quantise the theory on a spacetime of the form $S^1\times Y$,
and find topological and differential conditions placed on the background fields.
\item
In Section \ref{sec:KO} we construct specific quadratic refinements in differential KO theory, and by doing so we explicitly write a bulk action for the RR-fields of type I string theory.
We will see that the resulting RR fields cancel the anomaly of the fermions completely,
including both the perturbative and the global parts.

We also study some basic consequences of the bulk equation of motion and the boundary condition.
Namely, we will see that they force the real gauge bundle $V$ 
to have $\dim V=32$, $w_1(V)=0$ and $w_2(V)=0$,\footnote{%
See footnote~\ref{foot:disclaimer}.
}
so that the gauge group is $Spin(32)$. To show the generality of the construction, we also considered KO-theoretic RR fields in general $8k+2$ dimensions. 

\item 
In Section \ref{sec:other_strings}, we extend our analysis to two different types of string theories. 
We will discuss a quadratic refinement of the pairing of differential KSp theory describing the RR field in 10d Sugimoto string \cite{Sugimoto:1999tx}, 
and an alternative quadratic refinement (in comparison to that of type IIB) of the pairing of differential K-theory that describes the RR field in 10d Sagnotti string \cite{Sagnotti:1995ga}.
\item
Finally, in Section \ref{sec:reduction}, we discuss the circle reduction of the bulk invertible phase. Here, the anomaly cancellation of the 9D gravitational theory with a non-self-dual RR field is studied, resolving some puzzles regarding 9D anomaly cancellation of supergravity theories raised in \cite{Lee:2022spd}. 

We then study whether 4d Witten anomaly can be cancelled using the same technique. 
We conclude that this is not possible, since attempting such cancellation would produce sickness of the 4d theory which did not appear in the 10d or 9d case.
\end{itemize}
Some technical parts are collected in the appendices. 
\begin{itemize}
\item
In Appendix \ref{app:T2} we compute the square $\check T^2 \in \check E^2(S^1)$ of a generalised differential cohomology class $\check T \in \check E^1(S^1)$ lifting the generator of $E^1(S^1)\simeq \bZ$. We show that it produces $\frac{1}{2}$ for arbitrary $E$.
This plays a subtle but important role in the quantisation of the invertible Abelian Chern-Simons theory based on generalised differential cohomology theories.
We note that this computation was already done in the original paper \cite{CheegerSimons} of differential cohomology for $E=H(-;\bZ)$, but we need a KO version.
\item
In Appendix \ref{app:perfectness} we review the perfectness of ordinary and generalised differential cohomology theory, which is essential in establishing the invertibility of the bulk theory. 
\item
In Appendix \ref{app:expansions}, we collect expansions for various characteristic class that are useful for specific computations in differential K and KO theories.
\end{itemize}

\section{Invertible BF theories} \label{sec:BF}

We begin by giving a precise formulation of the BF theories, 
which roughly has an action of the form $\int B F = \int B dA$ at the level of differential forms.
This will be done by using differential cohomology classes\footnote{%
In this paper, we use $\check A$, rather than the more standard $\widehat A$,
to denote differential cohomology theories and differential cohomology classes.
This is because we will also need to use $\hat A$ to denote the A-roof genus,
and we wish to disambiguate the two easily.
}
and differential K or KO theory classes to mathematically represent $p$-form fields 
with integrally-quantised fluxes.

\subsection{With differential ordinary cohomology}
We start by recalling the concept of differential ordinary cohomology theory,
which is a precise mathematical formulation of form fields in quantum field theory.
Suppose we would like to consider an antisymmetric tensor field, or equivalently a form field, $\check C$ on a manifold $X$,
with the following physically motivated features:
\begin{itemize}
\item It has a field strength $R(\check C)$ valued in $\Omega^p(X)_{\cl,\bZ}$,
the space of closed $p$-forms with integral periods.
Here the subscript $\cl$ is to remind ourselves that the forms are closed,
and the subscript $\bZ$ is to show that the periods are integral.
\item A flat field, i.e.~a field with $R(\check C)=0$, can be identified with an element of $H^{p-1}(X;\bR/\bZ)$,
giving the holonomy of $\check C$ around $(p-1)$-cycles of $X$.
\item It has a topological flux $I(\check C)$ valued in $H^p(X;\bZ)$,
generalising the Dirac magnetic flux or equivalently the first Chern class of a Maxwell field.
\item A topologically trivial field, i.e.~a field with $I(\check C)=0$, is given by 
a globally-defined gauge potential taking values in $\Omega^{p-1}(X)$,
modulo the gauge transformations taking values in $\Omega^{p-1}(X)_{\cl,\bZ}$.
\end{itemize}
We would expect the following compatibility conditions:
\begin{itemize}
\item The de Rham cohomology class of $R(\check C)$ equals the topological flux $I(\check C)$ after tensoring by $\bR$, 
both taking values in the group $H^p(X;\bZ)_\text{free}\subset H^p(X;\bR)$.
Here $A_\text{free}$ for an Abelian group is its free quotient, $A/\mathrm{Tor}\ A$.
\item Elements in $H^{p-1}(X;\bR)$ give rise to fields that are flat and topologically trivial at the same time,
which is given by the natural homomorphisms $H^{p-1}(X;\bR)\to H^{p-1}(X;\bR/\bZ)$ and $H^{p-1}(X;\bR)\to \Omega^{p-1}(X)/\Omega^{p-1}(X)_{\cl,\bZ}$.
\end{itemize}
Finally, it should reduce to two special cases:
\begin{itemize}
\item $\check H^1(X)$ should be the space of $0$-form fields, i.e.~$S^1$-valued functions on $X$.
\item $\check H^2(X)$ should be the space of $1$-form fields, i.e.~$S^1$-bundles on $X$ with connections.
\end{itemize}

\begin{figure}
$$
    \begin{tikzcd}
    0 \arrow[rd]  &&&& 0\\
                         & H^{p-1}(X;\mathbb R / \mathbb Z) \arrow[rd] \arrow[rr, "-\beta"] &                          & H^{p}(X;\bZ) \arrow[rd, "\rho"] \arrow[ru]           &    \\
H^{p-1}(X ;\bR)  \arrow[ru] \arrow[rd] &                          & {\diff{H}}^{p}(X) \arrow[rd, "R"] \arrow[ru, "I"] &                          & H^{p}(X ;\bR)  \\
                    & \frac{ \Omega^{p-1}({X})}{\Omega^{p-1}(X)_{\cl,\bZ}} \arrow[ru,"a"] \arrow[rr, "d"] &                          & \Omega^p(X)_{\cl,\bZ} \arrow[ru, "r"] \arrow[rd]&\\
 0\arrow[ru]   &&&& 0
    \end{tikzcd}
$$
    \caption{The hexagon diagram for the differential cohomology theory.\label{fig:diag:E}}
\end{figure}
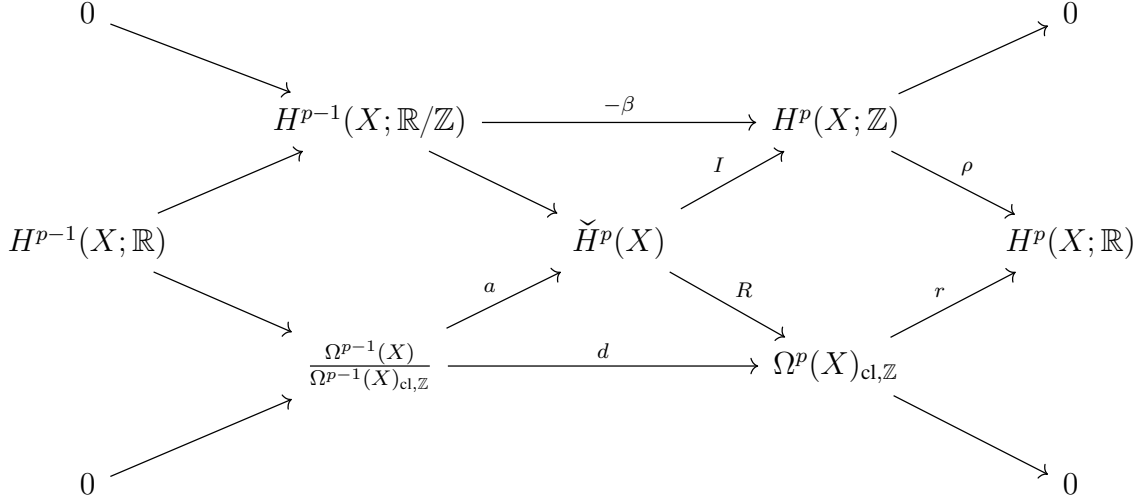

The Abelian group $\check H^p(X)$ of degree-$p$ differential cohomology classes does exactly fit this purpose.
Namely, it fits in the following commuting hexagon diagram
shown in Fig.~\ref{fig:diag:E}, 
where the diagonals are short exact sequences.

Here, the map $\rho$ is given by tensoring by $\bR$,
and the map $r$ is given by taking the de Rham cohomology class,
\begin{equation}\label{eq:rm}
    r: \Omega^p(X;\bR)_{\cl,\bZ}\to H^p(X;\bR)
\end{equation}
mapping $R(\diff A)$ to $[R(\diff A)]$.  
We have a compatibility condition \begin{equation}
r \circ R = \rho \circ I,
\end{equation}
and \begin{equation}
\rho(H^p(X;\bZ))= r(\Omega^p(X)_{\cl,\bZ}) \simeq H^p(X;\bZ)_\text{free}.
\end{equation}
This also means that
\begin{equation}
\frac{\check H^p(X)}{H^{p-1}(X;\bR/\bZ)+\Omega^{p-1}(X)/\Omega^{p-1}(X)_{\cl,\bZ}}
\simeq H^p(X;\bZ)_\text{free}.
\label{doublequoH}
\end{equation}

We here note that when $p+q=\dim X$,
the multiplication $\check H^p(X)\otimes \check H^{q}(X)\to \check H^{p+q}(X)$
followed by the integration $\check H^{n}(X)\to \check H^0(pt)\simeq \bZ$
gives a $\bZ$-valued pairing, \begin{equation}
\langle\check a,\check b\rangle := \int_X R(\check a)\wedge R(\check b)
=\int_X I(\check a) \cup I(\check b) \in \bZ.
\end{equation}
In this paper, we will use the following subtler pairing more often.

Assume 
$p+q=\dim X+1$ instead.
Then the multiplication $\check H^p(X)\otimes \check H^{q}(X)\to \check H^{p+q}(X)$
followed by the integration $\check H^{n+1}(X)\to \check H^1(pt)\simeq \bR/\bZ$
leads to a perfect pairing which we denote by \begin{equation}
(\check A,\check B) \in \bR/\bZ
\end{equation} for ${\check A}\in \check H^p(X)$ and ${\check B}\in \check H^q(X)$,
which reduces to $\int_X A_{p-1} dB_{q-1}$ when both $\check A$ and $\check B$ are topologically trivial.
This pairing is perfect in the sense that 
\begin{equation}
\int [D\check A] \exp(2\pi i (\check A,\check B)) = \delta(\check B)
\end{equation} where the delta function is with respect to the path integral measure $[D\check B]$.
For more details on the perfectness of the pairing, see Appendix~\ref{app:perfectness}.

This formalism allows us to set up the BF theory in a mathematically precise manner,
with the Euclidean action given by 
\begin{equation}\label{eq:bfh}
S=2\pi i k (\check a,\check b),
\end{equation} where $k$ is an integer.
Here and in the following, we often use Seiberg's convention that the path-integrated fields are written with lowercase letters.
This theory is topological in general, and is invertible when $|k|=1$.
We set $k=1$ in this paper.

This theory can couple to background fields $\check A$ and $\check B$ in the following manner: 
\begin{equation}
S=2\pi i \left[   (\check a,\check b) - (\check a,\check B) - (\check A,\check b) \right]
\label{eq:coho-BF-action}
\end{equation}
which makes sense on a manifold with boundary if we put the boundary condition 
${\check a}|_{\partial X}=0$ and ${\check b}|_{\partial X}=0$.
A standard computation \cite{Hsieh:2020jpj,GarciaEtxebarria:2024fuk} shows that the boundary hosts massless fields
with gauge-invariant field strengths $G_{p-1}$ and $G_{q-1}:=* dG_{p-1}$, satisfying \begin{equation}
d G_{p-1}=R(\check A), \qquad d G_{q-1}=R(\check B).
\label{eq:diff-trivialisation}
\end{equation}
Actually, the equations of motion resulting from \eqref{eq:coho-BF-action} sets 
$I(\check A)=0$ and $I(\check B)=0$ at the boundary, 
i.e.~$\check A$ and $\check B$ are topologically trivial there,
and the conditions \eqref{eq:diff-trivialisation} are part of this condition.

On a manifold without boundary, 
it is fairly easy to integrate out $\check a$ and $\check b$ from \eqref{eq:coho-BF-action}
using the perfectness of the pairing, resulting in \begin{equation}\label{eq:ABdiffH}
S=-2\pi i(\check A,\check B),
\end{equation} with no dynamical field in the bulk.
In this manner, we see that the theory of $p$-form fields with \eqref{eq:diff-trivialisation} has an anomaly 
described by $(\check A,\check B)$.
This is a precise version of the Green-Schwarz mechanism.

A small variant is to take the bulk action to be \begin{equation}
S=2\pi i \left[   (\check a,\check b) - (\check A,\check b) + W(\check a)\right],
\label{eq:actionA}
\end{equation}
where $W(\check A)$ is an arbitrary invertible phase for ${\check A}\in \check H^p(X)$.
In this case, integrating over $\check a$ is hard,
but the integral over $\check b$ is still easy to perform, which simply sets $\check a=\check A$ in the bulk,
resulting in \begin{equation}
S=2\pi i W(\check A).\label{eq:actionB}
\end{equation}
With a boundary, we still have \begin{equation}
d G_{p-1}=R(\check A) \label{eq:mod-b}
\end{equation} and  $\check A$ is topologically trivialised at the boundary.
The presence of $W(\check A)$ in the action makes the condition for $d*G_{p-1}$ harder to come by.
In any case, this construction shows that a form field satisfying \eqref{eq:mod-b} can have an arbitrary anomaly $W(\check A)$,
as was derived in \cite{Kobayashi:2019lep} using \eqref{eq:actionB} 
and in \cite{Lee:2022spd} using \eqref{eq:actionA}.

\subsection{With  differential K or KO theory} \label{subsec:BF_K}

\begin{figure}
    \begin{tikzcd}
    0 \arrow[rd]  &&&& 0
\\
    & E^{p-1}(X;\mathbb R / \mathbb Z) \arrow[rd] \arrow[rr, "-\beta"] && E^{p}(X) \arrow[rd, "\rho"] \arrow[ru]  &
\\
    H^{p-1}_E(X;\bR)  \arrow[ru] \arrow[rd] && {\diff{E}}^{p}(X) \arrow[rd, "R"] \arrow[ru, "I"] && H_{E}^{p}(X;\bR)  
\\
    & \frac{ \Omega^{p-1}_E({X})}{\Omega^{p-1}_{E}(X)_{\cl,\bZ}} \arrow[ru,"a"] \arrow[rr, "d"] && \Omega^p_{E}(X)_{\cl,\bZ} \arrow[ru, "r"] \arrow[rd]&
\\
    0\arrow[ru]  &&&& 0
    \end{tikzcd}
    \caption{The hexagon diagram for generalised differential cohomology theory $\check E$.
    We discuss the case $E=K$ and $E=KO$ more explicitly.
    \label{fig:diag:EE}}
\end{figure}
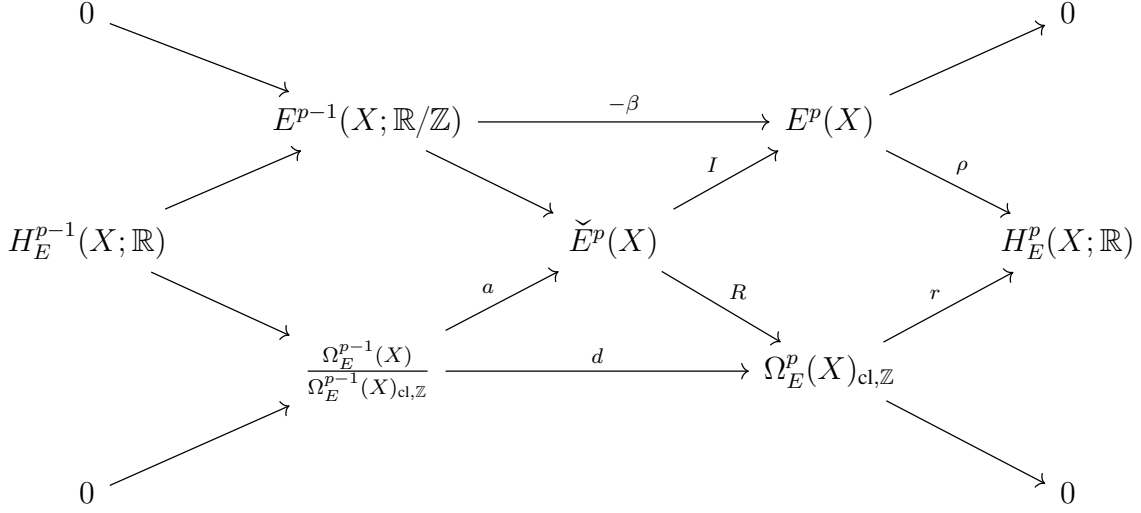

The story given above can be repeated with little change 
when we use any generalised differential cohomology. The differential $E$ cohomology group ${\diff{E}}^{p}(X)$ is a differential extension of the $E$ cohomology group $E^{p}(X)$. It is defined by the same commuting hexagon diagram as above, except that most occurrences of $H$ are replaced by $E$,
see Fig.~\ref{fig:diag:EE}.

\subsubsection{Defining features of differential K and KO theory classes}
In the remainder of this section, we focus on cases where the underlying generalised cohomology theory $E$ is either K theory or KO theory.
When there is a need to refer to both theories simultaneously, we use $E$ to stand for $K$ or $KO$.
In more detail, 
we first set
    \begin{equation}
    \Omega^p_{K}(X):= \bigoplus_n \Omega^{p+2n}(X), \qquad
    \Omega^p_{KO}(X):= \bigoplus_n \Omega^{p+4n}(X).
    \end{equation} 
We let $\Omega^p_{E}(X)_\cl$ to be the subspace of closed forms, and
we consider  the corresponding de Rham cohomology groups \begin{equation}
    H^p_{K}(X;\bR):= \bigoplus_n H^{p+2n}(X;\bR),\qquad
    H^p_{KO}(X;\bR):= \bigoplus_n H^{p+4n}(X;\bR).
\end{equation} 
The map \begin{equation}
r: \Omega^p_{E}(X)_\cl \to  H^p_{E}(X;\bR)
\end{equation} is simply the map to take the de Rham cohomology class,
the map $\rho$ is defined as the Chern character or the Pontryagin character\begin{equation}
           \rho= \ch : K^{p}(X) \to H^p_{K}(X), \qquad
            \rho=\ph : KO^{p}(X) \to H^p_{KO}(X),
\end{equation}
and 
the map $R$ is defined as the differential version of these maps $\ch$ or $\ph$,\begin{equation}
           R= \ch : \check K^{p}(X) \to \Omega^p_{K}(X)_\cl, \qquad
            R=\ph : \check{KO}^{p}(X) \to \Omega^p_{KO}(X)_\cl.
\end{equation} 
These maps satisfy the compatibility \begin{equation}
r\circ R = \rho \circ I \label{compati}
\end{equation}
Please see    Appendix \ref{app:expansions} for the explicit expression of $\ch$ and $\ph$.

In the hexagon diagram, to make the diagonal from the top left to the bottom right 
a short exact sequence,
we define \begin{equation}
	\Omega^p_{E} (X)_{\cl,\bZ} := \Im R \subset \Omega^p_{E}(X)_\cl,
\end{equation}
which is the subspace of forms whose periods are quantised by K or KO theory.
It is known that \begin{equation}
r (\Omega^p_{E} (X)_{\cl,\bZ} ) = \rho(E^p(X)) \subset H^p_{E}(X;\bR).
\end{equation}
This common subspace is isomorphic to the free quotient $E^p_\text{free}(X)$ of $E^p(X)$.
In other words, we have \begin{equation}
\frac{\check E^p(X)}{ E^{p-1}(X;\bR/\bZ) +  \Omega^{p-1}_{E}(X)/\Omega^{p-1}_{E}(X)_{\cl,\bZ}}
\simeq E^p(X)_\text{free}.
\label{doublequoK} 
\end{equation}

Similarly to the case of differential ordinary cohomology, a field $\breve A$ valued in 
the differential generalised cohomology theory $\diff E$  has the following features:
\begin{itemize}
    \item The map $I$ extracts its underlying topological $E$ theory class.
    \item It has a field strength $R(\diff A)$ valued in $\Omega_{E}^p(X)_{\cl,\bZ}$.
    \item A flat field can be identified with an element of $E^{p-1}(X;\bR/\bZ)$.
    \item A topologically trivial field is given by 
    a globally-defined gauge potential taking values in $\Omega_{E}^{p-1}(X)$,
    modulo the gauge transformations taking values in $\Omega_{E}^{p-1}(X)_{\cl,\bZ}$.
    \item The topologically trivial flat field lives in the space $\frac{ \Omega^{p-1}_{E}({X})_\cl}{\Omega^{p-1}_{E}(X)_{\cl,\bZ}}$.
\end{itemize}

Before proceeding, let us discuss a particular case of generalised differential cohomology class
whose features can already be determined just from the axioms.
Let us consider $\check E^1(S^1)$.
We know that there is a short exact sequence \begin{equation}
0 \to \Omega^0_E(S^1)/\Omega^0_E(S^1)_{\cl,\bZ} \xrightarrow{a} \check E^1(S^1) \xrightarrow{I} E^1(S^1) \to 0.
\end{equation} 
Here, $E^1(S^1)\simeq \bZ$.
Furthermore, $\Omega^0_E(S^1) = \Omega^0(S^1)$ is a real-valued function on $S^1$
and $\Omega^0_E(S^1)_{\cl,\bZ}$ is simply a constant integer.
We also have the sequence \begin{equation}
0 \to E^{0}(S^1;\bR/\bZ) \to \check E^1(S^1) \xrightarrow{R} \Omega_E^1(S^1)_{\cl,\bZ} \to 0.
\end{equation}
Here, $E^{0}(S^1;\bR/\bZ) \simeq \bR/\bZ$
and $\Omega_E^1(S^1)_{\cl,\bZ}=\Omega^1(S^1)_{\cl,\bZ}$ is simply the space of one-forms on $S^1$ whose integral is an integer.

Combined, an element of $\check E^1(S^1)$ is simply a function $\phi: S^1\to \bR/\bZ\simeq S^1$.
$I$ extracts the winding number of $\phi$,
and $R$ extracts its differential $d\phi$.
This holds for $E=H$, $K$ and $KO$ uniformly.

Let us now denote the generator of $E^1(S^1)\simeq \bZ$ by $T$,
we also introduce a coordinate $t$ on $S^1$ with $t\sim t+1$.
Take an inverse image of $dt$ by $R$ and call it $\check T$.
Then $I(\check T)=T$.
Note that this definition of $\check T$ depends on the choice of the coordinate function on $S^1$,
and also an ambiguity by $E^0(S^1;\bR/\bZ)\simeq \bR/\bZ$.

\subsubsection{Differential K or KO theory cocycles}
\label{sec:KOcocycle}

We need a more concrete model of differential K or KO theory elements to perform computations.
These are provided by differential K or KO theory cocycles.

We already noted the compatibility condition \eqref{compati}.
Let us study its consequence more explicitly when $p=0$.
Given $\diff A$, the topological class $I(\diff A)$ is given by  the class of a complex (real) virtual 
bundle $V$ for K (KO) theory, so that $I(\diff A)=[ V]$. 
We give a connection $B$ on $V$. Then 
\begin{equation}
[\rho(B)] = [R(\diff A)] \in H^p_{E}(X;\bR),
\end{equation}
where $\rho=\ch$ for $E=K$ and $\rho=\ph$ for $E=KO$.
This equality of the de Rham cohomology class means 
that there is $\underline C\in \Omega^{-1}_{E}(X)$ such that 
\begin{equation}
R(\diff A)= \rho(B)+ d\underline{C} .
\label{uC}
\end{equation}
It turns out more convenient for us to parameterise $R(\diff A)$ with $C\in \Omega^{-1}_{E}(X)$
such that the following relation \begin{equation}
\sqrt{\hat A(X)}\, R(\diff A)= \sqrt{\hat A(X)}\,\rho(B)+ dC 
\label{C}
\end{equation}
is satisfied, where ${\hat{A}(X)}$ is the A-roof genus. 

Note that one can introduce many connections \(B\) on $V$.   For two connections $B_0$ and $B_1$, let $B'(t)$ for $0 \leq t \leq 1$ be a homotopy between them. Then, $(B_0,C_0)$ and $(B_1,C_1)$ give the same $R(\diff A)$ if 
\[\label{eq:gaugeeq}
\int_{[0,1]} \sqrt{\hat{A}(X)} \, \rho(B') = C_0 - C_1 .
\]  
The left-hand side of this expression is commonly referred to as the Chern-Simons form, denoted as  
\[
CS(B') := \int_{[0,1]} \sqrt{\hat{A}(X)} \, \rho(B') .
\]  
Applying the exterior derivative, one obtains the transgression formula  
\[
d \, CS(B') =\sqrt{\hat{A}(X)} ( \rho(B_1) - \rho(B_0) ) .
\]  
We call $(B_0,C_0)$ and $(B_1,C_1)$ \emph{gauge equivalent} if they are related in the manner just described. We refer to  quantities invariant under this equivalence relation as \emph{gauge invariant}.

Combining the above considerations, we can define an element of ${\diff{E}}^{0}(X)$ as 
an equivalence class of triplets,
\[
    \diff A = (V, B, C) / \text{gauge equivalence}.
\]  
This formulation is known as the cocycle model of differential K- or KO-theory, originally introduced by Hopkins and Singer \cite{Hopkins:2002rd}.

For two differential cocycles $ (V_a, B_a, C_a)$ and $ (V_b, B_b, C_b)$,
we define its product to be 
\begin{multline}
(V_a, B_a, C_a) \cdot (V_b, B_b, C_b) :=
\Bigl(V_a \otimes V_b, (B_a \otimes 1 + 1 \otimes B_b),\\
     \left(\frac{1}{\sqrt{\hat{A}}} C_a \wedge dC_b + C_a \wedge \rho(B_b) + C_b \wedge \rho(B_a)\right) \Bigr) .
     \label{pro}
\end{multline}
It can be  shown that if $(V_a, B_a, C_a)\sim (V'_a, B'_a, C'_a)$,
then \begin{equation}
(V_a, B_a, C_a)\cdot  (V_b, B_b, C_b) \sim
 (V'_a, B'_a, C'_a)\cdot  (V_b, B_b, C_b),
\end{equation}
and similarly for $(V_b,B_b,C_b)$.
This means that the product of two differential classes $\check a$ and $\check b$
is well-defined,
making $\check K^0(X)$ and $\check{KO}{}^0(X)$ into a ring.

\subsubsection{Integration}

Take a fiber bundle $p: X \to Y$ whose fibers are closed manifolds. 
We have the differential K-theory or KO-theory integration map
\[
p_* : {\diff{E}}^{0}(X) \to {\diff{E}}^{-r}(Y) 
\]
with $r = \dim X - \dim Y$,
if the fibers have  $\text{spin}^c$ structure for K theory
or  a spin structure  for KO theory a spin structure.
We discuss two particular cases.

First, for $\dim X=8n+4$,  the differential integration map along \( {p_X}_*: X \to \text{pt} \) gives the homomorphism  
\[
{p_X}_* : {\diff{E}}^{0}(X) \to {\diff{E}}^{-8n-4}(\text{pt})  \cong \mathbb{Z} .
\]
It simply takes the index of the bundle, and is given for $\check A=(V,B,C)$ by \begin{equation}
{p_{X}}_*(\check A)=\int_X \hat A(X) \rho(B) = \kappa \int_X \hat A(X) R(\check A).
\end{equation}
where $\kappa=1$ for $E=K$ and $\kappa=1/2$ for $E=KO$.
This difference in the prefactor is because the map $KO^{-4}\cong \bZ$ to $K^{-4}\cong \bZ$ is given by multiplication by 2.

Second, for $\dim X=8n+3$,  the differential integration map along \( {p_X}_*: X \to \text{pt} \) gives the homomorphism  
\[
{p_X}_* : {\diff{E}}^{0}(X) \to {\diff{E}}^{-8n-3}(\text{pt})  \cong \mathbb{R}/\mathbb{Z} .
\]
For K theory it is given by
\[ \label{eq:holK}
    \chi^\bC(\diff A):={p_X}_*(V, B, C)=
    \eta \big(\mathcal{D}_X(B)\big)
    +\int_X\sqrt{\hat A}\wedge C .
\]
For KO theory, it is defined as 
\[ \label{eq:holKO}
    \chi^\bR(\diff A):={p_X}_*(V, B, C)=
    \frac{1}{2}\eta \big(\mathcal{D}_X(B)\big)
    +\frac{1}{2}\int_X\sqrt{\hat A}\wedge C ,
\]
which differs from the K theory case by a factor of $1/2$.
The reason is the same as above:
as the map $KO^{-4}\cong \bZ$ to $K^{-4}\cong \bZ$ is given by multiplication by 2,
the homomorphism from $\diff{KO}{}^{-3} \cong \bR/\bZ$ to
$\diff K{}^{-3} \cong \bR/\bZ$  involves a factor of $2$.

The quantity $\chi(\diff A)$  
is gauge invariant,
i.e.~it is independent of the cocycle representative for $\diff A$. 
This invariance follows from the APS index theorem as we now show for the case of KO theory. 
More concretely, we let $Z = X \times [0, 1]$ 
and use the coordinate $t$ to parameterise the interval $[0,1]$.
We regard the homotopy $B'$ between $B_0$ and $B_1$ 
as a connection on $Z$.
We also consider $C$ on $Z$ so that $dt \wedge \frac{\partial}{\partial t} C=\sqrt{\hat A(X)}\rho(B')$, realising
\eqref{eq:gaugeeq}.

On one hand, the APS index theorem for the $\eta$ invariant  gives:
\[
    \eta(\mathcal{D}_{X}(B_1)) - \eta(\mathcal{D}_{X}(B_0)) \equiv \int_Z \hat{A}(X) \ph(B') \mod 2.
\]
On the other hand, the difference  of $\sqrt{\hat A(X)} C$ on two boundaries can be re-written using Stokes' theorem as:
\begin{align}
 \int_X \sqrt{\hat{A}(X)} (C_1-C_0) &= \int_Z d_Z(\sqrt{\hat{A}(X)} C) \\
 &= \int_{[0, 1]} \underbrace{\int_X d_X (\sqrt{\hat{A}(X)}  C )}_{\text{vanish since } \partial X = 0} + \int_Z d_t (\sqrt{\hat{A}(X)} C) \\
 &= \int_Z \sqrt{\hat{A}(X)}  (\sqrt{\hat{A}(X)} \ph({B}')),
\end{align}
so we see that our holonomy function (\ref{eq:holKO}) for a differential KO class is indeed invariant under $(V, B_0, C_1) \sim (V, B_1, C_1)$.

For direct sums of bundles $V_1, V_2$ on the same manifold, it is straightforward to get linearity
\[
    \chi(\breve{a}_1 + \breve{a}_2) = \chi(\breve{a}_1) + \chi(\breve{a}_2)
\]
since both $C$ and the $\eta$-invariant are linear for direct sum of real vector bundles, and the $\sqrt{\hat{A}(R)}$ factor holds fixed on the same base manifold. Moreover, the reality condition gives us a $\mod$-2 identity as opposed to a $\mod$-1 identity, since the eigenvalue of the Dirac operator in $Z$ always comes in pairs. From this we know that the factor $\frac{1}{2}$ in the expression of $\chi(\breve{a})$ for a $\breve{KO}$ will still give $2\pi i \chi(\breve{a})$ as a phase that depends continuously on the fields.

Showing the same invariance in K theory under $(V, B_0, C_0) \sim (V, B_1, C_1)$ is essentially identical, once we replace $\ph$ with $\ch$ and mod 2 with mod 1.

\subsubsection{The $\bR/\bZ$-valued pairing and BF theories}

Using the holonomy function, we can define a bilinear pairing $(-,-)$ valued in $\bR/\bZ$ of two differential $K(O)$ theory classes as
\[\label{eq:pairingab}
    (\breve{a}, \breve{b}) := \chi(\breve{a} \cdot \breve{b}^*),
\]
where $b^*$ is a differential cocycle on the complex conjugate bundle $V^*$ of $V$, 
given by $(V^*, B^*,C^*)$. Here $C\mapsto C^*$ is the conjugation defined by \begin{equation}
C = \sum_n C_{2n-1} \mapsto C^*=\sum_n (-1)^n C_{2n-1}.
\end{equation} 
We note that for KO theory, $\check b=\check b^*$. 
This pairing is perfect, as we review in appendix \ref{app:perfectness}.
The linearity in each variable follows from the linearity of $\chi$,
and the symmetric-ness of the pairing can be checked by using the fact that $\chi$ satisfies $\chi(\check a)=\chi(\check a^*)$.

More explicitly, for two differential cocycles
$\diff a = (V_a, B_a, C_a)$ and $\diff b = (V_b, B_b, C_b)$, we have \begin{equation}
(\breve{a},\breve{b}) = \cdots + \int C_a d C_b^* \cdots \label{Kpair}
\end{equation} for  K theory and \begin{equation}
(\breve{a},\breve{b}) = \cdots + \frac12 \int C_a d C_b \cdots  \label{KOpair}
\end{equation} for KO theory,
as can be seen  using \eqref{pro}. 
The factor of $\sqrt{\hat A}$  was introduced in \eqref{C}, relative to a more naive definition \eqref{uC},
in order to eliminate a factor of $\hat A$ appearing in \eqref{Kpair} and \eqref{KOpair}.

One may define the BF theory based on differential K or KO theory in terms of the pairing, analogous to the construction for differential ordinary cohomology in (\ref{eq:bfh}), as  
\begin{equation}  
    S = 2\pi i (\diff a, \diff b).  
\end{equation}  
Then the same discussions apply here, \textit{mutatis mutandis}.  

\subsubsection{The $\bZ$-valued pairing and the quantisation of the fluxes}
\label{sec:normalization}
Before proceeding, let us make a comment on the interpretation $\Omega^p_{E}(X)_{\cl,\bZ}$,
which was introduced as the image of $R$ applied to $\diff{E}{}^p(X)$ in the general case.
For the ordinary cohomology $E=H$,
this space $\Omega^p_{H}(X)_{\cl,\bZ}$ has another interpretation as the space of 
closed forms with integral periods.
This interpretation arose because the integration of an integral cohomology class
on an integral homology class is an integer.

Let us discuss an analogous interpretation for the case $E=KO$.
Given a class $\diff a \in \diff{KO}{}^p(X)$
and a spin submanifold  $\iota: M\hookrightarrow X$, we can pull-back $\diff a$ to $M$.
The differential push-forward along $M\to pt$, \begin{equation}
	\diff{KO}{}^p(M)\to \diff{KO}{}^{p-\dim M}(pt),
\end{equation} then gives an integer when $p-\dim M=0$ or $4$ mod 8.
In terms of the curvature $R(\diff a)$, it is given by \begin{equation}
\kappa\int_M \hat A(M) \iota(R(\diff a)) \in \bZ, 
\end{equation}
where $\kappa$ is $1$ or $\frac12$ depending on whether $p-\dim M$ is $0$ or $4$ mod $8$.
This is what we mean by the `integrality' of the elements in $\Omega^p_{KO}(X)_{\cl,\bZ}$.

Note that this factor $\kappa=1/2$ is due to the pseudoreality of the spin bundle in dimension $8k+4$.
This leads to the quantisation law 
\begin{equation}
\int_{M_4} dC_3+\cdots \in 2\bZ,\qquad
\int_{M_8} dC_7 +\cdots \in \bZ,\qquad
\end{equation} and so on. 
This is different from a more conventional normalisation of $C_{p-1}$  in the literature, where \begin{equation}
\int_{M_p} dC_{p-1}{}^\text{more conventional} +\cdots  \in \bZ.
\end{equation}
This means that \begin{equation}
C_3{}^\text{ours} = 2 C_3{}^\text{more conventional},\qquad
C_7{}^\text{ours} =  C_7{}^\text{more conventional}.
\label{eq:relative-factors}
\end{equation}

Using the Poincar\'e dual cohomology class $[M]\in H^{\dim X-p}(X;\bZ)$, 
this translates to the integrality of \begin{equation}
\kappa \int_X \left(\sqrt{ \frac{\hat A(M)} {\hat A(N)} }[M]\right) 
\wedge \left(\sqrt{\hat A(X)} R(\diff a)\right),
\end{equation} where $N$ is the normal bundle to $M$ within $X$, 
so that $\hat A(N)\hat A(M)=\hat A(X)$.
Here we included a factor of $\sqrt{\hat A(X)}$ multiplying the elements of $\Omega^p_{KO}(X)_{\cl,\bZ}$.
Then this is paired, at the level of ordinary de Rham cohomology,
with the `brane current' $\sqrt{\hat A(M)/\hat A(N)} [M]$,
which is the well-known form of the RR-charge of the D-branes.
Therefore the integrality encoded in the definition of $\Omega^p_{KO}(X)_{\cl,\bZ}$
is exactly what we expect from the physics of D-branes.
See e.g.~\cite{Moore:1999gb} for more on this point.

\section{Invertible Abelian Chern-Simons theories} \label{sec:CS}

So far, we have seen how we can set up a precise version of BF theory
using ordinary and generalised differential cohomology theories.
They are based on two dynamical form fields in the bulk.
Here we move on to discuss Abelian Chern-Simons theories,
which have only one dynamical form field in the bulk.
\subsection{Quadratic refinements and invertible Abelian Chern-Simons theories}
\label{subsec:inv}
When $\dim X=4m-1$ and $\check c\in \check H^{2m}(X)$, 
one can consider the bulk action \begin{equation}
S = 2\pi i k' (\check c,\check c)
\end{equation} for an integer $k'$, 
which reduces to $2\pi i k'\int_X cdc$ when the fields are topologically trivial.
This is therefore (a precise version of) Abelian Chern-Simons theory,
but turns out not  to be an invertible theory and is not useful in anomaly cancellation.

To have an invertible theory we need an action which reduces to $2\pi i \frac12 \int_X cdc$ 
when the fields are topologically trivial.
For this we need a quadratic refinement $Q(\check C)$, satisfying the following properties:
\begin{equation}
Q(\check A+\check B)-Q(\check A)-Q(\check B)+Q(0)=(\check A,\check B),
\label{eq:qr}
\end{equation}
where we allow $Q(0)$ to depend on the background metric etc., for later convenience. 
We note that a more general-looking formula \begin{equation}
Q(\check A+\check B+\check C)-Q(\check A+\check C)-Q(\check B+\check C)+Q(\check C)=(\check A,\check B)
\end{equation}
follows from \eqref{eq:qr}.
The discussion in the rest of this subsection works with $\check H$, $\check K$ and $\check{KO}$,
as long as the degree of $\check C$ and the dimension of $X$ is appropriate 
such  that $(\check C,\check C)$ and its quadratic refinement are defined.

Let us first show that the bulk theory \begin{equation}
S= 2\pi i k Q(\check a) \label{eq:kQ}
\end{equation} is invertible when $k=1$. To show this, we consider another theory \begin{equation}
S'= - 2\pi i  k Q(\check b)
\end{equation}  and show that the theory $S+S'$ is completely trivial, when $k=1$.
For this, we set $\check a=\check a'+\check b$, and rewrite \begin{equation}
S+S'=2\pi i k ( Q(\check a') - Q(0) + (\check a',\check b) ).
\end{equation}
When $k=1$, the perfectness of the pairing allows us to integrate over $\check b$ and set $\check a'=0$,
resulting in $Q(0)-Q(0)=0$, 
so the partition function of the combined system is simply $1$ on arbitrary manifolds.
This means that the bulk theory $S$ and the bulk theory $S'$ are inverse to each other,
and the bulk theory $S$ is invertible.

Before proceeding, let us discuss a couple of general properties of quadratic refinements
of a perfect pairing.
Let $Q(\check C)$ be a quadratic refinement of $(\check A,\check B)$.
Then it is easy to check that 
\begin{equation}
Q'(\check C):=Q(-\check C) = (\check C,\check C)-Q(\check C)+2Q(0)
\end{equation}
is also a quadratic refinement, where we used \eqref{eq:qr} with  $\check A=-\check B=\check C$
in \eqref{eq:qr} to show the second equality.
The difference of two quadratic refinements $Q_0(\check C)$ and $Q_1(\check C)$ with $Q_0(0)=Q_1(0)$ is linear in $\check C$.
Applying this to $Q(\check C)$ and $Q'(\check C)$,
we find $Q'(\check C)-Q(\check C)$ is also linear in $\check C$.
By perfectness, there is an element $\check\nu$ canonically associated to $Q$ such that \begin{equation}
Q'(\check C)=Q(\check C)+(\check C,\check \nu)\label{eq:nu}
\end{equation} or in other words \begin{equation}
2Q(\check C) = (\check C,\check C) - (\check C,\check \nu) + 2Q(0).
\label{eq:nudef}
\end{equation}
Therefore, morally speaking, $Q(\check C)$ is not simply $\frac12 (\check C,\check C)$
but has a linear and a constant piece.

\subsection{Path integral over $Y\times S^1$}
\label{sec:bulk-path-integral}

Here we plan to quantise the aforementioned classical quadratically-refined action. 
We choose to perform this analysis on the manifold of the form $Y \times S^1$, since later when we discuss manifolds with boundary, we have the option to cut $S^1$ and make it an open interval $[0, 1]$, which will then have two boundaries, each isomorphic to one copy of $Y$. In addition, this analysis will also be relevant when we study the circle reduction of the quadratic refinement in Sec.~\ref{sec:reduction}. 

The wavefunction of this invertible theory was investigated in detail in \cite{Belov:2006jd,Belov:2006xj}.
Let us recall the parts of their arguments which are relevant for us in our language.
We consider the theory \eqref{eq:kQ} with $k=1$ on the space $X=Y\times S^1$.
In the case of the ordinary Chern-Simons theory in three dimensions,
the path integral localises to \emph{flat} connections on $Y$. It is a classical result that such a quantisation eventually involves a subtlety concerning the path integral over $F \cong H^1(Y, \bZ_2)$, which is interpreted as choosing a spin structure. 

Here we do the analogous analysis for the differential version of generalised cohomology theory, which we denote as $\check{E}^p(X)$. In our generalised case, we will see that the path integral localises on
(a torsor over) topologically trivial and flat fields on $Y$. 
As this subsection  will be lengthy,
here we provide the detailed summary of our procedure.

 Our strategy is to decompose the field space depending on whether the field configuration 
 $\check A\in \check E^p(X)$
 has a `leg' along $S^1$ or not. 
We use the symbol $t$ as the coordinate along $S^1$,
and $T$ as a generator of $E^1(S^1)=\bZ$.\footnote{%
This holds as long as $E^1(pt)=0$, which holds for $E=H$, $K$ or $KO$.
We assume this property throughout in this paper.} 
The curvature $R(\check A)$ is a differential form,
and takes values in a space with a direct sum decomposition 
into the $dt$-having part and  the $dt$-independent part. 
The topological class $I(\check A)$ takes values in $E^p(Y\times S^1) \simeq E^p(Y) \oplus E^{p-1}(Y)T$,
which again has a nice decomposition
 into the $T$-having part and the $T$-independent part.
 
Denote by $K$ the subspace of $\check E(X)$ generated by
(the preimage by $R^{-1}$ of) the $dt$-having parts of $R(\check A)$ and
(the preimage by $I^{-1}$ of) the $T$-having parts of $I(\check A)$.
This subspace $K$ can be called the $dt$- or $T$-having part of $\check E(X)$.

We organise the path integral over the entirety of $\check E(X)$
as a combination of the path integral over $K$ and the path integral over the coset $\check E(X)/K$. 
More specifically,
\begin{itemize}
\item We first consider the integral over \emph{flat}  part of $K$, which we call $V$,
in Sec.~\ref{3.2.1}.
\item We then consider the integral over \emph{topologically trivial}  part of $K$, which we call $W$,
in Sec.~\ref{subsubsec:dt}.
\end{itemize}
Each path integral produces a delta function fixing the topological class on $Y$ to be
$\mu_\text{top}\in E^{p}(Y)$
and the curvature on $Y$ to be $\mu_\text{diff}\in \Omega^p_E(Y)_{\cl,\bZ}$, respectively,
which are compatible when sent to $H^p_E(Y)$.
We will see that they satisfy $2\mu_\text{top}=I(\check \nu)$
and $2\mu_\text{diff}= R(\check \nu)$,
where $\nu$ was introduced in \eqref{eq:nu}.

After these two steps, 
the still-to-be integrated parts of $K$ are over (the preimage by $I^{-1}$ of)
$U T$, where $U=E^{p-1}(Y)_\text{free}$.
We also need to perform the path integral over $\check E(X)/K$,
not yet restricted by the delta functions generated so far.
This is a path integral over a motion parametrised by $t\in S^1$
in the moduli space of topologically trivial flat fields on $Y$.

It turns out that, out of the path integral over $U T$, 
a sum over $2U T$ is possible,
leaving the sum over $U/2U$.
This sum over $2UT$
localises the integral over the motion in the moduli space of topologically-trivial flat fields
 to a sum over its 2-torsion points.
The final and the most subtle part of the sum is then a double sum over \begin{equation}
U/2U
\end{equation}
and
\begin{equation}
\text{2-torsion points of the moduli space of topologically-trivial flat fields}.
\end{equation}
These two Abelian groups are actually both naturally isomorphic to \begin{equation}
E^{p-1}(Y)_\text{free}\otimes \bZ_2,
\end{equation}
which of the form $(\bZ_2)^r$ for some $r$,
and the sum involves a quadratic refinement on it.
This will be done in Sec.~\ref{3.2.3},
and this is where the spin structure appears as the quadratic refinement of $H^1(Y;\bZ_2)$
in the case of the standard level-1 spin Abelian Chern-Simons theory in three dimensions.

\subsubsection{Path integral fixes the topological class on $Y$}
\label{3.2.1}
The space of fields $\check E^p(X)$ contains a subspace $E^{p-1}(X;\bR/\bZ)$.
We have an injective map $E^{p-2}(Y;\bR/\bZ)\to E^{p-1}(Y\times S^1;\bR/\bZ)$ 
given by tensoring by the generator $T$ of $\tilde E^1(S^1)\simeq \bZ$.
Using this we regard $V:=E^{p-2}(Y;\bR/\bZ)$ as a subspace of $\check E^p(X)$.

We decompose the path integral into two steps, first along the orbit by the addition of $V$,
and then over the coset $\check E^p(X)/V$.
To analyse the first step, take a basepoint $\check c \in \check E^p(X)$,
and consider $\check c  + \check T \check\alpha  \in \check E^p(X)$,
where $\check\alpha \in  E^{p-2}(Y;\bR/\bZ)$.
We now show that the function $Q(\check c  + \check T\check\alpha )$ is affine linear in $\check\alpha \in \check E^{p-2}(Y;\bR/\bZ)$.
For this, note that \begin{equation}
Q(\check c  + \check T(\check\alpha  +\check\beta) ) 
-Q(\check c  + \check T\check\alpha  ) 
-Q(\check c  + \check T\check\beta) 
+Q(\check c ) 
= (\check T\check\alpha ,\check T \check\beta)
= \int_{Y\times S^1} \check T^2\check\alpha \check\beta.
\end{equation}
We note that $\check T^2$ is a class in $\check E^2(S^1)$ and $\check\alpha \check\beta$ is in $\check E^{2(p-1)}(Y)$,
and therefore \begin{equation}
= (\int_{S^1} \check T^2 ) (\int_Y \check\alpha  \check\beta)
\end{equation} where the first factor is the $\bR/\bZ$ valued pairing 
and the second factor is a $\bZ$-valued pairing.\footnote{%
For example, when $\dim Y=10$, $E=KO$ and $p=0$, $\check\alpha \in \check{KO}{}^{-1}(Y)$
and the integral is in $\check{KO}{}^{-12}(pt)\simeq \bZ$.
}
In particular $\int_Y \check\alpha \check\beta$ can be computed with the differential forms $R(\check\alpha )$ and $R(\check\beta)$. 
But we assumed that $\check\alpha $ and $\check\beta$ are from $E^{p-2}(Y;\bR/\bZ)$ 
and therefore $R(\check\alpha )=R(\check\beta)=0$, 
meaning further that $\int_Y\check\alpha \check\beta=0$,
meaning that $Q(\check c +T\check\alpha )$ is indeed affine linear in $\check\alpha $.
From the perfectness of the pairing, there is an element $M(\check c ) \in E^{p}(Y)$ such that
\begin{equation}
Q(\check c +T\check\alpha )=Q(\check c )+(M(\check c ),\check\alpha )_Y.
\label{eq:alphaintegral}
\end{equation}

To determine $\check c $ dependence of $M(\check c )$,
we use the quadratic refinement property again to find
\begin{equation}
(M(\check c) - M(0),\check\alpha )_Y=(\check c,T\check\alpha )_X
=(\pi(I(\check c)),\check\alpha)_Y \label{eqn:X_Y_pairing}
\end{equation}
where  $\pi:E^p(X)\to E^p(Y)$ is the projection.
Therefore, \begin{equation}
M(\check c )= \pi(I(\check c )) -\mu_\text{top}
\end{equation} where $\mu_\text{top} := -M(0)\in E^p(Y)$.\footnote{%
As the quadratic refinement on $X=Y\times S^1$ in general depends on the choice of the spin structure along $S^1$,
it is not immediate that $\mu_\text{top}$ thus defined is independent of this choice.
To show its independence, let us write 
$Q^\text{NS}(T\check \alpha)=(\mu_\text{top}^\text{NS},\check \alpha)_Y$ and 
$Q^\text{R}(T\check \alpha)=(\mu_\text{top}^\text{R},\check \alpha)_Y$,
where the superscript specifies the spin structure around $S^1$.
To compare $Q^\text{NS}(T\check\alpha)$ and $Q^\text{R}(T\check \alpha)$,
we use the discussion of Appendix~\ref{app:EKO}, in particular \eqref{eq:Wfig}--\eqref{YYYY},
but with $S^1_1$ and $S^1_2$ in the R-sector while $S^1_3$ in the NS sector,
furthermore taking $\check\beta=0$ there.
This shows that $Q^\text{NS}(T\check \alpha)-Q^\text{R}(T\check \alpha)=Q^\text{R}(0)$,
which is a constant independent of $\alpha$.
Therefore $\mu_\text{top}^\text{NS}=\mu_\text{top}^\text{R}$.
}

Now, we integrate over $\alpha$, under which the action changes by \eqref{eq:alphaintegral}.
This means that the integral localises to $M(\check c)=0$,
i.e.
\begin{equation}
\pi(I(\check c))=\mu_\text{top}.\label{eq:top-cond}
\end{equation}

Before proceeding, we show that $2\mu_\text{top} = \pi(I(\check\nu))$. 
Taking $\check{c}=T\check\alpha $ in (\ref{eq:nu}), we have
\[
2Q(T\check\alpha)=-(T\check\alpha,\check \nu)_X+2Q(0) .
\]
Now, setting $\check{c}$ to 0 in (\ref{eq:alphaintegral}) and multiplying by a factor of 2, we have
\[
2Q(T\check\alpha )=2Q(0)+(2M(0),\check\alpha )_Y .
\]
Comparing the last two equations and using the last identity in (\ref{eqn:X_Y_pairing}), we get the desired result that $2\mu_\text{top} = \pi(I(\check\nu))$ using perfectness of the relevant pairing. 

\subsubsection{Path integral fixes the field strength on $Y$}
\label{subsubsec:dt}
We now repeat the analysis above, by taking a different subspace $W\subset \check E^p(X)$.
The path integral can similarly decomposed into two steps,
first along the orbits under the shift by elements of $W$, and then along the coset $\check E^p(X)/W$.\footnote{%
More precisely we are doing the integral along $(W+V)/V\subset \check E^p(X)/V$,
and then along the coset $(\check E^p(X)/V) / ((W+V)/V)$. }

To choose $W$, consider the space of topologically trivial fields $\Omega^{p-1}_E(X) / \Omega^{p-1}_{E}(X)_{\cl,\bZ}$.
On $X=Y\times S^1$, the differential forms here can be split into those having $dt$ in it, and those not having $dt$ in it,
where $t$ is the coordinate along $S^1$.
We take $W$ to be the subspace having $dt$ as part of it.

Given a basepoint $\check c\in \check E^p(X)$ and shifts $ \check \alpha = a(\alpha)$
and $\check \beta = a(\beta)$ in $W$,
we can perform a similar analysis as above. We start with
\begin{align}
\begin{split}
& Q(\check c  + a(\alpha) + a(\beta) ) 
-Q(\check c  + a(\alpha)  ) 
-Q(\check c  +   a(\beta)) 
+Q(\check c ) 
= (a(\alpha), a(\beta)) .
\end{split}
\end{align}
For topologically trivial fields $\check A$ and $\check B$,
the pairing is of the form $\int_X \mathcal{A}_E(X) R(\check A) R(\check B)$
for some differential form $\mathcal{A}_E(X)$.
Here,  $R(a(\alpha))$ and $R(a(\beta))$ both contain a leg of $dt$,
so the pairing vanishes. 
This means that  $Q(\check c+a(\alpha))$ is affine linear in $\alpha$,
and therefore we should have the following relation
\begin{equation}\label{eq:diffshift}
Q(\check c+a(\alpha))=Q(\check c)+\int_X (\pi(R(\check c)) - \mu_\text{diff} ) \wedge \alpha.
\end{equation} 
Here, $\pi $ is the projection of the differential forms not having $d$ in it,
$\mu_\text{diff}$ is an element of $\Omega^p_{E}(Y)_{\cl,\bZ}$.
Performing the path integral over $\alpha$, we see that the integral localises to the configuration 
\begin{equation}
\pi(R(\check c))=\mu_\text{diff} \label{eq:diff-cond}
\end{equation} at every $t\in S^1$.

We note that $\mu_\text{diff}$ satisfies $2\mu_\text{diff}=R(\check\nu)$,
where $\check\nu$ was introduced in \eqref{eq:nu}.
To see this, 
take $\check{c}=a(\alpha)$ in (\ref{eq:nu}) to get
\[
2Q(a(\alpha))=-(a(\alpha),\nu)_X+2Q(0) .
\]
Then, comparing the latter with twice of (\ref{eq:diffshift}) for $\check{c}=0$ gives the desired results using perfectness of the relevant pairing. 
Here $\mu_\text{diff} \in \Omega^p_{E}(Y)_{\cl,\bZ}$ and $\mu_\text{top} \in E^p(Y)$
determine the same element in $E^{p-1}_{E, \mathrm{Free}}(Y;\bZ)$, which follows from commutativity of Figure~\ref{fig:diag:EE}, by starting from $\pi(\check c) \in \check E^{p-1}(Y)$. 

Recall that we saw the relation $2\mu_\text{top}=I(\check \nu)$ above.
Then it is tempting to assume the existence of a preferred differential element $\check \mu$
so that $2\check\mu=\check\nu$, $\mu_\text{top}=I(\check \mu)$, and
$\mu_\text{diff}=R(\check \mu)$.
But the situation is slightly more subtle, as we will see below.

\subsubsection{The rest of the path integral}
\label{3.2.3}

Recall we separated the path integral over  the subspace $K$ of $\check E(X)$,
i.e.~the $dt$- or $T$-having parts
and over the coset $\check E(X)/K$, i.e.~non-$dt$- or $T$-having parts.
As for the integral over $K$,
we just did the path integral for $V$ and $W$.
The only  remaining sum is over the part of $K$ which is neither topologically trivial nor flat.
In other words,
we still have to sum over $U\check T$, where 
\begin{equation}
U=E^{p-1}(Y)_\text{free} = E^{p-1}(Y)/\mathrm{Tors}\ E^{p-1}(Y).
\end{equation}

As for the latter, non-$dt$ non-$T$-having parts, 
the two conditions \eqref{eq:top-cond}
and \eqref{eq:diff-cond} means that 
(once we fix a basepoint solving these two conditions)
the remaining path integral is over a function on $t\in S^1$
taking values in a topologically trivial flat fields on $Y$.

This latter path integral is what leads to the case of the standard 3d Chern-Simons theory 
as the wavefunction on the moduli space of flat connections.
Indeed, for $d=3$, $p=2$ and $E=H$, this is the standard Abelian Chern-Simons theory,
for which the wavefunctions are the higher-genus theta functions,
which are indeed functions on a dual torus, 
parameterising flat $U(1)$ connections on the Riemann surface on which the Chern-Simons theory resides.

Let us now try to sum over  $U\check T$ introduced above.
Given two elements $\alpha \check T,\beta \check T$ from it, it is easy to check that 
\begin{equation}
Q(c+\alpha \check T+\beta \check T)
- Q(c+\alpha \check T)
- Q(c+\beta \check T)
+ Q(c)
= (\int_{S^1} \check T^2) \int_{Y} \alpha\beta .
\end{equation}
Here, $\int_{S^1}\check T^2=\pm1$ due to degree reasons, 
but it is actually $-1$, for $E=H(-;\bZ)$ and $E=KO$, see Appendix~\ref{app:T2}.
Then a sum over $2UT$ is possible.
Due to the perfectness of the pairing,
this sum over $2UT$ restricts the path integral over the motion on the topologically trivial flat 
configurations to a torsor over two-torsion points.
We also have a remaining sum over $(U/2U)T$.

After all these procedures, what remains on the $T$-having part is a sum over $(U/2U)T$.
In contrast, what remains on the non-$T$-having part is a sum over (a torsor over) the 2-torsion points 
on the topologically trivial flat configurations on $Y$,
i.e.~the 2-torsion points of the kernel of the Bockstein map $E^{p-1}(Y;\bR/\bZ)\to E^{p}(Y)$.
By the perfectness of the pairing on $Y$, this set of 2-torsion points is also actually isomorphic to $U/2U$.
Let $\upsilon$ be the map implementing the isomorphism
\begin{equation}
  U/2U \xrightarrow[\upsilon]{\sim} \text{2-torsion points of}\ \Ker (\beta: E^{p-1}(Y;\bR/\bZ)\to  E^{p}(Y)).
\end{equation}

As $U$ is a free Abelian group, $U/2U\simeq (\bZ_2)^r$ for some $r$.
We have a natural mod-2 pairing on it given by $\int_Y \alpha\beta$ mod 2.
Therefore the final sum is of the form \begin{equation}
\sum_{c,\alpha \in U/2U} e^{2\pi iQ(\upsilon(c) + \alpha T)}
\end{equation} where we have the relation \begin{equation}
Q(\upsilon(c)+\alpha T+\beta T)-Q(\upsilon(c)+\alpha T)-Q(\upsilon(c)+\beta T)+Q(\upsilon(c)) = \frac12\int_Y \alpha \beta
\end{equation} modulo 1.
We also have \begin{equation}
Q(\upsilon(c)+\upsilon(c')+\alpha T) - Q(\upsilon(c)+\upsilon(c')) -Q(\upsilon(c)+\alpha T)- Q(\upsilon(c)) = \int_X c' \alpha T = \int_Y c'\alpha
\end{equation} modulo 1.

For the standard case of $d=3$, $p=2$ and $E=H$, $U/2U=H^1(Y;\bZ_2)$,
 $Q$ is (up to a shift by $Q(0)$) the standard quadratic refinement induced by the spin structure of $Y$.
$c\in H^1(Y;\bZ_2) \subset H^1(Y;\bR/\bZ)$ specifies the classical holonomy of the $U(1)$ background on $Y$,
and the summation over $\alpha$ is over global gauge transformations. 

With this intuition, 
we can now evaluate the expectation value of the `Wilson loop' $L_m$ around $m\in U/2U$,
which is \begin{equation}
\langle L_m\rangle := \frac{1}{Z_0}\sum_{c,\alpha \in U/2U}  (-1)^{\int_Y mc}  e^{2\pi iQ(\upsilon(c) + \alpha T)}
\end{equation}
We can easily show that $\langle L_m\rangle=\pm1$, but this is not linear in $m$ but rather is a quadratic refinement: \begin{equation}
\langle L_{m+n}\rangle = \langle L_{m}\rangle \langle L_{n}\rangle (-1)^{\int_Y mn}.
\end{equation}

This is indeed what happens to the expectation value of  the `fermion line' 
$e^{2\pi i\int_m A}$ of the invertible Abelian Chern-Simons theory in 3d:
naively it looks like linear in $m$, but not quite due to the subtlety in the quantisation.
The quadratic refinement term is necessary to turn it into a line following the fusion rule of a fermion line.

When applied to the K-theory based RR fields of Type IIB theory, it gives somewhat of a surprise,
since what replaces $e^{2\pi i\int_m A}$ is $e^{2\pi i \int_m F_5}$.
This means that, on $S^5_A\times S^5_B$, we have \begin{equation}
\langle e^{2\pi i \int_{S^5_A+S^5_B} F_5}\rangle
=(-1)
\langle e^{2\pi i \int_{S^5_A} F_5}\rangle
\langle e^{2\pi i \int_{S^5_B} F_5}\rangle.
\end{equation}
This naively means that when $F_5$ is integrally quantised on both $S^5_A$ and $S^5_B$,
it is half-integrally quantised on $S^5_A+S^5_B$.

Such minus sign on quantisation of flux periods has been explained in general in \cite{Moore:1999gb} and exemplified on a bundle of $2k$-dimension sphere bundle over $(10-2k)$-dimensional sphere in the case of Type IIA. 
Here we obtained the corresponding results for Type IIB in the case of the spacetime $S^5 \times S^5$.
\subsection{With boundary}

Our strategy is to consider a Chern-Simons theory at level one on a spacetime with boundary
with a suitable boundary condition to realise a self-dual field on the boundary. 
For this purpose we need to consider the quadratic refinement on spacetimes with boundary.
The definition of quadratic refinement is subtle already on a space without boundary.
It is therefore not immediate to define it on a space with boundary. 

It would be very interesting to analyse this point in detail, but the authors' impression is that 
most of the subtleties already appear in the case of the standard spin Abelian Chern-Simons theory 
in three dimensions, and are not particular to our  Chern-Simons theory based on generalised differential cohomology theories.
Therefore we leave the detailed discussions to a future study, 
and only summarise the properties we will need in other parts of the paper.

\subsubsection{Choice of the action and the boundary condition}

Very schematically, one possible path integral realising the self-dual field is of the form 
\begin{equation}
\int D\check a\, e^{2\pi i Q_1(\check a)}, 
\end{equation} where $\check a$ is the field to be path-integrated
and $Q_1(\check a)$ is a quadratic refinement.
When we have a background field $\check A$, we can also consider a different theory
\begin{equation}
\int D\check a\, e^{2\pi i [Q_1(\check a) + (\check a,\check A)]}. \label{bulkaction}
\end{equation} 
Note that $Q_2(\check a):=Q_1(\check a)+(\check a,\check A)$ is also a quadratic refinement.
Therefore it is still of the form \begin{equation}
\int D\check a\, e^{2\pi i Q(\check a) }
\end{equation} for a quadratic refinement $Q(\check a)$.  
We therefore need to ask \emph{which} quadratic refinement is allowed in this construction.

We also need to specify the boundary condition. 
We choose the boundary condition of the form
\begin{equation}\label{eq:bca0}
\diff a|_\text{boundary}={\check A}|_\text{boundary},
\end{equation}
where $\check A$ is a background field, possibly zero. 
On the boundary $\check a $ has the gauge degrees of freedom,
schematically given by $\diff a\to \diff a+d\diff b$. 
The gauge degrees of freedom $\check b$ on the boundary are interpreted as boundary modes.
It is known that this boundary mode $\check b$ is a self-dual field; we will recall this fact in some detail below.

Note that for our purpose, the boundary condition needs to be such that,
when $\check a$ is not yet path-integrated, the boundary does not have any extra degrees of freedom.
Otherwise, after the path integral over $\check a$,
we will have a self dual field plus this additional degrees of freedom.
For example, if we use $Q_3(\check a):=Q_1(\check a) + W_\text{chiral fermion}$ in the construction above,
where $W_\text{chiral fermion}$ is the invertible phase for a chiral fermion on the boundary,
we would have a self dual field plus a chiral fermion on the boundary.
This is not what we want. 
 
To have no boundary modes \emph{before} the path integral over $\check a$,
a necessary condition on the choice of the boundary condition is such that, 
when the same condition is also placed in the entirety of the bulk,
the bulk invertible phase is trivial.
For example, ${\check a}|_\text{boundary}=0$ for \eqref{bulkaction} with $Q_1(0)=0$ is such an example,
and $Q_3(\check a)$ above is correctly disallowed.

A general form of the action satisfying this property is to take an arbitrary $ Q(\check a)$
and  take the path integral to be \begin{equation}
\int D\check a\, e^{2\pi i [Q(\check a) - Q(\check A)]}, 
\qquad {\check a}|_\text{boundary}={\check A}|_\text{boundary}.
\label{secondbulkaction}
\end{equation} 
Setting $\check a=\check a'+\check A$, $Q'(\check a'):=Q(\check a')-Q(0)$,
and dropping the primes, we see that this is actually equivalent to the set of actions \begin{equation}
\int D\check a\, e^{2\pi i [Q(\check a) + (\check a,\check A)]},
\qquad  {\check a}|_\text{boundary}=0,
\end{equation}
with the additional condition that $Q(0)=0$.
In our discussions below, we opt  the  form \eqref{secondbulkaction},
since the bulk analysis in Sec.~\ref{sec:bulk-path-integral} can be taken word-for-word.

\subsubsection{Non-topological part of the boundary modes}

Now that we have set up the action,
let us quickly recall how to see the chiral boundary modes,
while for the full details, we refer the reader to \cite{Hsieh:2020jpj,GarciaEtxebarria:2024fuk}. 
For definiteness, we take the case $E=KO$.
We consider a differential class $\diff a$ and denote its cocycle representative by $(V,B,C)$.
We temporarily write $F(\check a):=\sqrt{\hat A} R(\check a)=dC+\cdots$.
We need the $C$-dependent and differential form part of $Q(\diff a)$, 
which can be determined from \eqref{pro} and \eqref{eq:nudef} to be \begin{equation}
Q(\diff a) \sim \frac12\int_X (  C F(\check a) -C F(\check \nu)).\label{3.40}
\end{equation}
We then need to  add to the action a kinetic term for $\check a$ 
\[
S=
   \int_X(\frac{2\pi i}{2g^2} F(a)\wedge *F(a))+Q(\check a),
\]
with very large $g \gg 0$.
Taking the variation with respect to $C$,
one finds the equation of motion 
\[\label{eq:eomgeneral}
    \frac{1}{g^2}d*_{X} F(\check a)+F(\check a)-\frac12F(\check \nu)=0 ,
\]
where we moved to the Minkowski signature.
Let us now parameterise a local neighbourhood of the boundary by $Y\times (-\infty,0]$, with $\tau \in (-\infty,0]$.
Then the term $F(\check \nu)$ is pulled back from $Y$, and has no components along $d\tau$.
We now take the ansatz for $F(\check a)$ of the form \begin{equation}
\begin{aligned}
F(\check a)&= d(1-(e^{mg^2\tau})) \wedge H  + F(\check A)+ O(1/g^2)\\
&= -mg^2 e^{mg^2\tau}  d\tau \wedge  H + (1-e^{mg^2\tau}) dH + F(\check A)  + O(1/g^2),
\end{aligned}
\end{equation} 
where the constant piece $F(\check A)$ was chosen to satisfy the condition 
$F(\check a)|_\text{boundary}=F(\check A)|_\text{boundary}$ at $\tau=0$, and $H$ is a formal sum of differential forms living in $Y$.
Plugging this in to \eqref{eq:eomgeneral}
and comparing the term of order $O(g^2)$, one finds $m*_YH +H=0$.
As the eigenvalue of $*_Y$ is $\pm1$, and we want $m>0$ for the modes to exponentially localise,
we are forced to take $m=+1$, i.e.~we found \begin{equation}
*_Y H + H =0.
\end{equation}
Furthermore, by comparing the components of \eqref{eq:eomgeneral} along $Y$ at $\tau=0$, we find \begin{equation}
d*_Y H =F(\check A) - \frac12 F(\check\nu).\label{boundary-eom}
\end{equation}

\subsubsection{Some basic consequences}
\label{sec:basic}

From the discussions above, we can infer some basic consequences concerning the self-dual field on the boundary.
First, its anomaly is characterised by \begin{equation}
\int D\check a\, e^{2\pi i [Q_0(\check a) - Q_0(\check A)]}
=(\int D\check a\, e^{2\pi i Q_0(\check a)} )  e^{-2\pi i Q_0(\check A)}, \label{eqn:grav_anomaly}
\end{equation} where the first factor only depends on the metric.
In other words, the pure gravitational part of the anomaly of the boundary self-dual field
is \emph{uniquely} determined by the fact that the bulk action is given by the
quadratic refinement. 

In the next Section, we will see this fact in explicit examples, where we compute 
the gravitational anomaly of  form fields arising as the boundary modes
of abelian Chern-Simons theory whose action is the quadratic refinement of the pairing of differential KO theory. 
In particular, when the bulk is 11-dimensional and the boundary is 10-dimensional,
we will find that  the gravitational anomaly of the form fields
is exactly the right amount to cancel  the gravitational anomaly of dilatino and gravitino.

Second, the analysis of the bulk in Sec.~\ref{sec:bulk-path-integral} carries over,
and in particular the topological class $I({\check a})|_Y$ is fixed to be $\mu_\text{top}$.
As the boundary condition is ${\check a}|_\text{boundary} = {\check A}$,
we have 
\begin{equation}
0= I(\check A) - \mu_\text{top} \label{bc-top}
\end{equation}
at the boundary.
At the level of differential forms, this means that $R(\check A)$ and $\mu_\text{diff}$
differ only by an exact term,
so there is a differential form $H$ such that
\begin{equation}
dH= F(\check A)-F(\diff\mu), \label{bc-diff}
\end{equation}
which is exactly what we expect. (Recall that $F(\check A)$ and $R(\check A)$ are related 
by the multiplication by $\sqrt{\hat A}$, as we discussed above \eqref{3.40}.)
We already saw this from the analysis of the classical equation of motion in \eqref{boundary-eom}.
More concretely, when we use the differential KO theory,
 the bulk is 11-dimensional, and the boundary is 10-dimensional,
 we will see that the rank of the KO background field is set to be $32$ via \eqref{bc-top},
 and we will reproduce $dG_3$ and $dG_7$ given in the introduction 
 as \eqref{eq:introX4} and \eqref{eq:introY8} from this consideration in Sec.~\ref{ex:type1}.

\subsubsection{Shifted quantisation condition on the boundary modes}

A standard expectation is that the flux of $H$ \emph{would} satisfy a (possibly shifted) quantisation condition.
This \emph{would} happen if there is a preferred differential class $\check \mu$ 
which descends to both $\mu_\text{top}$ and $\mu_\text{diff}$ 
such that the two conditions \eqref{bc-top} and \eqref{bc-diff}
both come from a single conditions that 
\begin{equation}
\text{$\check H$ is a trivialisation of $\check A-\check \mu$}.
\end{equation}
The entirety of such $\check H$ then form a torsor over $\check E^{p-1}(Y)$,
but with a self-duality constraint on $R(\check H)$ imposed by the bulk equation of motion.
Actually, there does not seem to be a truly preferred choice of $\check \mu$ as discussed below,
but  let us first recall what would happen if there actually were such a preferred choice.

Take  the case of spin Abelian Chern-Simons theory based on $\check A \in \check H^2(Y;\bZ)$.
Suppose  the preferred choice were given by $\check\mu=0$.
Recall that $\check A$ is a $U(1)$ bundle over $Y$ together with a connection.
For simplicity assume $\check A$ is flat.
The condition \eqref{bc-diff} alone allows an arbitrary closed 1-form $H$.
But geometrically a trivialisation of a flat bundle 
is a section of the $U(1)$ bundle. In particular, on a closed path $C$, we have 
\begin{equation}
e^{2\pi i\int_C H} = e^{2\pi i\int_C \check A},
\end{equation} 
explaining the shifted quantisation condition of the 1-form $H$.
When $\check A$ is nonzero, this in particular means 
that $\check H$ is not quite in $H^1(Y;\bZ)$,
but rather an element of a torsor over $H^1(Y;\bZ)$.

This view is a bit too simplistic, unfortunately,
due to the subtleties discussed around the end of Sec.~\ref{sec:bulk-path-integral}.
 $\check\mu\in \check E^{p}(X)$'s satisfying 
 $I(\check\mu)=\mu_\text{top}$ and $R(\check\mu)=\mu_\text{diff}$
form a torsor over flat and topologically trivial classes.
Flat and topologically trivial classes have a finite subgroup 
consisting of 2-torsion points,  given by
\begin{equation}
F\simeq E^{p-1}(Y)_\text{free}\otimes \bZ_2 \label{mu-ambig}
\end{equation}
As we saw in Sec.~\ref{sec:bulk-path-integral},
$\check\mu$ can be determined up to shifts by $F$,
but the elements within it are summed using a quadratic refinement
of a natural pairing over $F$.
Therefore there is no natural canonical element within it,
and no canonical choice of $\check\mu$ is available.

This is actually as it should be.
Consider the case of the boundary of the invertible spin Abelian Chern-Simons theory in 3d,
when the $U(1)$ background $\check A$ is zero.
In this case, $\check\mu$ is ambiguous by a shift by $F\simeq H^1(Y;\bZ_2)$,
the elements of which are summed using a quadratic refinement on $F$.
This is sensible, since the boundary self-dual field is a chiral fermion
(associated to a quadratic refinement)
and not a circle-valued boson (which has a definite winding number).

Summarising, the boundary degrees of freedom $\check H$ on $Y$ arising from
the bulk Chern-Simons theory on $X$ 
whose action is given by the quadratic refinement $Q$ 
of the natural pairing on $\check E^p(X)$
with a background field $\check A\in \check E^p(X)$
is such that they give the boundary conditions \eqref{bc-top} and \eqref{bc-diff}
for the background field $\check A$,
where $\mu_\text{top}\in E^p(Y)$ 
and $\mu_\text{diff}\in \Omega^p_{E}(Y)$ are canonically determined from 
the property of the quadratic refinement $Q$.
This means that $\check H$ is \emph{almost} a trivialisation 
of a differential  class $\check A-\check \mu$,
\emph{except} that the class $\check\mu$ which gives $\mu_\text{top}$ and $\mu_\text{diff}$
is only defined up to a shift by a subgroup $E^{p-1}(Y)_\text{free}\otimes \bZ_2$ 
given in \eqref{mu-ambig}.
What is well-defined is rather a quadratic refinement over it,
much as in the prototypical case of the 2d boundary of the 3d spin Chern-Simons theory,
where a chiral fermion rather than a chiral boson appears on the boundary.
Note also that $2\mu_\text{top}=I(\check\nu)$
and $2\mu_\text{diff}=R(\check \nu)$ as we discussed above,
where $\check \nu$ was introduced in \eqref{eq:nu}.
This is compatible that $2\check\mu =\check\nu$ is canonically defined 
in terms of the quadratic refinement.

\section{Quadratic refinement and anomaly cancellation} \label{sec:KO}

In the last section, we discussed the structure of the  invertible Abelian Chern-Simons theory
based on generalised differential cohomology theories,
simply assuming the existence of a quadratic refinement.
To actually use it, we then need to construct the quadratic refinement.
In this section, we will first review the quadratic refinement for the case $E=K$,
and then extend it for the case $E=KO$.
This will allow us to construct the anomaly theory for RR fields in type I string theory,
and show that not only the perturbative anomaly
but also the global anomaly cancels between massless fermions and RR fields.

\subsection{Review: Quadratic refinement of K theory} \label{subsec:Q_K}

We first review the construction in the differential K theory case in \cite{Hsieh:2020jpj} for $8k+3$ dimensional bulk $Y$, which is applicable to type II string theory.
We would like to define a quadratic refinement of the pairing (\ref{eq:pairingab}) in K theory,
\[
    (\breve{a}, \breve{b}) := \chi^\mathbb{C}(\breve{a} \cdot \breve{b}^*).
\]
Morally speaking, we would like to divide the above phase by 2. 
But $\chi^\bC(\check a)$ for a generic $\check a$ is valued in $\bR/\bZ$.
Therefore, dividing the phase $\chi(\breve{a})$ by 2 would result in an ambiguity by $\pm1/2$.

However, the self-pairing for a complex bundle $\breve{a}$ uses the argument $\breve{a} \cdot \breve{a}^*$, which is real, i.e.~an element of $\check{KO}$. 
Therefore, we have the holonomy function $\chi^{\mathbb{R}}$, which is automatically one half of the holonomy function of $\breve{a} \cdot \breve{a}^*$ viewed as a complex bundle.
That is, if we define
\[
Q(\breve{a}) := \chi^\mathbb{R}(\breve{a} \cdot \breve{a}^*) \label{KUQ},
\] we have 
\begin{equation}
\chi^{\mathbb{C}}(\breve{a} \cdot \breve{a}^*)= 2\chi^{\mathbb{R}}(\breve{a} \cdot \breve{a}^*).
\end{equation}
This therefore is a quadratic refinement.

We can directly check that $Q(\breve{a})$ thus defined refines the bilinear pairing $(-,-)$, defined by $\chi^\mathbb{C}$, using the linearity of $\chi^\mathbb{R}$:
\begin{align}
& Q(\breve{a}_1 + \breve{a}_2) - Q(\breve{a}_1) - Q(\breve{a}_2) + Q(0) \\
= \quad &\chi^\mathbb{R}((\breve{a}_1 + \breve{a}_2) \cdot (\breve{a}_1 + \breve{a}_2)^*) - \chi^\mathbb{R}(\breve{a}_1 \cdot \breve{a}_1^*) - \chi^\mathbb{R}(\breve{a}_2 \cdot \breve{a}_2^*) 
\\
= \quad & \chi^\mathbb{R}(\breve{a}_1 \cdot \breve{a}_2^* + \breve{a}_2^* \cdot \breve{a}_1) = \chi^\mathbb{C}(\breve{a}_1 \cdot \breve{a}_2^*)
\end{align}

A more explicit expression of $Q(\breve{a})$ in terms of the components $(V, B, C)$ can be given as:
\begin{multline}
Q(\breve{a}) = -\frac{1}{2} \eta(D_Y(B \otimes 1 + 1 \otimes B^*)) \\
+ \frac{1}{4} \int_{Y}\left(C \wedge dC^* + \sqrt{\hat{A}(R)} \wedge (C \wedge \ch(B) + \ch(B)\wedge C^*)\right)
\end{multline}
Specialising to a bulk of dimension $8k + 3 = 11$, we have thus reproduced the bulk invertible phase for RR field in type IIB string theory from \cite{Hsieh:2020jpj}, see (4.36) there.

In more detail, the bulk action is $Q(\check  a)-Q(\check  A)$,
with the boundary condition that  $\check  a|_\text{boundary}=\check  A|_\text{boundary}$,
where $\check  a$ is path-integrated,
while $\check  A$ is a background field.
To use in Type IIB string theory, we identify the background field $\check  A$
to be given at the boundary by the $U(r)$ gauge bundle on $r$ D9-branes,
or more generally the formal difference of $U(r_+)$ and  $U(r_-)$ gauge bundles 
on $r_+$ D9 branes and $r_-$ anti-D9 branes.

The eta invariant part in the definition of the quadratic refinement, \eqref{KUQ},
is exactly the anomaly of the adjoint fermion on $r$ D9-branes,
or the charged fermions on the stack of $r_+$ D9-branes and $r_-$ anti D9-branes.
This means that the pure gauge and mixed gauge-gravity anomalies 
are cancelled between the bulk RR contribution from $Q(\check  a)$
and the background fermion contribution from $Q(\check  A)$.
If we unwind the definition of $Q(\check a)$ carefully,
one finds that this cancellation came from the fact 
$Q(\breve{a})$ is well-defined as a function of the differential K theory class,
i.e.~it is unchanged when the cocycle $(V_a,B_a,C_a)$ is varied within the same class.
So, in a sense, this formulation of the RR field is designed from the start
to cancel the gaugino part of the anomaly completely, including both the perturbative and the global parts.
The cancellation of pure gravitational anomaly is a different matter, to which we come back later in this section.

We note that the top-degree $C$-field $C_{11}$ appears in the above quadratic refinement via the coupling
\[
    \frac{1}{2} C_{11} \wedge \dim V
\]
 Since $C_{11}$ does not participate in the kinetic coupling $C \wedge dC$ for degree reasons, the equation of motion for $C_{11}$ will set 
\[
\dim V = r_+-r_- = 0.
\]
We have reproduced the fact that type IIB string theory contains  the same number of D9 branes
and anti D9-branes.

\subsection{Quadratic refinement of KO theory}
\label{sec:quadraKO}
We now proceed in constructing the quadratic refinement using differential KO theory, focusing on $8k+3$ dimensional bulk. 
We start from a differential class $\check a$ realised by $(V,B,C)$,
where $V$ is a real bundle (or a formal difference of such bundles)
and $B$ is a connection on it.\footnote{Later the dimension of the bundle will be fixed by the equation of motion, but at the level of bulk action, we can keep the dimension arbitrary for the moment.}

The bilinear pairing whose quadratic refinement we need to construct is
\[
(\breve{a}, \breve{b}) = \chi^{\mathbb{R}}(\breve{a} \cdot \breve{b}).
\]
Therefore, the trick back in the complex K theory case of refining 
the holonomy  function $\chi^\bC(\breve{a} \cdot \breve{a}^*)$  for the differential K theory class
by the holonomy function $\chi^\bR(\breve{a} \cdot \breve{a}^*)$ for the differential KO theory class
cannot be used here. 
Instead, we rely on a different trick. 

Note that $\breve {a}\cdot \breve{a}$ is based on $V\otimes V$.
This can be decomposed as the direct sum of the second-rank symmetric bundle $S^2 V$ and the second-rank antisymmetric bundle $\Lambda^2 V$,
so we can consider either $S^2V$ or $\Lambda^2 V$ as a `half' of $V\otimes V$.
The operations 
$V\mapsto S^2 V$ and 
$V\mapsto \Lambda^2 V$ are  `quadratic refinements' of the tensor product,
in the sense that 
\[
S^2(V_1 \oplus V_2) = S^2 V_1 \oplus S^2 V_2 \oplus V_1 \otimes V_2, \quad \Lambda^2(V_1 \oplus V_2) = \Lambda^2 V_1 \oplus \Lambda^2 V_2 \oplus V_1 \otimes V_2,
\]and therefore \begin{equation}
\Lambda^2(V_1\oplus V_2) - \Lambda^2 V_1 -\Lambda^2 V_2 = V_1\otimes V_2,
\label{qr-tensor}
\end{equation} when we allow virtual bundles.
Plugging in $V_2=-V_1$, we also find $\Lambda^2 (-V)=S^2 (V)$.

Using $S^2$ or $\Lambda^2$, 
we can then attempt to construct two quadratic refinements with the following form:
\begin{align}
    &Q_{S^2}(\breve{a}) =  
    -\frac{1}{2}\eta(D_Y(B_{S^2 V})) + \frac{1}{4} \int_Y C_{S^2 V}, \\ 
    &Q_{\Lambda^2}(\breve{a}) = 
    -\frac{1}{2}\eta(D_Y(B_{\Lambda^2 V})) + \frac{1}{4} \int_Y C_{\Lambda^2 V}, \label{eq:qrtype1}
\end{align}
where
$B_{\Lambda^2 V}, B_{S^2 V}$ are the restriction of the induced connection $B \otimes 1 + 1 \otimes B$   via the decomposition $V \otimes V = S^2 V \oplus \Lambda^2 V$,
$C_{S^2V}$ and $C_{\Lambda^2 V}$ are defined by
\begin{align} 
C_{S^2 V} &= C \wedge dC + 2 C \wedge X_{S^2 V},\\ 
 C_{\Lambda^2 V} &= C \wedge dC + 2 C \wedge X_{\Lambda^2 V},
\end{align}
and $X_{S^2 V}$ and $X_{\Lambda^2 V}$ are formal sums of degree-$4i$ differential forms.\footnote{
The $\eta$ invariant in the KO action will eventually describe the anomaly of the gaugino carrying a respective representation of the $SO$-type gauge group.}
As we just mentioned, we have $\Lambda^2(-V)=S^2(V)$, and therefore 
the property of $Q_{S^2}(\check a)$ can be easily derived from that of $Q_{\Lambda^2}(\check a)$.
For this reason, we will only consider $Q_{\Lambda^2}$ below,
and denote it simply by $Q$.

Our aim is to pick a suitable 
$X_{\Lambda^2V}$ so that the expressions above
are well-defined as functions of the differential KO classes.
To study this, 
we now consider a 1-parameter family $Z = Y \times [0, 1]$ where the differential cocycle continuously vary along $t \in [0, 1]$, with an interpolating connection $\tilde{B}_{S^2 V}$ or $\tilde{B}_{\Lambda^2 V}$ on $Z$. Using APS index theorem, the variation of the eta-invariant part can be written as
\begin{align}
\begin{split}
& \frac{1}{2}\eta(D_{Y \times 0 + \overline{Y \times 1}}(\tilde{B}_{\Lambda^2 V})) = \frac{1}{2}\int_Z \hat{A} \ph(\tilde{B}_{\Lambda^2 V}) = \frac{1}{2}\int_Z \hat{A} \frac{1}{2}(\ph(\tilde{B})^2 - \ph(2\tilde{B})).
\end{split}
\end{align}

Motivated by the result of the K-theory case where $X= \sqrt{\hat{A}} \ch(\tilde{B})$, 
and comparing the properties of  
$\Lambda^2 V$ with $V \otimes V^*$, we use the following ansatz:
\[
    X = \sqrt{\hat{A}} \ph(\tilde{B}) -  f, \label{eqn:X_ansatz}
\]
and \begin{equation}
f \in \bigoplus_i \Omega^{4i}(Z)
\end{equation}
is a differential form constructed from the Pontryagin classes of $Z$.
Our $f$ here is related to the object $\check \nu$ introduced in \eqref{eq:nudef}.
Indeed, by comparing the term linear in $C$,
we get the relation that 
\begin{equation}
2
f =  \sqrt{\hat{A}} R(\check \nu).
\end{equation}
If we make a choice of $\check \mu$ so that $2 \check \mu = \check \nu$,
we have \begin{equation}
f = \sqrt{\hat{A}} R(\check \mu).\label{f-mu-conversion}
\end{equation}

Using the fact that $C$ varies by $\sqrt{A}\ph(B)$, 
the condition determining $f$ is that \begin{equation}
f \sqrt{\hat{A}} \ph (\tilde {B}) \Bigm|_{8k+4}= \frac12\hat A \ph(2\tilde B) \Bigm|_{8k+4},
\label{eq:cancellation-key}
\end{equation}
The components of $f=\sum_i f_{4i}$ can be iteratively determined by starting from $f_0$:
\[
\begin{split}
    f_{0}  &=  2^{4k+1}, \quad \\
    f_{4l}  &= 2^{4k+1-2l} \hat{A}_{4l} - \sum_{j = 0}^{l-1} f_{4j} \sqrt{\hat{A}}_{4(l-j)} \qquad (l = 1, \dots, 2k+1).
\end{split}
\]
In this way, we have identified an explicit form of $\chi^{\mathbb{R}}(\breve{a}_{\Lambda^2 V})$ or $\chi^{\mathbb{R}}(\breve{a}_{S^2 V})$ that is invariant under continuous change of differential cocycle. 
Note that the quadratic refinement contains $f_{4l}$ only up to $l=2k$,
but we will also have a use for $f_{8k+4}$ later.

To check that our construction satisfies the definition of the quadratic refinement, we  use the
property \eqref{qr-tensor} satisfied by the operation $V\mapsto \Lambda^2 V$.
Then the defining property
\begin{align}
& Q
(\breve{a}_1 +\breve{a}_2) -  Q
(\breve{a}_1 ) - Q
(\breve{a}_2) + Q
(0) = (\breve{a}_1, \breve{a}_2)
\end{align}
can be checked by using the the linearity of the holonomy $\chi^{\bR}(\breve{a})$. In other words, 
$Q_{\Lambda^2}(\breve{a})$ is indeed a  quadratic refinement of the  bilinear form $(\cdot, \cdot)$.

Imposing the equation of for the $C$ field of highest degree will give us,
for the bulk with dimension $8k+3$,
\[
    \dim V = 2^{4k+1}. 
\]

In particular, for the type I string theory with $k = 1$, we indeed get $SO(32)$. This case will be discussed in more detail later.

\subsection{Pure gravitational anomaly: generalities}
\label{subsec:general-pure-grav}

We stress that a well-defined quadratic refinement in differential KO theory will produce a \textit{gravitational anomaly} of the boundary RR-field.
By the arguments in Sec.~\ref{sec:basic}, the pure gravitational part of the anomaly is given by \begin{equation}
(\int D\check a e^{2\pi i Q_0(\check a)})e^{-2\pi i Q_0(\breve{A})}.
\end{equation}
when the bulk dynamical field $\breve{a}$ is subject to the boundary condition of ${\check a}|_\text{boundary}=\diff A$.
An honest computation of this path integral should be possible, but here we can 
use a far quicker alternative when the bulk dimension is either $8k+3=3$ or $8k+3=11$.

Note that our general argument in Sec.~\ref{subsec:inv} guarantees that the resulting action is a spin invertible phase in eleven dimensions,
which is classified by 
$(I_\bZ \Omega^\text{spin})^{4}(pt)$ or
$(I_\bZ \Omega^\text{spin})^{12}(pt)$. 
This group does not have a torsion part,
because  $\mathrm{Tor}(\Omega^{\text{spin}})_{3,11}(pt) = 0$.
Therefore, the resulting invertible phase can be determined just by looking at its perturbative part.
This quicker route does not exist in $8k+3=19$ or above, due to the existence of the torsion parts in the bordism groups,
but this suffices for our purposes here.

Let us now perform the perturbative analysis. 
As we saw in \eqref{eq:nudef}, at the perturbative level the quadratic refinement is \begin{equation}
Q(\check a)=\frac12 \Bigl((\check a,\check a)-(\check a,\check \nu)\Bigr),
\end{equation} where we used the fact that our $Q$ satisfies $Q(0)=0$.
The path integral over $\check a$ sets $\check a=\mu$, where $\mu=\frac12\nu$.
and effectively converts the action into \begin{equation}
Q(\check\mu) = -\frac12(\check \mu,\check\mu).
\end{equation} 
Therefore the perturbative part of the anomaly polynomial is \begin{equation}
g_{8k+4}
:= -\frac14 \hat A R(\check\mu) \wedge R(\check\mu)|_{8k+4},
\end{equation}
or in other words
\[
    g_{8k+4} := - \frac{1}{4} [f \wedge f]_{8k+4} = -\frac{1}{2} \sum_{l = 0}^{k} f_{4l}  f_{8k+4-4l} 
    \label{eq:grav-formula}
\]
after we use the relation \eqref{f-mu-conversion} derived above.

As we already discussed, this is enough to determine the full anomaly
in the case $k=0$ and $k=1$, but not for $k\ge 2$.
Explicit values will be given below when we discuss examples concretely.

We remark that, in case a middle degree component $C_l d C_l$ appear in the kinetic term, 
we need to add the contribution that describes the perturbative gravitational anomaly of the anti-self-dual field, as obtained in \cite{Alvarez-Gaume:1983ihn}. 
For example, a full treatment of the gravitational anomaly in type II string theory would involve re-casting the gravitational anomaly of the $C_4$ field in our language, which was discussed in \cite{Hsieh:2020jpj}.
Luckily this complication does not arise in the case under consideration.

Before proceeding, let us discuss the issue of the structure imposed on the spacetime manifolds
in our analysis.
In the existing papers on self-dual fields,
it is often the case that the structure called a Wu structure is introduced,
in addition to spin structure on the spacetime which we always assume to exist on the spacetime.
Then, to discuss the global anomaly cancellation,
it is necessary to analyse the bordism groups of manifolds with Wu structure.

In this paper,  we tried to use only the spin structure, without adding anything to it.
This way, the tangential structure is still given by spin structure, under which the Anderson dual of bordism group $(I_\bZ \Omega^{\text{spin}})^{12}(pt)$ suitable for the type I string theory contains no torsion. Therefore, to check the full anomaly cancellation in type I string theory, it sufficed to consider only perturbative anomalies.
The burden of the construction was rather in a careful construction of the RR fields as a theory depending on the spin structure.

In our analysis, we saw that the differential class $\check \nu$ is naturally defined on the bulk spacetime,
but there are several places where $\mu_\text{top}$ and $\mu_\text{diff}$ appeared,
which would correspond to $\check\mu$ satisfying $2\check\mu = \check \nu$.
But there seems no natural preferred choice of $\check\mu$.
We also note that there \emph{was} a preferred $\mu_\text{top}$ \emph{on the boundary},
but already the existence of a preferred choice of $\mu_\text{top}$ in the bulk is not clear to the authors.
In the analysis in \cite{Freed:2000ta}, 
it seemed that 
a topological class $\mu_\text{top}$ 
together with (a choice of) its differential refinement $\check \mu$
was introduced in the 11d bulk. It is unclear whether the bordism analysis would be modified with this additional condition,
and if so, whether the global anomaly cancellation still works.
These issues might be worth studying further. 
The authors still want to stress here that the global anomaly is cancelled 
if we stay on the realm of spacetimes with spin structure only.

\subsection{Pure gravitational anomaly: examples}

Here we discuss the structure and the pure gravitational anomaly of the KO-based RR fields in 2D, 10D and 18D.
This analysis will also allow us to fix $\dim V$ of the KO theory background.

\subsubsection{In 2D}

In this case, the holonomy of a KO-class will have no kinetic term in the action. Taking the bundle to be in the 2-index antisymmetric representation for now, the action can be written as
\begin{align}
    Q(\check a)
    &=  \frac{1}{2} \eta(D_{\Lambda^2 V}) + \int_{Y} C_3 \wedge X_0
\end{align}

$f_0$ and $f_4$ can be determined as follows:
\begin{align}
    f_0 &=  2^1,  \\
    f_4 &=  2^{-1} \hat{A}_4 - f_0 \sqrt{\hat{A}}_4 = -\frac{1}{48} p_1
\end{align}
Here, $X_0 = \dim V - 2$ and the equation of motion for the background field $C_3$ fixes $\dim V = 2$ via the $C_3 \wedge X_0$ coupling. So the total gravitational anomaly of the RR field is
\[
    g_4(R) = -\frac{1}{2} f_0 f_4 = -\frac{1}{48} p_1
\]

\subsubsection{In 10D: Type I string theory}
\label{ex:type1}

In 10D,  we may rewrite the action (\ref{eq:qrtype1}) in terms of the components $C=(C_3,C_7,C_{11})$, as
\[ \label{eq:type1a}
    Q(\diff a)=& -\frac{1}{2} \eta(D_{\Lambda V^2})
    +\frac{1}{2}\int_{Y} 
    C_3\wedge dC_7+ C_3\wedge X_8+ X_4\wedge C_7+ X_0\wedge C_{11} ,
\]
where \begin{equation}
    f_0 =32,
    \qquad
    f_4 =  \frac{p_1}{3}, \qquad
    f_8 = - \frac{p_1^2}{320} + \frac{7}{720}p_2 
\end{equation}
and \begin{equation}
f_{12} = \frac{79p_1^3}{1935360}-\frac{5 p_1p_2}{32256}+\frac{31p_3}{120960}.
\end{equation}
Then we see that $X_{4l}=(\sqrt{\hat{A}(R)} \ph(\tilde{B}))_{4l} - f_{4l}$ are given by 
\begin{align}
    X_0 &=  \dim V-32,\\
    X_4 &=   p_1(F) - p_1(R),\\
    X_8 &=  \frac{1}{64}p_1(R)^2 - \frac{1}{48} p_2(R) - \frac{1}{48} p_1(R) p_1(F) + \frac{1}{12} p_1(F)^2 - \frac{1}{6} p_2(F).
\end{align}
Comparing this to the $X_4^{GS}, Y_8^{GS}$ given in (\ref{eq:introX4}) and (\ref{eq:introY8}), we get:
\begin{equation}
    X_4^{GS} = \frac{1}{2} X_4, \qquad Y_8^{GS} = X_8.
\end{equation}
The relative factor of $1/2$ for $X_4$ was explained in Eq.~\ref{eq:relative-factors} in Sec.~\ref{sec:normalization}.

From the equation of motion we must have $X_0=0$, this implies that for type I string theory $\dim V=32$.
The total gravitational anomaly of RR forms can be computed using the formula \eqref{eq:grav-formula},
and the result is
\[
    g_{12} = -(f_0 f_{12} + f_4 f_8)/2 = -\frac{p_1^3}{7560}+\frac{13 p_2 p_1}{15120}-\frac{31 p_3}{7560}
    \label{typeI-bulkanomaly}
\]
This reproduces the result of Green-Schwarz anomaly cancellation of type I string theory, in that the above anomaly exactly cancels the anomalies of a gravitino and a dilatino:
\[
I_{\text{gravitino}} + I_{\text{dilatino}} = \left[\frac{1}{2}\hat{A}(R)(\ph(R) - 4)\right]_{12} = \frac{p_1^3}{7560}-\frac{13 p_2 p_1}{15120}+\frac{31 p_3}{7560}
\]

The standard presentation of Green-Schwarz anomaly cancellation in type I theory is slightly different. 
We first sum up  three types of fermion anomalies 
\[
    I_{12}(R, F) = I_{\text{gravitino}}(R) + I_{\text{dilatino}}(R) + I_{\text{gaugino}}(R, F),
\]
where we keep all $R, F$ dependences for all degree-$4k$ class until the end of this example, to emphasise the terms that depends on the gauge bundle.
Now we notice that there is a factorisation \[
    I_{12} = X_4^{GS}(R, F) Y_8^{GS}(R, F) = \frac{1}{2} X_4(R, F) X_8(R, F)
\] when $\dim V=32$.
This makes it possible to cancel $I_{12}$ by the anomaly of RR forms:
\[
    dH_3 = X_4^{GS}(R, F) = \frac{1}{2} X_4(R, F), \quad dH_7 = Y_8^{GS}(R, F) = X_8^{GS}(R, F).
\]
In contrast, we are organising the terms in a slightly different fashion. 
Namely, the differential KO-theory based formalism of the RR fields guarantees that 
it automatically cancels the gaugino anomalies, with the correct $X_4$ and $X_8$.
Then the remaining part,
\[
    g_{12}(R) = I_{\text{gaugino}}(R, F) - \frac{1}{2} X_4(R, F) X_8(R, F) = -(I_{\text{gravitino}}(R) + I_{\text{dilatino}}(R)),
\]
is purely gravitational, and which is then found to cancel the gravitational anomaly of the gaugino and the dilatino.

\subsubsection{In 18D}

To illustrate the generality of the quadratic refinement, we also discuss the $k = 2$ case with a 18 dimensional spacetime. There, the quadratic refinement for the differential KO class corresponding to the RR field is well-defined, although we do not know any use of it at present.
The $f_{4i}$ as defined in (\ref{eqn:X_ansatz}) can be computed explicitly:\begin{align}
\begin{split}
    f_0 &=  512, \\
    f_4 &=  \frac{p_1}{3}, \\
    f_8 &=   -\frac{1}{20}p_1^2 + \frac{7}{45} p_2,\\
    f_{12} &=  \frac{79 p_1^3}{120960}  - \frac{5 p_1 p_2}{2016}  + \frac{31 p_3}{7560} ,\\
    f_{16} &=  -\frac{2339 p_1^4}{232243200}+\frac{1367 p_2 p_1^2}{29030400}-\frac{113 p_3 p_1}{1814400}+\frac{127 p_4}{1209600}-\frac{419 p_2^2}{14515200} ,\\
    f_{20} &=\frac{677 p_1^5}{3892838400}-\frac{29873 p_2 p_1^3}{30656102400}+\frac{8989 p_3 p_1^2}{7664025600}+\frac{229 p_2^2 p_1}{204374016}, \\
    &-\frac{6029 p_4 p_1}{3832012800}+\frac{73 p_5}{27371520}-\frac{389 p_2 p_3}{273715200}.
\end{split}
\end{align}
From this, we can find \begin{equation}
X_4=p_1(F)-16 p_1(R)
\end{equation} for example.
Finally, the gravitational anomaly $g_{20}$ for both choices of quadratic refinement is
\begin{align}
    g_{20} &= -(f_0 f_{20} + f_4 f_{16} + f_8 f_{12})/2 \\
    &= -\frac{p_1^5}{748440}+\frac{83 p_2 p_1^3}{7484400}-\frac{79 p_3 p_1^2}{2494800}+\frac{919 p_4 p_1}{7484400}-\frac{127 p_2^2 p_1}{7484400}+\frac{p_2 p_3}{22275}-\frac{73 p_5}{106920}.
\end{align}

\subsection{Topological restriction on the gauge bundle}
\label{subsec:o-so-spin}
So far we have often referred  the gauge bundle of the Type I theory to have an $SO$ gauge group,
but this was simply because it was customary to do so.
The differential KO theory formalism allows the background field $\check A$ to be constructed from
any real bundle  $V$ with gauge group $O(n)$, not necessarily 
reduced to $SO(n)$ and additionally liftable to $Spin(n)$.
In this subsection, we will show that $V$ satisfies $w_1(V)=0$ and $w_2(V) = 0$,
and therefore the gauge group is actually $Spin$.

As the KO theory class of the underlying real bundle $V$ of the background field $\check A$ 
equals $\mu_\text{top}$ by the boundary condition $I(\check A)=\mu_\text{top}$,
what we need to establish is that the conditions  $w_1(\mu_\text{top})=0$
and $w_2(\mu_\text{top})=0$ follow from the bulk equation of motion.

This is in a sense analogous to the way the condition $\dim V=32$ arose 
in the last section from the condition that the constant piece of $\mu_\text{diff}$ was $32$.
We can summarise what we would like to establish as follows.
Let $V$ be the common orthogonal bundle underlying both $\mu_\text{diff}$ and $\mu_\text{top}$.
Then, the bulk equation of motion not only determines 
that $\dim V=32$,  but also $w_1(V)=0$ and $w_2(V)=0$,
so that the gauge bundle is forced to be $SO$ and further liftable to $Spin(32)$.

More precisely, this is what the authors believe. 
Although we will establish $w_1(V)=0$,
we can only partially establish $w_2(V)=0$, in the sense that
we will only be able to show $\int_{M_2} w_2(V)=0$ for \emph{orientable}
two-dimensional submanifolds $M_2\subset Y$.
The authors believe that we can similarly show $\int_{M_2} w_2(V)=0$ even when 
$M_2$ is non-orientable by solving the bulk equation of motion, concluding $w_2(V)=0$.
The authors would hope to come back to this question in the future.

The content of this subsection is not used later.
It also relies on somewhat heavier use of  machineries of algebraic topology than the other sections.
Therefore, readers who trust the authors
can skip the rest of the subsection without causing problems in the understanding of later sections.

\subsubsection{Conditions in terms of cohomology}
Let us now study the analysis of the bulk equation of motion.
We start from the following observation. 
If $w_1(V)$ is zero, $\int_{M_1} w_1(V)=0$ for any one-dimensional submanifold.
If $w_1(V)$ is nonzero, there is a one-cycle $M_1$ (with $\bZ_2$ coefficient) for which $\int_{M_1} w_1(V)\neq 0$.
Any such one-cycle is represented by an actual submanifold.

Similarly, if $w_2(V)$ is zero, $\int_{M_2} w_2(V)=0$ for any two-dimensional submanifold.
If $w_2(V)$ is nonzero, there is a two-cycle (with $\bZ_2$ coefficient) $M_2$ for which $\int_{M_2} w_2(V)\neq 0$.
Any such two-cycle is represented by an actual submanifold.
Therefore, $w_1(V)=0$ and $w_2(V)=0$ if and only if \begin{equation}
\int_{M_1}w_1(V)=0,\qquad
\int_{M_2} w_2(V)=0.
\end{equation}

\subsubsection{Translation of the conditions to KO theory}
As our formalism is based on KO theory rather than $\bZ_2$-coefficient cohomology, we would like to express this condition in terms of KO theory.
We regard $V$ as an element of $KO^0(Y)$.
As all one-dimensional manifolds are automatically spin, we can equip $M_1$ with a spin structure.
The mod-2 index of $V$ on $M_1$ is \begin{equation}
\int_{M_1}V = (\dim V) [M_1] + \int_{M_1} w_1(V) \in KO^{-1}(pt)=\bZ_2,
\end{equation}
where the left hand side is the KO-theory integral,
$[M_1]\in KO^{-1}(pt)$ is the spin bordism class of $M_1$,
and $\int_{M_1}w_1(V)$ is the integral of the cohomology class.

Two-dimensional manifolds are either orientable or non-orientable.
When it is orientable, it is automatically spin, and we equip $M_2$ with a spin structure.
When it is non-orientable, it is automatically pin$^-$,
and we can equip $M_2$ with a pin$^-$ structure.\footnote{%
There is actually a case where the two-dimensional submanifold $M_2$ of $Y$ 
representing a nontrivial class $H_2(Y;\bZ_2)$ in an oriented spin manifold $Y$
is non-orientable.
An explicit example is to take $Y=\mathbb{RP}^3$.
Then the generator of  $H^2(Y;\bZ_2)=\bZ_2$ is Poincar\'e dual to $\mathbb{RP}^2$,
which is non-orientable.
Note $\mathbb{RP}^3$ is spin but $\mathbb{RP}^2$ is only pin$^-$.
}

When $M_2$ is spin,
the mod-2 index of $V$ on a spin manifold $M_2$ is a KO-theoretic integral,
and is given by  \begin{equation}
\int_{M_2}V = (\dim V) [M_2] + q_{M_2}(w_1(V)) + \int_{M_2} w_2(V) 
\in KO^{-2}(pt)=\bZ_2.
\label{mod-2-spin}
\end{equation}
Here,  $[M_2]\in KO^{-2}(pt)$ is the spin bordism class of $M_2$,
or equivalently the value of the Arf invariant; $q_{M_2}$ is the quadratic refinement
on $H^1(M_2;\bZ_2)$ associated to the spin structure;
and $\int_{M_2}w_2(V)$ is the integral of the cohomology class.

When $M_2$ is only pin$^-$, and the analogue of \eqref{mod-2-spin} is
\begin{equation}
\int_{M_2} V:=(\dim V)[M_2] + q_{M_2}(w_1(V))+\int_{M_2} w_2(V) \in \bZ_8\subset \bR/\bZ.
\end{equation}
As there is no KO theoretic fundamental class when $M_2$ is not spin,
the left hand side does not have an independent definition as a KO-theoretic pushforward,
but rather is defined by the right hand side.
There,  $[M_2]$ is the pin$^-$ bordism class of $M_2$ in $\Omega^\text{pin$-$}_2(pt)=\bZ_8$,
$q_{M_2}$ is the $\bZ_4$-valued quadratic refinement of the intersection product on $H^1(M_2;\bZ_2)$,
and $\int_{M_2} w_2(V)$ is the $\bZ_2$-valued pairing;
we identified $\bZ_8$, $\bZ_4$ and $\bZ_2$ as the corresponding subgroup of $\bR/\bZ$,
respectively.

Therefore, assuming that $\dim V$ is even, 
demonstrating $w_1(V)=0$ and $w_2(V)=0$ reduces to demonstrating
\begin{equation}
\int_{M_1} V=0
\end{equation}
for all one-dimensional spin submanifolds $M_1\in Y$,
and
\begin{equation}
\int_{M_2} V=0
\end{equation}
 for all two-dimensional spin and pin$^-$ submanifolds $M_2\in Y$. Here, the vanishing of $q_{M_2}(w_1(V))$ is a consequence of $w_1(V) = 0$, and getting $\int_{M_2} V=0$ for $M_2$ that is pin$^-$ requires $\dim V$ to be a multiple of 8, which is satisfied by type I string with $\dim V = 32$. 

Now, for elements $\check a\in \check{KO}{}^0(Y)$, the maps
\begin{equation}
\check a\mapsto \int_{M_1} V \in \bR/\bZ,\qquad
\check a\mapsto \int_{M_2} V \in \bR/\bZ
\end{equation}
are homomorphisms. From the perfectness of the pairing, 
there are  elements $\check x(M_1), \check x(M_2)\in\check{KO}{}^{-1}(Y)$ 
such that \begin{equation}
\int_{M_1} V=(\check a,\check x(M_1))_Y,\qquad
\int_{M_2} V=(\check a,\check x(M_2))_Y.
\label{4.70}
\end{equation}
As the left hand side only takes discrete values, 
only the topological part $V=I(\check a)$ matters, and
$\check x(M_1)$ and $\check x(M_2)$ are flat.
Summarising,
to establish $w_1(\mu_{top})=0$ and $w_2(\mu_{top})=0$,
we need to show \begin{equation}
(\mu_\text{top}, \check x(M_1))=(\mu_\text{top},\check x(M_2))=0 \label{eqn:mu_vanishing_pairing}
\end{equation} for all $M_1$ and $M_2$.

\subsubsection{The bulk equation of motion}
Now recall that $\mu_\text{top}$ was determined in Sec.~\ref{3.2.1} by the equation of motion \begin{equation}
Q(\iota(a) \check T) - (\mu_\text{top},\check a\check T)_X =0
\end{equation} for an arbitrary flat element $\check a\in \check{KO}{}^{-1}(Y)$.
Recall also that $(\mu_\text{top},\check a\check T)_X=(\mu_\text{top},\check a)_Y$.
As $\check x(M_1) $ and $\check x(M_2)$ are flat, 
what we need to establish (by taking a particular $\check a$ and using \eqref{eqn:mu_vanishing_pairing}) is \begin{equation}
Q_X(\check x(M_1) \check T) =0
\qquad\text{and}\qquad
Q_X(\check x(M_2)\check T) =0
\label{QQ}
\end{equation}
for arbitrary one- and two-dimensional submanifolds $M_1$ and $M_2$.

Let us denote by $\upeta$ the generator of $KO^{-1}(pt)=\bZ_2$, as is conventional in homotopy theory.
Then \begin{equation}
\int_{M_1} V= \int_Y V\PD(M_1) =(V,\PD(M_1)\upeta^2)_Y,
\end{equation}
where $\PD(M_1)\in KO^9(Y)\simeq KO^1(Y)$ is the KO-theoretic Poincar\'e dual of $M_1$,
$\PD(M_1)\upeta^2 \in KO^{-1}(Y)$,
and the last expression is the differential KO-theory pairing.
Similarly, \emph{when $M_2$ is orientable and therefore spin}, we have \begin{equation}
\int_{M_2} V= \int_Y V\PD(M_2) =(V,\PD(M_2)\upeta)_Y,
\label{important}
\end{equation}
where $\PD(M_2)\in KO^8(Y)\simeq KO^0(Y)$ is the KO-theoretic Poincar\'e dual of $M_2$.
In such cases, therefore, we have \begin{equation}
\check x(M_1)=\PD(M_1)\upeta^2,\qquad
\check x(M_2)=\PD(M_2)\upeta.
\end{equation}

This decomposition when $M_1$ or $M_2$ are spin allows us to compute $Q_X(\check x \check T)$.
For this, we use the fact that $\upeta T\in KO^0(S^1)$ is represented by the virtual bundle $1- L$,
where $L$ is a real 1-dimensional bundle over $S^1$ whose total space is a M\"obius strip.
Therefore, for an element $V\in KO^0(Y)$,
$Q_X(V\upeta T)$ is the differential pushforward on $Y\times S^1$ of \begin{equation}
\Lambda^2 (V (1- L) )
=  V^2 - V^2 L.
\label{4.62}
\end{equation} 
Now, this element is a sum of the elements of the form $ab$ where $a\in KO^0(Y)$
and $b\in KO^0(S^1)$.
In this case, we have \begin{equation}
\int_{Y\times S^1} ab=\int_Y a \int_{S^1} b,
\end{equation} where \begin{equation}
\int_Y a \in KO^{-2}(pt)\simeq \bZ_2 \eta^2,\qquad
\int_{S^1} b \in KO^{-1}(pt)\simeq \bZ_2 \eta,
\end{equation} and the product \begin{equation}
KO^{-2}(pt)\times KO^{-1}(pt) \to \check{KO}{}^{-3}(pt)\simeq \bR/\bZ
\end{equation}
is such that the product of $\eta^2$ and $\eta$ is $1/2\in \bR/\bZ$.
Then \begin{equation}
Q_X(V\eta T)=
\int_{Y\times S^1}(V^2-V^2L)
=\int_{Y} V^2 \int_{S^1} L 
= \eta\int_Y V^2 \in \bR/\bZ.
\label{VV}
\end{equation}

We now apply this formula for $V=\PD(M_1)\eta$ or 
$V=\PD(M_2)$.
Then $V$ can be realised as $V=[E]$,
where $E$ is a virtual orthogonal bundle which is nontrivial
only in a small neighbourhood of $M_1$ or $M_2$.
We can now move $M_1$ or $M_2$ slightly to $M_1'$ or $M_2'$,
so that $V=[E']$ where $E'$ is nontrivial only in a small neighbourhood of $M_1'$ or $M_2'$.
Then $V^2=[E\otimes E']$.
But as twice the dimension of $M_1$ and $M_2$ is sufficiently smaller than $\dim Y=10$,
we can arrange so that $M_1$ and $M_2$ do not intersect.
Then $E\otimes E'$ is a trivial zero-dimensional bundle, and therefore 
\begin{equation}
V^2=0 \in KO^0(Y). \label{4.67}
\end{equation}
Plugging this in to \eqref{VV},
we learn \eqref{QQ}, which in turn shows that \begin{equation}
w_1(\mu_\text{top})=0,
\end{equation}
and that \begin{equation}
\int_{M_2} w_2(\mu_\text{top})=0\label{spinspin}
\end{equation}
when $M_2$ is orientable and spin.

Before concluding this subsection, we repeat that we have not been able to show \eqref{spinspin}
when $M_2$ is non-orientable and only pin$^-$.
It should be possible to do a similar computation above,
using the fact that $M_2$ still has a Poincar\'e dual in twisted KO-theory.
The authors hope to come back to this question in the future.

\section{Two other string theories}\label{sec:other_strings}

In this section, we generalise our analysis to two other 10-dimensional string theories. 
We discuss Sugimoto's string theory with gauge group $usp(32)$ \cite{Sugimoto:1999tx}, 
and Sagnotti's string theory with gauge group $u(32)$ \cite{Sagnotti:1995ga}. 
In both cases, we will find that the quadratic refinement comes from straightforward generalisation of our analysis in type I case.
We note that the perturbative anomaly cancellation of these two cases were treated in parallel in \cite{Schwarz:2001sf},
and that our discussion is a natural extension of the analysis there to include the global anomaly cancellation.

Before proceeding, we note that the global anomaly cancellation of these two types of string theories
was studied in \cite{Basile:2023knk} from the point of view using twisted string structure.
In our approach, no computations of such twisted string bordism classes are required,
and the burden is instead placed in the construction of an appropriate quadratic refinement 
in the version of K-theory appropriate for each string theory considered.

\subsection{Quadratic refinement of KSp theory and Sugimoto's string}
\label{sec:sugimoto}

To begin with, we construct a quadratic refinement in differential KSp theory. The procedure is mostly parallel to the differential KO case due the Bott periodicity of $KO^{n}(X) \cong KSp^{n \pm 4}(X)$. However, some subtle distinction between real vs pseudo-real vector bundles used in the general construction will lead to some subtle factors of 2 or 1/2 when comparing the two cases. 
Such factors reproduce the result of the analysis by \cite{Larotonda:2024thv} on anomaly cancellation in Type I string and Sugimoto string \cite{Sugimoto:1999tx} based on anomaly inflow onto D1 and D5 branes. 

\subsubsection{Green-Schwarz anomaly cancellation in type I versus Sugimoto String}

Sugimoto's string is a non-supersymmetric string theory obtained by an $O9^+$ orientifold projection on IIB, having gauge group $usp(32)$. 
In contrast, type I string theory with gauge group $so(32)$ is  obtained from type IIB from a $O9^-$ orientifold projection.

These two theories both have one gravitino and one dilatino from the closed string sector.
The gaugino\footnote{%
Calling the charged fermions in the Sugimoto theory as gauginos is actually a misnomer,
since this is not a superpartner of the $USp(32)$ gauge field.
We call them gauginos purely out of convenience in the presentation.
}
in both Type I string and Sugimoto string  have dimension 496:
\begin{align}
\text{Type I}:& \Lambda^2 V = \mathbf{496} \\
\text{Sugimoto}:& \Lambda^2 V = \mathbf{495} + \mathbf{1},
\end{align} 
The standard type I theory is supersymmetric, while the $usp(32)$ Sugimoto string is non-supersymmetric, since the gauge boson is in the adjoint representation of dimension $n(2n+1) = 528$.

We define the Pontryagin character of a symplectic bundle $V$ by $\ph(V):=\ch(V_\bC)$,
where $V_\bC$ is the underlying complex bundle of $V$ by forgetting the quaternionic structure.
Then the decomposition of Pontryagin character of the  bundle  $\Lambda^2 V$ into Pontryagin character of the bundle  $V$  is formally the same for both cases: $\ph(B_{\Lambda^2}) = (\frac{1}{2}\ph(B)^2 - \ph(2B))$. 
Therefore, the total 12-form anomalies $I_{12}$ in both cases are equal as polynomials
of Pontryagin characters.
As such, the chiral anomalies are both cancelled by the Green-Schwarz mechanism via the pair of RR field $C_3, C_7$, with $dC_3 = H_4$ and $dC_7 = H_8$:
\begin{equation}
    X_4^{\text{type I}} X_8^{\text{type I}} = X_4^{\text{Sugimoto}} X_8^{\text{Sugimoto}}
    =\frac 12 X_4^\text{ours}X_8^\text{ours}.
\end{equation}

From this, one might be tempted to conclude that the two case has the identical $X_4, X_8$. 
However, recall $d C_3 = X_4$ and $d C_7 = X_8$, and therefore the normalisation of $X_4, X_8$ is tied to those of $C_3, C_7$. 
The latter is sensitive to whether the corresponding branes carries gauge group that are real or quaternionic.
Indeed, by studying the anomaly inflow onto D1 and D5 branes, Larotonda and Lin found in \cite{Larotonda:2024thv}  a relative coefficient between the two cases by factors of 2.
Including our notations, the relations are
\begin{align}
X_4^\text{type I} &= \frac12 X_4^\text{ours},&
X_8^\text{type I} &= \phantom{\frac12} X_8^\text{ours},\\
X_4^\text{Sugimoto} &= \phantom{\frac12} X_4^\text{ours},&
X_8^\text{Sugimoto} &= \frac12 X_8^\text{ours}.
\label{eqn:2_and_half}
\end{align}

\subsubsection{The RR field for Sugimoto theory}

Let us now study the anomaly cancellation of Sugimoto's string theory using our formalism.
The gauge bundle is $Sp$ rather than $SO$,
therefore we are going to use a class in $\check{KSp}{}^{0}(X)$ rather than $\check{KO}{}^0(X)$
to represent the field in the 11d bulk.
A cocycle for such a class is given by a triple $(V,B,C)$ exactly as in Sec.~\ref{sec:KOcocycle},
but with an $Sp$ bundle $V$ and an $Sp$ connection $B$.

One point to be noted is that the product of two classes $\check a,\check b \in \check{KSp}{}^{0}(X)$
is not in $ \check{KSp}{}^{0}(X)$ but in $ \check{KO}{}^{0}(X)$.
Similarly, $\check a_{\Lambda^2V}$ is also in $\check{KO}{}^0(X)$,
which allows us to construct a quadratic refinement just as in Sec.~\ref{sec:quadraKO}.

Let us now look at the issue of the quantisation of $C_3$ and $C_7$, repeating the analysis in Sec.~\ref{sec:normalization}. 
For a class $\check a\in \check{KSp}{}^{0}(X)$ and a spin submanifold $ \iota: M\hookrightarrow X$,
the quantisation condition is that \begin{equation}
\kappa\int_M \hat A(M) \iota(R(\diff a))  \in \bZ
\end{equation}
where $\kappa$ is $1$  when $\dim Y=4$ mod $8$
while $\kappa$ is $1/2$ when $\dim Y=0$ mod $8$.
Note that this is opposite to the case of a class in $\check{KO}{}^0(X)$,
and is due to the pseudoreality of the product of the spinor bundle and the gauge bundle.
This leads to the quantisation law 
\begin{equation}
\int_{M_4} dC_3+\cdots \in \bZ,\qquad
\int_{M_8} dC_7 +\cdots \in 2\bZ,\qquad
\end{equation} 
leading to the identification  \begin{equation}
C_3{}^\text{ours, $Sp$} =  C_3{}^\text{conventional},\qquad
C_7{}^\text{ours, $Sp$} = 2 C_7{}^\text{conventional},
\end{equation} 
where the conventional normalisation was that $\int_{M_n} dC_{n-1} + \cdots \in \bZ$.
Compare  this with the result \eqref{eq:relative-factors} for the $SO$ case, which was 
\begin{equation}
C_3{}^\text{ours, $SO$} = 2 C_3{}^\text{conventional},\qquad
C_7{}^\text{ours, $SO$} =  C_7{}^\text{conventional}.
\end{equation}
Together, this explains the differences of the factor $1/2$ in \eqref{eqn:2_and_half}.

\subsection{Alternative quadratic refinement of K-theory and Sagnotti's string}

We now further extend our analysis to the Sagnotti's $u(32)$ string. 
We first review the construction of Sagnotti's string and how the $u(32)$ gauge group arises. 
We then explain that the D-branes in the Sagnotti's string were known to follows the (complex) $K$-theory classification, identical to that of Type IIB string. 
At the end, we give the construction of the bulk action in Sagnotti's string using a  quadratic refinement of differential K-theory different from the one used for Type IIB string, and we show that it correctly reproduces the exotic Green-Schwarz anomaly cancellation involving multiple pairs of RR fields.

\subsubsection{Construction of Sagnotti's string and D-brane charges}

The construction of the Sagnotti string theory \cite{Sagnotti:1995ga} follows the following two-step procedure. First of all, we change the GSO projection used in type IIB string theory into a diagonal one
\begin{equation}
    P_{GSO}^{0B} = \frac{1 + (-1)^{F}}{2}
\end{equation}
to construct oriented Type 0B theory. 
Here, the 0 in 0B means that it has zero supersymmetry,
and $(-1)^F = (-1)^{F_L + F_R}$ is the total fermion parity operator.
The second step is to perform a further gauging by $\Omega' = \Omega (-1)^{F_L}$, 
i.e., projecting to states that are invariant under $\Omega'$.
Here $\Omega$ is the standard orientation reversal of the worldsheet with $\Omega^2=+1$, 
and we have \begin{equation}
(\Omega')^2 = \Omega (-1)^{F_L} \Omega (-1)^{F_L} = \Omega^2 (-1)^{F_L} (-1)^{F_R} = (-1)^F.
\end{equation}
Notice that Type 0B theory has closed string tachyons, but it was projected out upon performing the $\Omega'$ projection.

The original description of the projection in \cite{Sagnotti:1995ga} by Sagnotti was done by expressing various 1-loop amplitudes in terms of characters of the $so(8)$ little group, which we do not review here.
We give a heuristic explanation below, and we refer the reader to \cite{Sagnotti:1995ga,Sagnotti:1996qj} for a complete treatment.  

In the intermediate setup of oriented type 0B string, there are two copies of the type IIB brane spectrum classified by $K^{0}(M) \oplus K^{0}(M)$.  
In particular, there are two types of $D9$ branes, which we can denote by $D9$ and $D9'$.
Note that$D9'$ is distinct from the anti-$D9$ brane.
If we were to only gauge the worldsheet orientation $\Omega$, we would then change the two copies of complex gauge bundles to two copies of real gauge bundles, thus obtaining a string theory with brane spectrum $KO^0(M) \oplus KO^{0}(M)$,
where the projection changes both factors of $K^0$ into $KO^0$.

The projection by $\Omega'$ instead equates two copies of $K^0$.
Therefore, the two types of 9-branes back in type 0B, $D9$ and $D9'$, need to come in pairs in Sagnotti's string,
such that $N$ pairs  host a $u(N)$ gauge group, as opposed to $so(N) \oplus so(N)$. 
In fact, tadpole cancellation condition tells us that we would need 32 pairs of spacetime-filling $D9$ and $D9'$ branes, so we have a gauge group of $u(32)$. 
The same phenomenon happens for lower-dimensional D-branes, so we have pairs of D7, D5, D3, D1 branes,
each hosting unitary gauge group,
see \cite{Dudas:2000sn} for explicit case-by-case treatments for the Chan-Paton factors and anomaly inflow analysis of these D-branes.\footnote{In \cite{Kaidi:2019pzj,Kaidi:2019tyf}, these types of unoriented closed string theories were re-analysed from the perspective of string worldsheet. 
There,  Sagnotti's string was understood to have $\mathrm{Pin}^+$ structure on the worldsheet, 
whereas the other type of string theory with the brane spectrum $KO^0(M) \oplus KO^0(M)$ was shown to have $\mathrm{Pin}^-$ structure on the worldsheet.} 

\subsubsection{Exotic Green-Schwarz anomaly cancellation}

Sagnotti's string has two distinctive features: it  has both RR fields whose degrees jump by 2,
and it has a non-trivial gauge group of $u(32)$ in the bulk 10-dimensional spacetime.
Therefore, the total fermion anomaly has the following schematic form:
\begin{equation}
    I_{12}^{\text{sagnotti}} \sim \hat{X}_2 \hat{X}_{10} + \hat{X}_4 \hat{X}_8 + \frac{1}{2} \hat{X}_6^2.
\label{eqn:sagnotti_GS}
\end{equation}

We now explain our construction of a quadratic refinement of differential K theory
suitable for Sagnotti's string rather than Type II string,
 by using a bilinear pairing different from that of Type II string.
As usual, the top-degree form $C_{11}$ as a Lagrange multiplier is automatically included, 
whose equation of motion fixes the rank of the gauge bundle. 
We will see that this alternative construction will correctly reproduce the rank 32, 
as opposed to the rank 0 for the type IIB string. 
We will see in the end that the above pattern is correctly reproduced by our construction of quadratic refinement,
ensuring the anomaly cancellation not just perturbatively but also globally.

Back in type IIB string, we considered the pairing of
\begin{equation}
    (\check a, \check b)_\text{IIB} = \chi^\bC(\check a \cdot \check b^*).
    \label{IIBpairing}
\end{equation}
A quadratic refinement of the right hand side was obtained by taking $\check a \cdot \check a^*$, which is a real bundle, and considering the real holonomy function of the latter:
\begin{equation}
    Q(\check a)_\text{IIB} = \chi^\bR (\check a \cdot \check a^*)
\end{equation}
where $\chi^\bC (\check a \cdot \check a^*) = 2 \chi^\bR (\check a \cdot \check a^*)$. This gives gauge rank 0 by equation of motion, as we reviewed in Section \ref{subsec:Q_K}.

We now write down the alternative option of the bilinear pairing in Sagnotti's  string:
\begin{equation}
    (\check a, \check b)_\text{sagnotti} = \chi^\bC(\check a \cdot \check b).
\end{equation}
Note the absence of the complex conjugation in the second factor, compared to \eqref{IIBpairing}.
The quadratic refinement can be taken to be 
\begin{equation}
    Q(\check a)_\text{sagnotti} = \chi^\bC(\Lambda^2 \check a),
\end{equation}
in analogy with the KO quadratic refinement for Type I string.
This in particular means that the coefficient of the quadratic term will be 1, as opposed to 1/2 for the IIB case where $\chi^{\mathbb{R}}$ was used.

To fix the part of this quadratic refinement other than the $\eta$-invariant, we could consider the variation of the $\eta$-invariant
\begin{equation}
    \eta(D_{Y \times 0 + \overline{Y \times 1}}) = \int_Z \hat{A} \ch(\tilde{B}_{\Lambda^2 V}) = \int_Z \hat{A} \frac{1}{2} (\ch(\tilde{B})^2 - \ch(2\tilde{B}))
\end{equation}
and demand that this part is cancelled by the variation of the rest. Therefore, $Q(\check a)_\text{sagnotti} $ can be written down as:
\begin{equation}
    Q(\check a)_\text{sagnotti} =-\eta(D_{\Lambda^2 V}) + \frac{1}{2} \int (C \wedge dC + 2C \wedge X)
    \label{sagnotti-action}
\end{equation}
where 
\begin{equation}
    X = \sqrt{\hat{A}(R)} \ch(\tilde{B}) - f .\label{eqn:X_shifting}
\end{equation}
Here $\tilde{B}$ is the connection of the bundle of the differential K-theory class, 
and $\ch(\tilde{B}) \in \Omega^{2i}(Y)$ is its total Chern character. 
We now proceed by showing that this quadratic refinement correctly reproduces the anomaly cancellation in Sagnotti strings.

Here, $f$ should again be obtained by the following identity:
\begin{equation}
    f \sqrt{\hat{A}} \ch(\tilde{B})|_{12} = \frac{1}{2} \hat{A} \ch(2\tilde{B})|_{12}.
\end{equation}
Here, $\ch(\tilde{B})$ is non-trivial in every even degree, 
but $\hat{A}$ and $\sqrt{\hat{A}}$ are non-trivial
only  in degrees that are a multiple of 4. So the recursion relation gives
\begin{align}
\begin{split}
    f_0 &= \frac{1}{2} 2^6 = 32,\\ f_2 &= 0, \\
    f_{4l} &= 2^{4k+1-2l}\hat{A}_{4l} - \sum_{j = 0}^{l-1} f_{4j} \sqrt{\hat{A}}_{4(l-j)} \\
    f_{4l+2} &= 0 - \sum_{j = 0}^{l-1} f_{4j+2} \sqrt{\hat{A}}_{4(l-j)} \label{eqn:f_sagnotti}
\end{split}
\end{align}
so we indeed find a positive gauge rank of $f_0 = 32$. We also have $f_2 = 0$, and using the recursion in the last line we can see that
\begin{equation}
    f_2, f_6, f_{10} = 0.
\end{equation}
The remaining $f$ variables turn out to be exactly the same as in the type I case:
\begin{equation}
    f_4 = \frac{p_1}{3}, \quad f_8 = -\frac{p_1^2}{320} + \frac{7 p_1}{720}, \quad f_{12} = -\frac{79 p_1^3}{1935360} + \frac{5 p_1 p_2}{32256} - \frac{31 p_3}{120960}.
\end{equation}
This is because both this and the answer from type I are a consequence of matching coefficients of a pair of formal power series $g = \sum_{n} g_{2n}$ and $g' = \sum_n 2^{n} g_{2n}$, 
and the precise result is insensitive to the specific form of $g_{4n}$, given  by $\ch(\tilde{B})_{2n}$ for complex bundles, or $(1 - (-1)^n) \ph(\tilde{B})_{2n}$ for real bundles. 
In this way, we can reproduce the perturbative Green-Schwarz anomaly cancellation in the literature, see \cite{Schwarz:2001sf}.\footnote{%
The factorization  $I_{12} = X_2 X_{10} + X_4 X_8 + X_6^2/2$ does not uniquely determine $X_{2,4,6,8,10}$.
Here, they were uniquely fixed by the formalism, once we use the total characteristic classes as in (\ref{eqn:X_shifting}). 
We remind the reader that, this result is different from the ones given in older literature in terms of the explicit form of $X_6$ and $X_{10}$. 
In these old papers, they were fixed  by imposing the vanishing of the $\tr (F^3)$ terms, 
which is unnatural from the perspective of total characteristic classes.}

To determine the anomaly carried by the RR fields, we need to find the bulk invertible phase
obtained by path-integrating the action $Q(\check a)_\text{sagnotti}$.
This can be found as in Sec.~\ref{subsec:general-pure-grav} by simply computing the 
perturbative part.
Note that our action $Q(\check a)_\text{sagnotti}$ not only contains the $C\wedge X$ 
part, but also the kinetic term for $C_5$ with $+\frac12\int C_5 dC_5$.
The part $C\wedge X$ contributes by \begin{equation}
- (f_0 f_{12}+f_4 f_8) = - \frac{p_1^3}{3780} + \frac{13 p_2 p_1}{7560} - \frac{31 p_3}{3780}
\end{equation}
as before, where the factor-of-two difference with respect to \eqref{typeI-bulkanomaly}
is due to the factor-of-two difference between the actions \eqref{eq:type1a} and \eqref{sagnotti-action}.
The perturbative path-integral of the $C_5 dC_5$ term, in contrast, contributes by 
\begin{equation}
+\frac18L_{12} = \frac{p_1^3}{3780}-\frac{13 p_2 p_1}{7560}+\frac{31 p_3}{3780},
\end{equation}
where $L$ is the Hirzebruch's $L$ genus, according to \cite{Alvarez-Gaume:1983ihn,Hsieh:2020jpj}.
In total, the alternative quadratic refinement of diffrential $K$-theory produces the anomaly of 
\begin{equation}
+\frac18L_{12}  -(   f_0 f_{12} + f_4 f_8)=0.
\end{equation}

We have thus confirmed that our quadratic refinement for Sagnotti's string produced
exactly zero gravitational contribution upon the path integral, and 
it correctly reproduced the exotic Green-Schwarz anomaly cancellation. This concludes our discussion of the global anomaly cancellation of Sagnotti's string.

Let us end this section by comparing anomaly cancellations in Type IIB, Type I and Sagnotti's string.
\begin{itemize}
\item In Type IIB, we have two dilatinos and two gravitinos from the closed string NSNS sector.
The RR fields are described by the bulk Abelian Chern-Simons theory described by differential
K-theory, whose pairing is $(\check a,\check b^*)$  and whose quadratic refinement is \[
\chi^\bR(\check a\cdot \check a^*) = -\frac12\eta(D_{V\otimes V^*}) + \frac{1}4 \int C\wedge dC^*.
\]
The bulk equation of motion leads to an equal number of $D9$ and anti-$D9$-branes.
The bulk action contains the Chern-Simons-type kinetic term $\propto C_5dC_5$,
leading to a self-dual 4-form field on the 10-dimensional boundary.
The bulk path integral gives an invertible phase representing the gravitational anomaly
of the RR 4-form fields, which cancels that of the NSNS fermions.
\item In Type I, we have one dilatino and one gravitino from the closed string NSNS sector.
The RR fields are described by the bulk Abelian Chern-Simons theory described by differential
KO-theory, whose pairing is $(\check a,\check b)$  and whose quadratic refinement is \[
\chi^\bR(\Lambda^2\check a) = -\frac12\eta(D_{\Lambda^2 V}) + \frac{1}4 \int (C\wedge dC + 2C\wedge X),
\label{qux}
\]
leading to an $so(32)$ gauge algebra.
The bulk action only contains the BF-type kinetic terms,
but also a background-field contributions denoted schematically by $X$ above.
The bulk path integral gives an invertible phase representing the gravitational anomaly
of the gauginos together with the Green-Schwarz term $\propto X^2$,
which cancels the anomaly of the NSNS fermions.
\item Finally in the case of Sagnotti's string, we have no fermions from the closed string NSNS sector.
The RR fields are described by the bulk Abelian Chern-Simons theory described again by differential
K-theory, but with the pairing given by  $(\check a,\check b)$  without the complex conjugate in the second factor.
The quadratic refinement is then \[
\chi^\bC(\Lambda^2\check a) = -\eta(D_{\Lambda^2 V}) + \frac{1}2 \int (C\wedge dC + 2C\wedge X).
\label{fred}
\]
The bulk equation of motion  leads to a $u(32)$ gauge algebra.
The bulk path integral gives an invertible phase, representing the gravitational anomaly
of the gauginos, the Green-Schwarz contribution $\propto X^2$, and the contribution from the $C_5dC_5$ term.
The background $X$ term turns out to be exactly the same as the background $X$ term of Type I case,
but leads to twice the anomaly because of the factor-of-two difference between the normalisation of 
the $CdC$ term between \eqref{qux} and \eqref{fred}.
This exactly cancels the contribution from the $C_5 dC_5$ term, leading to complete anomaly cancellation
within the fields described by the differential K-theory.
\end{itemize}

\section{Dimensional reduction on $S^1$} \label{sec:reduction}

Now that we understood the anomaly cancellation of 10d Type I string theory,
we would like to discuss what happens when we consider its compactifications. 
The simplest cases would be to study $S^1$ or $T^2$ compactifications.
In these examples, the massless fermions have mod-2 anomalies,
which should be cancelled by its coupling to the RR field. 
Let us focus on the $S^1$ case.

As we have discussed at length,
the 10d RR field arises from the bulk 11d KO class $\check{KO}{}^0(Y)$.
Then the boundary 10d field is, roughly speaking,
specified by a differential KO class $\check{KO}{}^{-1}(Y)$ with a self-duality constraint.
By a dimensional reduction on $S^1$ of the spacetime of the form $Y=\tilde Y\times S^1$, 
an element of $\check{KO}{}^{-1}(S^1\times \tilde Y)$
would be specified by  $\check a\oplus \check b \in \check{KO}{}^{-1}(\tilde Y)\oplus \check{KO}{}^{-2}(\tilde Y)$,
and the self-duality in 10d now relates the two components $\check a$ and $\check b$.
This should mean that, in 9d, we can simply use a non-self-dual field $\check a$ alone.
A non-self-dual field should be easier to study,
and we might hope that the anomaly of the boundary RR field can be understood with less complications. 
Below, we will study the anomaly cancellation in 9d in two ways.
First, in Sec.~\ref{sec:bottom-up}, we try to guess how this might be achieved, by trying to come up with the 10d bulk theory
for the 9d RR fields, in a bottom-up approach.
Second, in Sec.~\ref{subsec:last}, we will make a more direct, top-down analysis of the dimensional reduction of the 11d bulk theory on $S^1$,
to write down the actual 10d bulk theory provided by the Type I string theory.
Finally, in Sec.~\ref{app:4dWitten},
we try to cancel the 4d Witten anomaly using the lessons we would have learned to this point
by introducing an $H_3$-field described by a differential KSp class,
and see what happens.

\subsection{Bottom-up approach}
\label{sec:bottom-up}
\subsubsection{Using ordinary differential cohomology}
In \cite{Lee:2022spd} the following approach for the mod-2 anomaly cancellation was considered.
(The paper considered the situation in 8d, but the 9d analysis goes completely in parallel with little change.)
On the 9d boundary, we consider a 3-form field satisfying \begin{equation}
dH=A_4,
\end{equation} where $A_4$ is a closed 4-form constructed from the gauge field and/or the metric.
Assuming that $A_4$ can be upgraded to an element  $\check A\in \check H^4(Y)$,
they considered  a 10d bulk theory with dynamical fields $\check a\in \check H^4(X)$ and $\check b\in \check H^7(X)$,
with the bulk action 
\begin{equation}
iS= 2\pi i [  (\check a,\check b)  - (\check A,\check b) + W(\check a) ], \label{9d-action-1}
\end{equation}  where at the boundary $\check a=\check b=0$ is imposed,
where $W(\check a)$ is an invertible phase which satisfies $W(0)=0$.
Performing a path integral over $\check b$, we can replace this action with an even simpler one \begin{equation}
iS=2\pi i W(\check A) \label{9d-action-2}
\end{equation} without any bulk dynamical field,
so that the boundary 3-form field appears as the gauge mode allowed by the boundary condition.

By a computation using an explicit knowledge of the generators of bordism groups etc and of the eta invariants,
the authors of \cite{Lee:2022spd} showed that 
there is a suitable invertible phase $W(\check A)$
which can cancel each of the following sets of fermions given below,
by choosing $\check A$ appropriately:
\begin{itemize}
\item the gauge anomaly of  an $SO(32)$ gaugino is cancelled by taking 
$R(\check A)=\frac12{p_1(\text{gauge)}}$,
\item the anomaly of  a gravitino plus dilatino (i.e.~the massless fermions in the gravity multiplet) is cancelled by taking  $R(\check A)=\frac12{p_1(\text{gravity})}$
\item and the anomaly of the massless fermion in the $S^1$  compactification of the type I string is cancelled by taking $R(\check A)=\frac12{p_1(\text{gauge})}-\frac12{p_1(\text{gravity})}$.
\end{itemize}
This allowed the construction an $H$ field in 9d which cancels the anomalies of massless fermions mentioned above,
as the boundary mode of the field $\check a$,
while the background field $\check A$ is constructed from the $SO$ gauge field and the metric connection.

\subsubsection{Using differential KO theory}
\label{sec:non-selfdual-differentialKO-RR}
If we think that the Type I theory or its compactifications prefer the use of KO theory rather than ordinary cohomology,
we might be tempted to replace the ordinary differential cohomology used above
by differential KO theory.
The bulk actions \eqref{9d-action-1} and \eqref{9d-action-2} work exactly as before,
so we can easily have a non-self-dual differential KO-based RR field
with arbitrary anomaly $W(\check A)$, where $Q$ is an arbitrary invertible phase depending 
on the background field $\check A\in \check{KO}{}^{0}(X)$
such that $W(0)=0$.

This makes it extremely easy to cancel the gauge anomaly of fermions arbitrarily charged under an $SO(n)$ gauge field $F$.
For this purpose, regard the $SO$ gauge field as an element $n+\check A\in \check{KO}{}^0(X)$,
where $n$ represents the trivial gauge field.
Let $I(\check A)$ be the invertible phase for the anomaly of this set of fermions;
then $I(0)$ is the pure gravitational part of the anomaly.
We simply use $W(\check A):=I(\check A)-I(0)$ in the construction above,
and we have an RR field whose anomaly cancels this set of $SO$-charged fermions.
Note that, in this construction, no explicit computation of the bordism groups or the eta invariant was necessary;
the differential KO-based RR-field is almost designed on purpose to achieve this task.
We also note that  the general boundary condition $dH=R(\check A)$  includes  \begin{equation}
dH_3= \frac{p_1(\text{gauge})}2 + \text{contribution from the metric},
\label{eq:dH3general}
\end{equation} as the 4-form part of it.

It is unclear, however, how the anomaly of the gaugino together with that of the gravitino and the dilatino can be cancelled 
by a \emph{single} KO-based RR-field in this manner.
This is because we need to come up with a suitable invertible $Q(\check A)$
and a way to embed both a given $SO(n)$ gauge field and the spacetime metric into a single background field $\check A\in \check KO^0(X)$.

That said, as we already saw that the KO-based self-dual RR-field in 10d correctly cancels 
the anomaly of massless fermions of the Type I theory,
we can expect that its $S^1$ compactification would achieve exactly this construction, or a slight variant of it.
We will see this in the last subsection, Sec.~\ref{subsec:last}.

\subsection{Top-down approach}
\label{subsec:last}

In the 10d case, we already have a bulk action on $X$ for the boundary  $Y$.
To carry out the dimensional reduction, we let $X=S^1\times \tilde X$ and $Y=S^1\times \tilde Y$.
We now decompose the field $\check C\in E^p(X)$ using the product structure $X=S^1\times \tilde X$.
Our analysis is mostly analogous to what we did in Sec.~\ref{sec:bulk-path-integral},
where we considered $X=S^1\times Y$ instead.
(There was a difference in the interpretation, however: $S^1$ was a direction perpendicular to the boundary there,
whereas $S^1$ is internal to the bulk spacetime used in the dimensional reduction here.)
We call the direction along $S^1$ as $t$.

Our bulk action is \begin{equation}
Q_X(\check c) - Q_X(\check A)
\end{equation} in the 11d bulk. 
We first consider the subgroup $W$ of $E^p(X)$ consisting of topologically trivial fields whose field strength has $dt$.
Following our analysis in Sec.~\ref{subsubsec:dt},
the entirety of $W$ can be path-integrated,
but for our purpose of dimensional reduction here, we make the following modification.
The field strength of fields in $W$ can be KK-decomposed.
We take a further subgroup $W'$ for which KK momentum is nonzero.
By performing a path integral over $W'$,
we simply generate a delta function which says 
that the not-$dt$-having part of $R(\check C)$ is independent of $t$.

As a representative of a coset by $W'$, we can  take
an element such that the $dt$-having part of $R(\check C)$ is independent of $t$.
Combining, the remaining integral is over the subspace $Z$ of $\check C$ such that $R(\check C)$ is independent of $t$.

Now $I(\check C)$ can be decomposed as $\alpha  + \beta T$ where $\alpha\in E^{p}(\tilde X)$, 
$\beta \in E^{p-1}(\tilde X)$, and $T$ is the generator of $E^1(S^1)\simeq \bZ$ as before.
We pick  arbitrary differential lifts $\check \alpha$ and $\check \beta$,
and consider \begin{equation}
\check D:=\check C-(\check\alpha +\check \beta \check T).
\end{equation}
By construction, $I(\check D)=0$, 
and $R(\check D)$ is independent of $t$.
Therefore $\check D$ can be chosen to be a differential form of the form $\alpha'  + \beta' dt$,
where $\alpha'\in \Omega^{p-1}_E(\tilde X)$
and $\beta'\in \Omega^{p-2}_E(\tilde X)$.
Regarding $\alpha'$ and $\beta'$ as differential $E$ classes,
we find \begin{equation}
\check C = (\check \alpha+\alpha')  + (\check \beta+\beta')\check T.
\end{equation} 
In other words we have that the subspace $Z$ of differential $E$ classes such that $R(\check C)$ is independent of $t$
has a decomposition
\begin{equation}
Z \simeq \check E^{p}(\tilde X)  \oplus  \check E^{p-1}(\tilde X) \check T,
\end{equation} although the decomposition is not canonical, as the choice of $\check T$ itself is not canonical.

In any case, parameterising elements of $Z$ by $\check a+\check b \check T$,
our bulk action is now of the form \begin{equation}
Q_X( \check a + \check b \check T) - Q_X(\check A)
= Q_X(\check a) +(\check a,\check b)_{\tilde X} +Q_X(\check b\check T) -Q_X(\check A)
\label{9d-true-action}
\end{equation}
where the boundary condition is 
\begin{equation}
{\check a}|_\text{boundary}={\check A}|_\text{boundary},
\qquad 
{\check b}|_\text{boundary}=0
\end{equation}
and $Q_X(\check a)$ is evaluated by first pulling back $\check a\in \check E^p(\tilde X)$
to $\check E^p(X)$.
One major difference from the action \eqref{9d-action-1} or \eqref{9d-action-2}
is that we have a quadratic term not only for $\check a$ but also for $\check b$.
As we discussed in Sec.~\ref{3.2.3}, $Q(\check b\check T)$ has a discrete dependence on $\check b$
due to the fact that $\int_{S^1}\check T^2 = 1/2$.

The analysis of this path integral in the presence of the boundary is left to the future.
For now, let us consider the case without boundary, 
which suffices for the purpose of determining the bulk invertible phase characterising the 9d anomaly.

The trick is to perform the path integral over the flat part of $\check b$ as before.
This generates a delta function setting $I(\check a)$ to be $\mu_\text{top}$.
Now, the remaining integral over $\check a$ is along the topologically trivial part,
of which $Q_X(\check a)$ is independent;
this is because $Q_X(\check a)$ is purely given by the eta invariant part from degree reasons,
and the eta invariant part is the mod-2 index.
The sole dependence of the action on this topologically trivial part of $\check a$
is through $(\check a,\check b)_{\tilde X}$,
which then sets $R(\check b)=0$, so the field $\check b$ is gone.
We find that the final bulk action is simply \begin{equation}
Q_X(\mu_\text{top}) - Q_X(\check A).
\end{equation}
From this we also see that $Q_X(\mu_\text{top})$ equals the anomaly of the gravitino plus dilatino.

If there \emph{were} a nice differential class $\check\mu$ lifting $\mu_\text{top}$
determined solely by the underlying manifold, the metric and the spin structure,
we can consider the bulk action to be \begin{equation}
Q_X(\check \mu)-Q_X(\check A),
\end{equation}
which would have made sense on a manifold with boundary with the boundary condition
saying that \begin{equation}
\text{$\check \mu-\check A$ is trivialised at the boundary.}
\end{equation}
A non-chiral RR field would have then arisen as the  degrees of freedom in the trivialisation.
As it is unclear if such a lift $\check\mu$ necessarily exists,
we need to make do with the action \eqref{9d-true-action} at this stage.

\subsection{Cancelling 4d Witten anomaly via differential KSp theory} \label{app:4dWitten}

In Sec.~\ref{sec:non-selfdual-differentialKO-RR},
we saw that it is easy to cancel the anomaly of fermions
arbitrarily charged under an $SO$ gauge field
by a non-chiral RR field.
By replacing $KO^0$ with $KSp^0=KO^4$,
it is equally easy to cancel the anomaly of fermions
arbitrarily charged under an $Sp$ gauge field,
just as we generalised our discussion from $KO^0$ to $KSp^0=KO^4$
in the case of chiral RR fields in Sec.~\ref{sec:sugimoto}.
This involves the introduction of a form field with the modified Bianchi identity \eqref{eq:dH3general}.

This begs the question:
what happens if we apply this construction
to a 4d gauge theory with $Sp(n)$ gauge group
with a fermion in the fundamental,
so that it has a Witten anomaly?
Our construction guarantees that the gauge anomaly is cancelled. 

The non-chiral RR field is (a torsor over) $\check{KSp}{}^{-1}(Y_4)$,
and its topology is characterised by $KSp^{-1}(Y_4)$.
At a very rough level, a generalised cohomology group $E^{d}(Y_n)$
is constructed out of \begin{equation}
H^p(Y_n;E^q(pt))
\end{equation}
 with $p+q=d$.
Therefore, 
$KSp^{-1}(Y_4)$ is a combination of $H^3(Y_4;\bZ)$
and $H^4(Y_4;\bZ_2)$,
where the former came from $KSp^{-4}(pt)=\bZ$ and the latter from $KSp^{-5}(pt)=\bZ_2$.
So we are not only introducing a $B$-field
with a three-form field strength $H$, corresponding to the $H^3(Y_4;\bZ)$ part,
but also a discrete $\bZ_2$ higher-form gauge field 
characterised by $H^4(Y_4;\bZ_2)$.

By a bordism analysis, it is already known that
the Witten anomaly cannot be cancelled by just adding a $B$-field \cite{Saito:2024iiu}.
Therefore the $\bZ_2$ field should be the key to cancelling the Witten anomaly.
How is this achieved?

In some sense it is simply cheating.
Any continuous anomaly in a $d$-dimensional theory can be `cancelled'
by introducing a $d$-form gauge potential $C_d$ with the coupling $\int_{Y_d} C_d$,
by assigning a suitable anomalous variation for $C_d$.
Similarly, any $\bZ_n$-valued anomaly can be `cancelled'
by introducing a discrete $\bZ_n$ gauge field $C_d$ characterised by $H^d(Y_d;\bZ_n)$
with the coupling $\int_{Y_d} C_d$,
by assigning a suitable gauge transformation law.
But this is not really a cancellation, as the path integral over $C_d$ 
makes the entire partition function to vanish, making the theory empty.

To see this more concretely, note that 
the topology of the KSp-valued RR field is characterised by a homotopy class of the map
$[Y_4,Sp(\infty)]$, where $Sp(\infty)$ is the classifying space of $KSp^{-1}$.
It is as if we have an $Sp(\infty)$ sigma model,
 although we only care about the topological class of maps here.
Let us now recall that there is a discrete Wess-Zumino-Witten term
reproducing the Witten anomaly on such a sigma model \cite{Witten:1983tw,Witten:1983tx},
so we are simply using that.
On $S^4$, for example, there are two homotopy classes of maps,
one of which has the phase  $+1$ and the other of which has the phase $-1$.
Unfortunately, the total partition function vanishes when we sum over these possibility,
and therefore the theory becomes somewhat sick.

We note that the Type I anomaly cancellation in 10d we discussed did not use this `cheat', 
since the top degree part of $KO^{-1}(Y_{10})$ comes from $H^{10}(Y_{10};KO^{-11}(pt))$,
where $KO^{-11}(pt)=0$.
In 9d or in 8d, the same argument might suggest that we are doing the same cheat,
since the top degree part is $H^{9}(Y_9;KO^{-10}(pt))$
and $H^{8}(Y_9;KO^{-9}(pt))$ respectively,
where $KO^{-10}(pt)=KO^{-9}(pt)=\bZ_2$.
So let us investigate this issue more generally, when the gauge group is $SO$.
The case when the gauge group is $Sp$ works analogously.

Such an issue would arise in a $d$-dimensional boundary theory if $KO^{-d-1}(pt)=\bZ_2$,
which means that there is a $\bZ_2$-valued form field $C_d$.
This field would have a coupling 
\begin{equation}
(-1)^{u \int_{Y_d} C_d}
\end{equation}
where $u=0$ or $1$. The theory is sick when $u=1$, but has no problem when $u=0$.

The $\bZ_2$ higher-form gauge field valued in $H^d(Y_d;\bZ_2)$ 
corresponds to $\pi_d(SO(\infty))=\bZ_2$.
Unwinding the definitions,
the coefficient $u$ of $\int_{Y_d} C_d$ is simply the phase associated by the $(d+1)$-dimensional
bulk differential-KO-based invertible theory $I_{d+1}$ to the background 
obtained by taking $[0,1]\times S^d$ and gluing the two ends via a nontrivial
global gauge transformation in $\pi_d(SO(\infty))$.
When $I_{d+1}$ is meant to cancel some anomalies coming from massless fermions,
this simply means that the coefficient $u$ of $\int_{Y_d} C_d$ 
is nontrivial if and only if the said fermions have a global anomaly associated to $\pi_d(SO(\infty))$.
This means that, when $SO$-charged fermions have the global anomaly under $\pi_d(SO(\infty))=\bZ_2$,
we can formally introduce the KO-valued RR field to cancel this anomaly,
although this introduces a higher-form $\bZ_2$ field valued in $H^d(Y_d;\bZ_2)$ 
and the path integral over this makes the theory sick.

In our case in 9d, the bulk theory $I_{d+1}$  cancels the anomaly of the gaugino of $SO(n)$, which is in the adjoint.
According to \cite{Lee:2022spd}, it has an anomaly associated to $\pi_9(SO(\infty))=\bZ_2$
if and only if $n$ is odd.
When $n=32$, this means that we are not cheating and the RR theory is not sick.

\section*{Acknowledgements}

The authors would like to thank J. Distler, D. Freed and G. W. Moore for correspondences,
and Mayuko Yamashita and Yi Zhang for discussions on Appendix~\ref{app:T2}.
The authors would also like to thank J. J. Heckman for asking the question 
which led us to the content of Sec.~\ref{subsec:o-so-spin}.
They also would like to thank H. Saleem for reporting the typos in an earlier version of the preprint.

YT is supported in part  
by WPI Initiative, MEXT, Japan at Kavli IPMU, the University of Tokyo
and by JSPS KAKENHI Grant-in-Aid (Kiban-C), No.24K06883. SSH and HYZ are supported by WPI
Initiative, MEXT, Japan at Kavli IPMU, the University of Tokyo.

\appendix

\section{The computation of $\check T^2$} \label{app:T2}

Here we compute $\check T^2 \in \check E^2(S^1)\simeq \bR/\bZ$,
where $\check T\in \check E^1(S^1)$ is such that $I(T)\in E^1(S^1)\simeq \bZ$ is a generator.
From the degree reasons $\check T^2=-\check T^2$, which means $\check T^2 =0$ or $1/2$.
We will show that it is actually $\check T^2=1/2$.
That it is $1/2$ instead of $0$ brings in some subtleties in our quantisation of the invertible Abelian Chern-Simons theory in Sec.~\ref{3.2.3}.

\subsection{$E=H(-;\bZ)$}

This case was already performed in the original paper \cite{CheegerSimons} on differential cohomology,
where an explicit cocycle model was used to compute it. See Example 1.16 there.
Here, we use a different approach, using mostly the axioms only.
First, we rewrite $\check T^2\in \check E^2(S^1)$ via a two-step process.
Consider the torus $S^1\times S^1$ with two projections $p_s$, $p_t$ to two $S^1$'s. 
We parameterise two $S^1$'s with $s$ and $t$.
Let $\Delta \subset S^1\times S^1$ be the diagonal.
Then we have $\check T^2 = \Delta^*(p_s^*(\check T) p_t^*(\check T))$.

Second, recall that an element of $\check H^2(M;\bZ)$ can be thought of as a $U(1)$ bundle with connection on $M$.
In this description, the identification $ \check H^2(S^1;\bZ)\simeq \bR/\bZ$ is given by taking the holonomy.
Therefore, $\check T^2$ as an element of $\check H^2(S^1;\bZ)\simeq \bR/\bZ$ is given by 
considering the $U(1)$ bundle $p_s^*(\check T) p_t^*(\check T)$ with connection,
and taking its holonomy along the diagonal.

Now, $R(p_s^*(\check T)p_t^*(\check T))=dsdt$. This does not uniquely fix the $U(1)$ bundle with connection on $S^1\times S^1$.
Correspondingly, even when we fix a particular $U(1)$ bundle with curvature $ds dt$, 
the holonomy along $s=t+\epsilon$ depends on $\epsilon$.
So we need to find a way to fix these issues.

To understand it, recall that elements of $\hat H^1(M;\bZ)$ can be thought of as an $S^1$-valued map on $M$.
Now, take a point $p\in S^1$ and consider $p^*(\check T)\in \hat H^1(pt;\bZ)\simeq \bR/\bZ$.
From the variational formula, $p^*(\check T)-q^*(\check T)$ is given by $\int_p^q dt$.
Therefore there is a unique point $p_0\in S^1$ where $p_0^*(\check T)=0$.

\begin{figure}
\centering
\begin{tikzpicture}[scale=2]  
  \draw[->] (0,-.2) -- (.3,-.2) node[right] {$s$};  
  \draw[->] (-.2,0) -- (-.2,.3) node[above] {$t$};  
  \draw (0,0)--(1,0)--(1,1)--(0,1)--(0,0);
  \draw[red] (0,0) -- (1,1);  
    \draw[dashed] (.2,0) -- (.2,1);  
  \draw[dashed] (0,.2) -- (1,.2);  
\end{tikzpicture}  
\caption{$p_s^*(\check T)p_t^*(\check T)\in \check H^2(S^1\times S^1;\bZ)$
and the diagonal. \label{fig:tt}}
\end{figure}
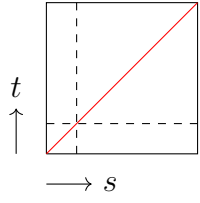

We now consider an inverse image $S^1\simeq p_s^*(p) \subset S^1\times S^1$.
When restricted on this $S^1$, the element $p_s^*(\check T)p_t^*(\check T)$ is given by $p^*(\check T) p_t^*(\check T)$,
which is given by a globally well-defined one-form $p^*(\check T) dt$.
So the holonomy around this $S^1$ is $p^*(\check T)$.
The same argument works with the inverse image $S^1\simeq p_t^*(p) \subset S^1\times S^1$.
This means that the diagonal $\Delta$ on $S^1\times S^1$ 
is such that, for a point $p\in \Delta$,
the holonomy along the two $S^1$'s, vertical and horizontal, is equal, see Fig.~\ref{fig:tt}.

To realise this, we simply take the $U(1)$ bundle over $S^1\times S^1$
to have the connection $sdt$, with a suitable gauge transformation between $s=0$ and  $s=1$.
This makes the holonomy along $S^1$ of constant $t$ to be $t$,
the holonomy along $S^1$ of constant $s$ to be $s$,
so that the holonomy along the diagonal $s=t$ to be unambiguously \begin{equation}
\int_0^1 tdt=\frac12.
\end{equation}

\subsection{$E=KO$}
\label{app:EKO}
Here the authors do not know how to evaluate it explicitly. 
Instead we rely on indirect methods.
We present two such methods, one very general but abstract,
another more concrete but computational.

One method using the general machinery of generalised differential cohomology is the following.\footnote{%
The authors thank Mayuko Yamashita for providing this argument.
}
We note that for any multiplicative generalised differential cohomology theory is a module over the differential sphere spectrum.
This element $\check T$ comes from the differential sphere spectrum.
Therefore its value $\check T^2 \in \check E^2(S^1)\simeq \bR/\bZ$ is universal, for any $E$ such that $E^0(pt)=\bZ$ and $E^{1}(pt)=0$.
As we computed this value for $E=H(-,\bZ)$, the result for $E=KO$ follows.

Another indirect method which is self-contained using the material covered in this paper is the following.
This  essentially follows \cite[Sec.~5.1]{Belov:2006jd},
and applies to any generalised differential cohomology theory with a suitable quadratic refinement.
This in particular means that it applies to $E=H(-;\bZ)$, $KU$, and $KO$.

We assume the dimension of $X$ and 
the degree $p$ of $\check A,\check B,\check C\in \check E^p(X)$ 
to be such that we have a $\bR/\bZ$-valued quadratic refinement  $Q_X(\check C)$
of a perfect $\bR/\bZ$-valued pairing $(\check A,\check B)_X$.
Assume $X=\partial Z$.
The pairing is then given by 
 \begin{equation}
(\check A,\check B)_X = \int_Z  \mathcal{A}(Z) R(\check A)R(\check B).
\end{equation}
where $\mathcal{A}(Z)$ is  a differential form constructed from the curvature of the metric,
depending on the choice of $E$.
The quadratic refinement should then have the general form 
\begin{equation}
Q_X(\check C)=\int_Z (\frac12 \mathcal{A}(Z)R(\check C)^2 +  R(\check C) D(Z)), \label{eq:var}
\end{equation} where $D(Z)$ is a differential form constructed from the metric of $Z$.

Now consider the case $X=S^1\times Y$,
and choose $\check\alpha,\check\beta\in \check E^{p-1}(Y)$.
Then we have 
\begin{align}
Q(\check\alpha \check T+\check\beta \check T)-Q(\check\alpha \check T)-Q(\check\beta \check T)&=\int_{S^1} \check T^2 \int_Y \check\alpha\check\beta\\
&=\int_{S^1} \check T^2 \int_Y \mathcal{A}(Y) R(\check\alpha)R(\check\beta).
\label{XXXX}
\end{align}

Now we take $Z = W \times Y$, where \begin{equation}
\label{eq:Wfig}
W=\vcenter{\hbox{
\begin{tikzpicture}
  \draw (0,0) ellipse (.5 and .25); \node (A) at (-1,0)   {$(S^1){}_3$};
  \draw (-1,2) ellipse (.5 and .25); \node (B) at (-2,2) {$(S^1){}_1$};
  \draw (1,2) ellipse (.5 and .25); \node (C) at (2,2) {$(S^1){}_2$};
  \draw (-.5,0) to[out=90,in=-90] (-1.5,2);
  \draw (.5,0) to[out=90,in=-90] (1.5,2);
  \draw (-.5,2) to[out=-90,in=-90] (.5,2);
\end{tikzpicture}}
}.
\end{equation}
For definiteness, we take all three $S^1$'s to be in the NS sector.
We now pick a class $\check T_1\in \check E^1(W)$ which restricts to $\check T_1$ on $(S^1)_1$, $0$ on $(S^1)_2$, $\check T_1$ on $(S^1)_3$.
We also pick a class $\check T_2\in \check E^1(W)$ which restricts to $0$ on $(S^1)_1$, $\check T_2$ on $(S^1)_2$, $\check T_2$ on $(S^1)_3$.
We now take the class $\check C:=\check\alpha \check T_1 + \check\beta \check T_2 \in \check E^{p}(Z)$.
Then \begin{align}
Q(\check\alpha \check T_1+\check\beta \check T_2)-Q(\check\alpha \check T_1)-Q(\check\beta \check T_2)
&= \int_{Z=W\times Y} (\frac12\mathcal{A}(Z) R(\check C)^2 +  R(\check C) D(Z))\\
&= \int_W R(\check T_1) R(\check T_2) \int_Y  \mathcal{A}(Y) R(\check\alpha) R(\check\beta),
\label{YYYY}
\end{align} where the term $R(\check C) D(Z)$ did not contribute in the second line due to degree reasons.
Comparing \eqref{XXXX} and \eqref{YYYY},  we have \begin{equation}
\int_{S^1} \check T^2 = \int_W R(\check T_1)R(\check T_2).\label{foo}
\end{equation}
The right hand side is determined solely by conditions on the differential form 
(such as how $R(\check T_1)$ and $R(\check T_2)$ restrict on the boundaries),
and does not in particular depend on whether $E=KO$ or $E=H$.
As we already computed $\check T^2$ to be $1/2$ above using a different method, this should also be the answer for $E=KO$.

For completeness here we provide a way to explicitly evaluate the right hand side of \eqref{foo}.
The trick is to note that there are many non-canonicality  in the construction.
Firstly $I(\check T)$ be a generator determines $\check T$ 
only up to an addition by an $\bR/\bZ$-valued function on $S^1$,
shifting $R(\check T)$ by $d\chi$.
Secondly,  we need to pick three diffeomorphisms $S^1\to S^1_{1,2,3}$ to talk about
the pull-backs of differential cohomology classes.
As our construction guarantees that the result does not depend on these non-canonicality,
we can make choices which are particularly convenient for an explicit computation.

We fix a particular $\check T\in \check E^1(S^1)$ such that $R(\check T)=dt$. 
We also pick a particular $p_0\in S^1$
such that $p_0^*(\check T)\in \check E^1(pt)\simeq \bR/\bZ$ is zero.
We set this $p_0$ to be the origin of the $t$ coordinate, i.e. $t=0$ at $p_0$.

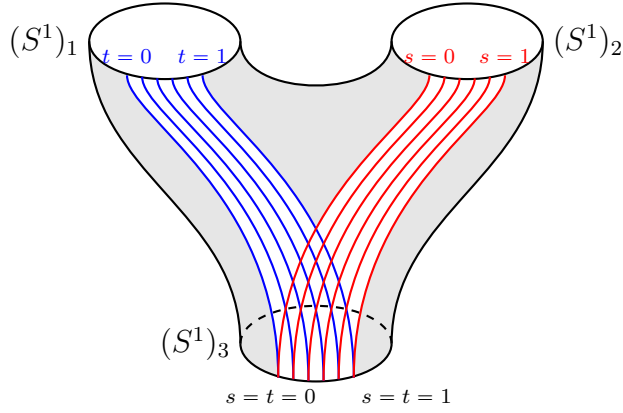
\begin{figure}
\centering
\begin{tikzpicture}[scale=2, thick]

\begin{scope}
  \pgfmathsetmacro{\xbluebot}{0.25}
  \pgfmathsetmacro{\xbluetop}{-0.75}
  \pgfmathsetmacro{\xredbot}{-0.25}
  \pgfmathsetmacro{\xredtop}{0.75}
  \pgfmathsetmacro{\offset}{0.01 * 2.5}
  \pgfmathsetmacro{\ybot}{-0.25 + \offset}
  \pgfmathsetmacro{\ytop}{1.75 + \offset}

  \begin{scope}
    \clip (-2,0.79) rectangle (2,3);

    \path[fill=gray!20]
      (\xbluebot,\ybot)
      .. controls (\xbluebot,1) and (\xbluetop,1.5) .. (\xbluetop,\ytop)
      -- (-0.5,2)
      to[out=-90,in=-90] (0.5,2)
      -- (\xredtop,\ytop)
      .. controls (\xredtop,1.5) and (\xredbot,1) .. (\xredbot,\ybot)
      -- cycle;
  \end{scope}

  \path[fill=white] (-1,2) ellipse (.5 and .25);
  \path[fill=white] (1,2) ellipse (.5 and .25);
\end{scope}

\begin{scope}
  \path[fill=gray!20]
    (-0.5,0)
    to[out=90,in=-90] (-1.5,2)
    arc[start angle=180, end angle=240, x radius=0.5, y radius=0.25, shift={(-1,2)}]
    .. controls (-1.25,1.5) and (-0.25,1) .. (-0.25,-0.2125)
    arc[start angle=240,end angle=180, x radius=0.5,y radius=0.25]
    -- cycle;
\end{scope}

\begin{scope}
  \path[fill=gray!20]
    (0.5,0)
    to[out=90,in=-90] (1.5,2)
    arc[start angle=0, end angle=-60, x radius=0.5, y radius=0.25, shift={(1,2)}]
    .. controls (1.25,1.5) and (0.25,1) .. (0.25,-0.2125)
    arc[start angle=-60,end angle=0, x radius=0.5,y radius=0.25]
    -- cycle;
\end{scope}

\draw[dashed] (-0.5,0) arc[start angle=180,end angle=0,x radius=0.5,y radius=0.25];
\draw         (-0.5,0) arc[start angle=180,end angle=360,x radius=0.5,y radius=0.25];

\draw (-1.0,2) ellipse (.5 and .25);
\draw (1.0,2) ellipse (.5 and .25);

\draw (-0.5,0) to[out=90,in=-90] (-1.5,2);
\draw (0.5,0)  to[out=90,in=-90] (1.5,2);
\draw (-0.5,2) to[out=-90,in=-90] (0.5,2);

\foreach \i in {0,...,5} {
  \pgfmathsetmacro{\xbot}{-0.25 + \i * (0.5/5)}
  \pgfmathsetmacro{\xtop}{-1.75 + \i * (0.5/5) + 0.5}
  \pgfmathsetmacro{\offset}{0.01 * abs(2.5 - \i)}
  \pgfmathsetmacro{\ybot}{-0.25 + \offset}
  \pgfmathsetmacro{\ytop}{1.75 + \offset}
  \draw[blue]
    (\xbot,\ybot)
    .. controls (\xbot,1) and (\xtop,1.5)
    .. (\xtop,\ytop);
}

\foreach \i in {0,...,5} {
  \pgfmathsetmacro{\xbot}{0.25 - \i * (0.5/5)}
  \pgfmathsetmacro{\xtop}{1.75 - \i * (0.5/5) - 0.5}
  \pgfmathsetmacro{\offset}{0.01 * abs(2.5 - \i)}
  \pgfmathsetmacro{\ybot}{-0.25 + \offset}
  \pgfmathsetmacro{\ytop}{1.75 + \offset}
  \draw[red]
    (\xbot,\ybot)
    .. controls (\xbot,1) and (\xtop,1.5)
    .. (\xtop,\ytop);
}

\node at (-0.8,0)   {$(S^1)_3$};
\node at (-1.8,2)   {$(S^1)_1$};
\node at (1.8,2)    {$(S^1)_2$};

\node at (-1.25,1.9) {\scriptsize\textcolor{blue}{$t=0$}};
\node at (-0.75,1.9) {\scriptsize\textcolor{blue}{$t=1$}};
\node at (0.75,1.9)  {\scriptsize\textcolor{red}{$s=0$}};
\node at (1.25,1.9)  {\scriptsize\textcolor{red}{$s=1$}};
\node at (-0.3,-0.35) {\scriptsize$s=t=0$};
\node at (0.6,-0.35)  {\scriptsize$s=t=1$};

\end{tikzpicture}
\caption{The choice of $\check T_1$ and $\check T_2$ on $W$.}
\label{fig:WW}
\end{figure}

We now pick a map $s,t: W\to S^1$ as shown in Fig.~\ref{fig:WW}.
For both $s$ and $t$, the gray shaded area is all mapped to $p_0\in S^1$.
The value $t$ varies from $0$ to $1$ continuously in the unshaded area,
and the blue lines show the lines of constant $t$.
The same goes with $s$,
where the red lines show the lines of constant $s$.
We then pick $\check T_1:=t^*(\check T)$ and $\check T_2:=s^*(\check T)$.
The right hand side of \eqref{foo} is simply $ds dt$ integrated 
in the `isosceles right triangle' region where the red lines and the blue lines meet perpendicularly.
So \begin{equation}
\int_{S^1} \check T_1 \check T_2 = \int_{t=0}^1\int_{s=0}^t ds dt= \frac12.
\end{equation}

\section{On the perfectness of the differential cohomology pairing} \label{app:perfectness}

To establish that our bulk theory at $k=1$ is an invertible theory, 
the perfectness of the differential cohomology pairing is essential. 
In this appendix, we review it in the case of differential ordinary cohomology,
and extend it to differential KO/KU theory.
A nice discussion can be found also in Appendix B of \cite{Freed:2006ya}.

Our interest in the perfectness of the pairing comes from the following consideration.
We start from an $(d+1)$-dimensional bulk theory $Q$ with background field $\check A \in \check E^{n+1}(X)$,
with the partition function $Z_Q[\check A]$.
We define a generalised version of Witten's S operation \cite{Witten:2003ya} via \begin{equation}
Z_{\sS Q}[\check B] = \int [D\check A] \exp(2\pi i(\check B,\check A) ) Z_Q[\check A]
\end{equation} where $\check B\in \check E^{m+1}(X)$, $n+m=d$.
We want this operation to essentially square to identity:
\begin{align}
Z_{\sS\sS Q}[\check C] &= \int [D\check A][D\check B] 
\exp(2\pi i(\check C,\check B)+(\check B,\check A) ) Z_T[\check A] \\
&=\int [D\check A][D\check B] 
\exp(2\pi i(\check B,\check A\pm \check C) ) Z_Q[\check A] .
\end{align}
So we want \begin{equation}
\int [D\check B] \exp(2\pi i (\check B,\check A)) = \delta(\check A)
\end{equation} where the delta function is with respect to the path integral measure $[D\check A]$.

\subsection{The case with differential ordinary cohomology}

To check this we need some idea on the space $\check H^{n+1}(X)$ in which $\check A$ takes values.
We first consider the subspace $R(\check H(X))$ of $H^{n+1}(X,\bR)$, which has the form \begin{equation}
0\to \textcolor{ForestGreen}{d\Omega^{n}(X)} \to R(\check H^{n+1}(X)) \stackrel{I}{\to} \textcolor{red}{ H^{n+1}(X,\bZ)/\tors}\to 0.
\end{equation} Then $\check H(X)$ itself has the form \begin{equation}
0\to H^{n}(X,U(1))\stackrel{\iota}{\to} \check H^{n+1}(X) \to R(\check H^{n+1}(X))\to 0,
\end{equation}  and finally $H^{n}(X,U(1))=\Hom(H_{n}(X,\bZ),U(1))$ has the form \begin{multline}
0\to \textcolor{blue}{\Hom(H_{n}(X,\bZ)/\tors, U(1))} \to H^{n}(X,U(1)) \\
\to \textcolor{purple}{\Hom(\Tors H_{n}(X,\bZ), U(1))}\to 0.
\end{multline}

Dually, $\check B\in \check H^{m+1}(X)$ has the decomposition \begin{equation}
0\to \textcolor{ForestGreen}{d\Omega^{m}(X)} \to R(\check H^{m+1}(X)) \to \textcolor{blue} {H^{m+1}(X,\bZ)/\tors}\to 0,
\end{equation} and \begin{equation}
0\to H^{m}(X,U(1))\to \check H^{m+1}(X) \to R(\check H^{m+1}(X))\to 0,
\end{equation}  then \begin{multline}
0\to \textcolor{red}{\Hom(H_{m}(X,\bZ)/\tors, U(1))} \to H^{m}(X,U(1)) \\
\to \textcolor{purple}{\Hom(\Tors H_{m}(X,\bZ), U(1))}
\to 0.
\end{multline}

Recall that $\dim X=n+m+1$.
Then the Poincar\'e duality says $H^{m+1}(X,\bZ)=H_n(X,\bZ)$ and $H^{n+1}(X,\bZ)=H_m(X,\bZ)$.
This means that the red parts have a natural pairing;
the differential cohomology pairing should exactly this,
and similarly for the blue parts.
The only infinite-dimensional parts of the pairing is the one between the green parts;
that is exactly where the most naive $ C d C'$ pairing lives,
although $C$ is considered up to exact forms. 

Finally, the universal coefficient theorem says that the purple parts,
\begin{equation}
\Hom(\Tors H_{n}(X,\bZ), U(1))=\Tors H^{n+1}(X,\bZ)
\end{equation}and
\begin{equation}
\Hom(\Tors H_{m}(X,\bZ), U(1))=\Tors H^{m+1}(X,\bZ),
\end{equation}
are dual to each other.
Again  the differential pairing should reduce to this pairing.

Then the following should happen upon the integration:
\begin{itemize}
\item  Integrating over the part of $\check B$  in $\textcolor{red}{\Hom(H_{m}(X,\bZ)/\tors, U(1))}$ should set the part of $\check A$ given by $\textcolor{red}{ H^{n+1}(X,\bZ)/\tors}$ to be zero.
To show that it actually works, we need to show that, 
given an element $\beta\in H^m(X,\bZ)\otimes U(1)$ and $\check A \in \check H^{n+1}(X,\bZ)$,
we have \begin{equation}
(\iota(\beta),\hat A)= (\beta , I(\hat A)). \label{AAA}
\end{equation}
At this point the differential pairing reduces to a pairing between \begin{equation}
\frac{\check H^{m+1}(X)}{\textcolor{red}{\Hom(H_{m}(X,\bZ)/\tors, U(1))}}
\longleftrightarrow
\Ker (\check H^{n+1}(X) \to \textcolor{red}{H^{n+1}(X,\bZ)/\tors}).
\end{equation}
\item We then integrate the part of $\check B$ in $\textcolor{ForestGreen}{d\Omega^{m}(X)}$,
which sets the part of $\check A$ given by $\textcolor{ForestGreen}{d\Omega^{n}(X)} $ to be zero.
This sets $R(\check A)$ to be entirely zero.
At this point the differential pairing reduces to a pairing between \begin{equation}
\frac{\check H^{m+1}(X)}{\textcolor{red}{\Hom(H_{m}(X,\bZ)/\tors, U(1))}+\textcolor{ForestGreen}{d\Omega^{m}(X)}}
\longleftrightarrow
\Ker (\check H^{n+1}(X) \to R(\check H^{n+1}(X))),
\end{equation}
or in other words a pairing between
\begin{equation}
H^{m+1}(X,\bZ)
\longleftrightarrow
H^{n}(X,U(1)) \simeq \Hom(H_n(X,\bZ),U(1))
\end{equation}
We should check, at this point, that for $\alpha \in H^n(X,U(1))$ and $\check B\in \check H^{m+1}(X,\bZ)$
we have \begin{equation}
(\check B, \iota(\alpha))=(I(\check B), \alpha) \label{BBB}
\end{equation} 
where the left hand side is defined by the differential pairing and the right hand side comes from a natural pairing of the non-differential cohomology theories.
This equality follows from the product formula \begin{equation}
\check B\cdot \iota(\alpha) = \iota(I(B)\alpha)
\end{equation} in the differential ordinary cohomology theory,
and the fact that the pairing is defined as the integral of the product.
\item We now integrate over the part of $\hat B$ in $\Hom(\Tors H_{m}(X,\bZ), U(1))=\Tors H^{m+1}(X,\bZ)$.
This sets the part of $\hat A$ in $\Hom(\Tors H_{n}(X,\bZ), U(1))=\Tors H^{n+1}(X,\bZ)$ to be zero.
At this point the last remaining pairing is the one between \begin{equation}
H^{m+1}(X,\bZ)/\tors
\longleftrightarrow
H^{n}(X,\bZ)\otimes U(1).
\end{equation}
\item Finally we integrate over the part of $\check B$ given by $\textcolor{blue} {H^{m+1}(X,\bZ)/\tors}$. 
This sets the final remaining part of $\check A$, $\textcolor{blue}{\Hom(H_{n}(X,\bZ)/\tors, U(1))} $, to be zero.
\end{itemize}
To summarise, to check that the computation works, we need to check \eqref{AAA} and \eqref{BBB}.
Actually, \eqref{AAA} is a subcase of \eqref{BBB} after exchanging $m$ and $n$.
So we only have to show \eqref{BBB} for arbitrary pairs $m,n$ satisfying $m+n=d+1$.

\subsection{The case with differential generalised cohomology}

Now let us try to generalise the considerations above to arbitrary generalised differential cohomology theory $\check E$.
For this the crucial input is that the kernel of the curvature map $R$,
which can be called the topological $E$-theory with $U(1)$ coefficient, is known to be given by \begin{equation}
E^n(X,U(1)) = \Hom((I_\bZ E)_n(X),U(1)).
\end{equation} It therefore has the structure \begin{equation}
0\to E^n(X)\otimes U(1)\to E^n(X,U(1)) \to \Tors E^{n+1}(X)\to 0
\end{equation} which is the same as \begin{multline}
0\to \Hom((I_\bZ E)_n(X)/\tors,U(1))  \to E^n(X,U(1)) \\
\to \Hom(\Tors (I_\bZ E)_n(X),U(1))\to 0.
\end{multline}

This allows us to do the following.
We first consider the subspace $R(\check E(X))$ of $E^{n+1}(X,\bR)$, which has the form \begin{equation}
0\to \textcolor{ForestGreen}{d\Omega_E^{n}(X)} \to R(\check E^{n+1}(X)) \to \textcolor{red}{ E^{n+1}(X,\bZ)/\tors}\to 0.
\end{equation} 

Then $\check E(X)$ itself has the form \begin{equation}
0\to E^{n}(X,U(1))\to \check E^{n+1}(X) \to R(\check E^{n+1}(X))\to 0,
\end{equation}  and finally $E^{n}(X,U(1))=\Hom((I_\bZ E)_{n}(X,\bZ),U(1))$ has the form \begin{multline}
0\to \textcolor{blue}{\Hom((I_\bZ E)_{n}(X,\bZ)/\tors, U(1))} \to E^{n}(X,U(1)) \\
\to \textcolor{purple}{\Hom(\Tors (I_\bZ E)_{n}(X,\bZ), U(1))}
\to 0.
\end{multline}

To have appropriate duals,  we let  $\widecheck B\in (\widecheck {I_\bZ E})^{m+1}(X)$, so that it has the decomposition \begin{equation}
0\to \textcolor{ForestGreen}{d\Omega^{m}_{I_\bZ E}(X)} \to R((\widecheck {I_\bZ E})^{m+1}(X)) \to \textcolor{blue} {(I_\bZ E)^{m+1}(X,\bZ)/\tors}\to 0,
\end{equation} 
and \begin{equation}
0\to (I_\bZ E)^{m}(X,U(1))\to (\widecheck {I_\bZ E})^{m+1}(X) \to R((\widecheck {I_\bZ E})^{m+1}(X))\to 0,
\end{equation}
where $(I_\bZ E)^m(X,U(1))=\Hom(E_m(X),U(1))$ and therefore
 \begin{multline}
0\to \textcolor{red}{\Hom( E_{m}(X,\bZ)/\tors, U(1))} \to (I_\bZ E)^{m}(X,U(1)) \\
\to \textcolor{purple}{\Hom(\Tors E_{m}(X,\bZ), U(1))}
\to 0.
\end{multline}
We note that the purple parts satisfy
\begin{align}
\Hom(\Tors (I_\bZ E)_{n}(X,\bZ), U(1)) &= \Tors E^{n+1}(X,\bZ),\\
\Hom(\Tors E_{m}(X,\bZ), U(1)) &= \Tors (I_\bZ E)^{m+1}(X,\bZ).
\end{align}

Let us now restrict to the case $E=KU$ or $E=KO$. 
Recall that $\dim X=n+m+1$, and assume $X$ is $E$-oriented.
Then the Poincar\'e duality says 
 $E^{n+1}(X)=E_m(X)$.
In this case, $I_\bZ KU= KU$ and $I_\bZ KO=\Sigma^4 KO$,
and therefore we also have $(I_\bZ E)^{m+1}(X)=(I_\bZ E)_n(X)$.
Together with the periodicity, we expect to have a natural pairing 
\begin{equation}
\widecheck{KU}^m(X) \leftrightarrow\widecheck{KU}^n(X)
\end{equation} with $n+m\equiv d+1  \mod 2$, or 
\begin{equation}
\widecheck{KO}^m(X) \leftrightarrow\widecheck{KO}^n(X)
\end{equation} with $n+m\equiv d+1+4 \mod 8$.

In either case, what needs to be checked is the version of \eqref{BBB}, namely, 
for $\alpha\in E^n(X,U(1))$ and $\check B\in \check E^m(X)$,
we have \begin{equation}
(\check B,\iota(\alpha)) = (I(\check B), \alpha).
\end{equation}
This equality follows from the general fact that the product of differential cohomology classes
satisfy \begin{equation}
\iota(\alpha) \check B=\iota(\alpha I(\check B)),
\end{equation}
and that the pairing is defined by first taking the product and then performing the pushforward.
For more details, we refer the reader to
\cite{BunkeSchick}  for differential KU,
and Appendix B of \cite{Freed:2006ya} for additional ingredient to generalise the proof to differential KO \footnote{The authors thanks M. Yamashita for these references.}.

\section{Expansions for characteristic classes} \label{app:expansions}

In this appendix, we collect notations and conventions necessary for expanding (total) characteristic classes as formal sum of homology classes at different degree. These expressions will be useful when we study the quadratic refinement for the differential KO in a particular dimension.

Let $B$ be a connection on the complex bundle $V_\bC$ with associated curvature $F$,
the Chern character is defined by the series
\[
   \ch(F)= \tr_R e^{iF}=\tr_R 
   \sum_{j=0}^{n}\frac{1}{j!}(iF)^{j}= \ch_0(F)+\ch_1(F)+\ch_2(F)+\cdots ,
\]
with
\[
    \ch_{2j}(F)= \frac{1}{j!}\mathrm{tr}_R(iF)^j .
\]

For the complex bundle $V$ of rank $r$, the Chern character may be expressed in terms of Chern classes as 
\[
    \ch(V)=&r + c_1 + \frac{1}{2}(c_1^2-2c_2)+\frac{1}{2}(\frac{1}{3}c_1^3-c_1 c_2+c_3)+\\
    &+\frac{1}{2\cdot 3}(\frac{1}{2^4}c_1^4-c^2 c_2+c_1 c_3+\frac{1}{2}c_2^2-c_4)+\cdots ,
\]
where
$\ch_0=r$, $\ch_2=c_1$ and so on.

For $V$ a real bundle of rank $r$, its complexification is a complex bundle $V_\bC$ of complex rank $r$ (real rank $2r$) denoted by $V\otimes \bC$.
When complexifying, the odd Chern characters of degree $(4k+2)$ vanish, and the Pontryagin classes may be defined in terms of the even Chern classes
\[
  p_k(V) = (-1)^k c_{2k}(V\otimes \bC) .
\]
Then, the Pontryagin character of the real bundle $V$ may be expressed in terms of the Chern character of its complexified bundle as
\[\label{eq:phi}
    \ph(V)&= \ch(V_\bC)\\
    &=\frac{1}{2}\tr_R (e^{iF}+e^{-iF})\\&=\tr_R \cosh (iF)\\
    &=\tr_R ( r + \frac{1}{2} (iF)^2+\frac{1}{4!} (iF)^4+\cdots)\\
    &= 
    r + p_1(V)  + \frac{p_1(V)^2 - 2p_2(V)}{12} + \frac{p_1(V)^3 - 3 p_1(V) p_2(V) + 3 p_3(V)}{360} + \cdots.
\]

The A-roof genus may be expressed in terms of Pontryagin classes as
\[
    \hat{A}= 1 - \frac{p_1}{24} +\frac{7p_1^2-4p_2}{5760} +\frac{-31p_1^3+44p_1 p_2 -16p_3}{967680}+\cdots  .
\]

\textbf{Conflicts of interest } The authors declare no conflict of commercial interest.

\textbf{Data availability statement} Data sharing
not applicable to this article as no datasets were generated or analysed during the current study.

 \bibliographystyle{ytamsalpha}
 \def\arxivfont{\rm}
 \baselineskip=.95\baselineskip
\bibliography{refs}

\providecommand{\bysame}{\leavevmode\hbox to3em{\hrulefill}\thinspace}
\providecommand{\MR}{\relax\ifhmode\unskip\space\fi MR }
\providecommand{\MRhref}[2]{%
  \href{http://www.ams.org/mathscinet-getitem?mr=#1}{#2}
}
\providecommand{\href}[2]{#2}
\providecommand{\doihref}[2]{\href{#1}{#2}}
\providecommand{\arxivfont}{\tt}
\begin{thebibliography}{KPMT19b}

\bibitem[AGW84]{Alvarez-Gaume:1983ihn}
L.~Alvarez-Gaume and E.~Witten, \emph{{Gravitational Anomalies}}, \doihref{http://dx.doi.org/10.1016/0550-3213(84)90066-X}{Nucl. Phys. B \textbf{234} (1984) 269}.

\bibitem[BDDM23]{Basile:2023knk}
I.~Basile, A.~Debray, M.~Delgado, and M.~Montero, \emph{{Global anomalies \& bordism of non-supersymmetric strings}}, \doihref{http://dx.doi.org/10.1007/JHEP02(2024)092}{JHEP \textbf{02} (2024) 092}, \href{http://arxiv.org/abs/2310.06895}{{\arxivfont arXiv:2310.06895 [hep-th]}}.

\bibitem[BE23]{berwick2023families}
D.~Berwick-Evans, \emph{{The families Clifford index and differential KO-theory}}, \href{http://arxiv.org/abs/2303.09091}{{\arxivfont arXiv:2303.09091 [math.AT]}}.

\bibitem[BM06a]{Belov:2006jd}
D.~Belov and G.~W. Moore, \emph{{Holographic Action for the Self-Dual Field}}, \href{http://arxiv.org/abs/hep-th/0605038}{{\arxivfont arXiv:hep-th/0605038}}.

\bibitem[BM06b]{Belov:2006xj}
D.~M. Belov and G.~W. Moore, \emph{{Type II Actions from 11-Dimensional Chern-Simons Theories}}, \href{http://arxiv.org/abs/hep-th/0611020}{{\arxivfont arXiv:hep-th/0611020}}.

\bibitem[BS10]{BunkeSchick}
U.~Bunke and T.~Schick, \doihref{http://dx.doi.org/10.1007/978-3-642-22842-1\_11}{\emph{Differential {K}-theory: a survey}}, Global differential geometry, Springer Proc. Math., vol.~17, Springer, Heidelberg, 2012, pp.~303--357. \href{http://arxiv.org/abs/1011.6663}{{\arxivfont arXiv:1011.6663 [math.KT]}}.

\bibitem[CS85]{CheegerSimons}
J.~Cheeger and J.~Simons, \href{https://doi.org/10.1007/BFb0075216}{\emph{Differential characters and geometric invariants}}, Geometry and topology ({C}ollege {P}ark, {M}d., 1983/84), Lecture Notes in Math., vol. 1167, Springer, Berlin, 1985, pp.~50--80.

\bibitem[DFM09]{Distler:2009ri}
J.~Distler, D.~S. Freed, and G.~W. Moore, \emph{{Orientifold Pr\'ecis}}, \href{http://arxiv.org/abs/0906.0795}{{\arxivfont arXiv:0906.0795 [hep-th]}}.

\bibitem[DM00]{Dudas:2000sn}
E.~Dudas and J.~Mourad, \emph{{D-branes in nontachyonic 0B orientifolds}}, \doihref{http://dx.doi.org/10.1016/S0550-3213(00)00781-1}{Nucl. Phys. B \textbf{598} (2001) 189--224}, \href{http://arxiv.org/abs/hep-th/0010179}{{\arxivfont arXiv:hep-th/0010179}}.

\bibitem[FH00]{Freed:2000tt}
D.~S. Freed and M.~J. Hopkins, \emph{{On Ramond-Ramond Fields and K Theory}}, \doihref{http://dx.doi.org/10.1088/1126-6708/2000/05/044}{JHEP \textbf{05} (2000) 044}, \href{http://arxiv.org/abs/hep-th/0002027}{{\arxivfont arXiv:hep-th/0002027}}.

\bibitem[FMS06]{Freed:2006ya}
D.~S. Freed, G.~W. Moore, and G.~Segal, \emph{{The Uncertainty of Fluxes}}, \doihref{http://dx.doi.org/10.1007/s00220-006-0181-3}{Commun. Math. Phys. \textbf{271} (2007) 247--274},
\href{http://arxiv.org/abs/hep-th/0605198}{{\arxivfont arXiv:hep-th/0605198}}.

\bibitem[Fre00]{Freed:2000ta}
D.~S. Freed, \emph{{Dirac charge quantization and generalized differential cohomology}}, \doihref{http://dx.doi.org/10.4310/SDG.2002.v7.n1.a6}{Surv. Differ. Geom. \textbf{7} (2000) 129--194}, \href{http://arxiv.org/abs/hep-th/0011220}{{\arxivfont arXiv:hep-th/0011220}}.

\bibitem[GEH24]{GarciaEtxebarria:2024fuk}
I.~Garc\'\i{}a~Etxebarria and S.~S. Hosseini, \emph{{Some aspects of symmetry descent}}, \doihref{http://dx.doi.org/10.1007/JHEP12(2024)223}{JHEP \textbf{12} (2025) 223}, \href{http://arxiv.org/abs/2404.16028}{{\arxivfont arXiv:2404.16028 [hep-th]}}.

\bibitem[GS84]{Green:1984sg}
M.~B. Green and J.~H. Schwarz, \emph{{Anomaly Cancellation in Supersymmetric D=10 Gauge Theory and Superstring Theory}}, \doihref{http://dx.doi.org/10.1016/0370-2693(84)91565-X}{Phys. Lett. B \textbf{149} (1984) 117--122}.

\bibitem[GS85a]{Green:1984ed}
\bysame, \emph{{Infinity Cancellations in SO(32) Superstring Theory}}, \doihref{http://dx.doi.org/10.1016/0370-2693(85)90816-0}{Phys. Lett. B \textbf{151} (1985) 21--25}.

\bibitem[GS85b]{Green:1984qs}
\bysame, \emph{{The Hexagon Gauge Anomaly in Type I Superstring Theory}}, \doihref{http://dx.doi.org/10.1016/0550-3213(85)90130-0}{Nucl. Phys. B \textbf{255} (1985) 93--114}.

\bibitem[GS18]{Grady:2018suz}
D.~Grady and H.~Sati, \emph{{Differential KO-theory: constructions, computations, and applications}}, \doihref{http://dx.doi.org/10.1016/j.aim.2021.107671}{Adv. Math. \textbf{384} (2021) 107671}, \href{http://arxiv.org/abs/1809.07059}{{\arxivfont arXiv:1809.07059 [math.AT]}}.

\bibitem[GY21]{Gomi:2021bhy}
K.~Gomi and M.~Yamashita, \emph{{Differential $KO$-theory via gradations and mass terms}}, \doihref{http://dx.doi.org/10.4310/ATMP.2023.v27.n2.a1}{Adv. Theor. Math. Phys. \textbf{27} (2023) 381--481}, \href{http://arxiv.org/abs/2111.01377}{{\arxivfont arXiv:2111.01377 [math.KT]}}.

\bibitem[HS02]{Hopkins:2002rd}
M.~J. Hopkins and I.~M. Singer, \emph{{Quadratic functions in geometry, topology, and M theory}}, \doihref{http://dx.doi.org/10.4310/jdg/1143642908}{J. Diff. Geom. \textbf{70} (2005) 329--452}, \href{http://arxiv.org/abs/math/0211216}{{\arxivfont arXiv:math/0211216}}.

\bibitem[HTY20]{Hsieh:2020jpj}
C.-T. Hsieh, Y.~Tachikawa, and K.~Yonekura, \emph{{Anomaly Inflow and p-Form Gauge Theories}}, \doihref{http://dx.doi.org/10.1007/s00220-022-04333-w}{Commun. Math. Phys. \textbf{391} (2022) 495--608}, \href{http://arxiv.org/abs/2003.11550}{{\arxivfont arXiv:2003.11550 [hep-th]}}.

\bibitem[Kne24]{Kneissl:2024zox}
C.~Kneissl, \emph{{Spin cobordism and the gauge group of type I/heterotic string theory}}, \doihref{http://dx.doi.org/10.1007/JHEP01(2025)181}{JHEP \textbf{01} (2025) 181}, \href{http://arxiv.org/abs/2407.20333}{{\arxivfont arXiv:2407.20333 [hep-th]}}.

\bibitem[KOT19]{Kobayashi:2019lep}
R.~Kobayashi, K.~Ohmori, and Y.~Tachikawa, \emph{{On Gapped Boundaries for SPT Phases Beyond Group Cohomology}}, \doihref{http://dx.doi.org/10.1007/JHEP11(2019)131}{JHEP \textbf{11} (2019) 131}, \href{http://arxiv.org/abs/1905.05391}{{\arxivfont arXiv:1905.05391 [cond-mat.str-el]}}.

\bibitem[KPMT19a]{Kaidi:2019pzj}
J.~Kaidi, J.~Parra-Martinez, and Y.~Tachikawa, \emph{{Classification of String Theories via Topological Phases}}, \doihref{http://dx.doi.org/10.1103/PhysRevLett.124.121601}{Phys. Rev. Lett. \textbf{124} (2020) 121601}, \href{http://arxiv.org/abs/1908.04805}{{\arxivfont arXiv:1908.04805 [hep-th]}}.

\bibitem[KPMT19b]{Kaidi:2019tyf}
\bysame, \emph{{Topological Superconductors on Superstring Worldsheets}}, \doihref{http://dx.doi.org/10.21468/SciPostPhys.9.1.010}{SciPost Phys. \textbf{9} (2020) 10}, \href{http://arxiv.org/abs/1911.11780}{{\arxivfont arXiv:1911.11780 [hep-th]}}. {With a mathematical appendix by Arun Debray.}

\bibitem[LL24]{Larotonda:2024thv}
V.~Larotonda and L.~Lin, \emph{{Anomaly Inflow and Gauge Group Topology in the 10d Sugimoto String Theory}}, \href{http://arxiv.org/abs/2412.17894}{{\arxivfont arXiv:2412.17894 [hep-th]}}.

\bibitem[LY22]{Lee:2022spd}
Y.~Lee and K.~Yonekura, \emph{{Global anomalies in 8d supergravity}}, \doihref{http://dx.doi.org/10.1007/JHEP07(2022)125}{JHEP \textbf{07} (2022) 125}, \href{http://arxiv.org/abs/2203.12631}{{\arxivfont arXiv:2203.12631 [hep-th]}}.

\bibitem[MW99]{Moore:1999gb}
G.~W. Moore and E.~Witten, \emph{{Selfduality, Ramond-Ramond Fields, and K Theory}}, \doihref{http://dx.doi.org/10.1088/1126-6708/2000/05/032}{JHEP \textbf{05} (2000) 032}, \href{http://arxiv.org/abs/hep-th/9912279}{{\arxivfont arXiv:hep-th/9912279}}.

\bibitem[Sag95]{Sagnotti:1995ga}
A.~Sagnotti, \emph{{Some properties of open string theories}}, {International Workshop on Supersymmetry and Unification of Fundamental Interactions (SUSY 95)}, 1995, pp.~473--484. \href{http://arxiv.org/abs/hep-th/9509080}{{\arxivfont arXiv:hep-th/9509080}}.

\bibitem[Sag97]{Sagnotti:1996qj}
\bysame, \emph{{Surprises in open string perturbation theory}}, \doihref{http://dx.doi.org/10.1016/S0920-5632(97)00344-7}{Nucl. Phys. B Proc. Suppl. \textbf{56} (1997) 332--343}, \href{http://arxiv.org/abs/hep-th/9702093}{{\arxivfont arXiv:hep-th/9702093}}.

\bibitem[ST24]{Saito:2024iiu}
S.~Saito and Y.~Tachikawa, \emph{{Cancelling mod-2 anomalies by Green-Schwarz mechanism with $B_{\mu\nu}$}}, \href{http://arxiv.org/abs/2411.09223}{{\arxivfont arXiv:2411.09223 [hep-th]}}.

\bibitem[Sto86]{StongAppendix}
R.~E. Stong, \emph{Calculation of {$\Omega^{{\rm Spin}}_{11}(K(\mathbb{Z},4))$}}. Appendix to \cite{Witten:1985bt}.

\bibitem[Sug99]{Sugimoto:1999tx}
S.~Sugimoto, \emph{{Anomaly Cancellations in Type I D9 -- Anti-D9 System and the USp(32) String Theory}}, \doihref{http://dx.doi.org/10.1143/PTP.102.685}{Prog. Theor. Phys. \textbf{102} (1999) 685--699}, \href{http://arxiv.org/abs/hep-th/9905159}{{\arxivfont arXiv:hep-th/9905159}}.

\bibitem[SW01]{Schwarz:2001sf}
J.~H. Schwarz and E.~Witten, \emph{{Anomaly analysis of brane - anti-brane systems}}, \doihref{http://dx.doi.org/10.1088/1126-6708/2001/03/032}{JHEP \textbf{03} (2001) 032}, \href{http://arxiv.org/abs/hep-th/0103099}{{\arxivfont arXiv:hep-th/0103099}}.

\bibitem[TY21]{Tachikawa:2021mby}
Y.~Tachikawa and M.~Yamashita, \emph{{Topological Modular Forms and the Absence of All Heterotic Global Anomalies}}, \doihref{http://dx.doi.org/10.1007/s00220-023-04761-2}{Commun. Math. Phys. \textbf{402} (2023) 1585--1620}, \href{http://arxiv.org/abs/2108.13542}{{\arxivfont arXiv:2108.13542 [hep-th]}}. [Erratum: Commun.Math.Phys. 402, 2131 (2023)].

\bibitem[Wit83a]{Witten:1983tw}
E.~Witten, \emph{{Global Aspects of Current Algebra}},
\doihref{http://dx.doi.org/10.1016/0550-3213(83)90063-9}{Nucl. Phys. \textbf{B223} (1983) 422--432}.

\bibitem[Wit83b]{Witten:1983tx}
\bysame, \emph{{Current Algebra, Baryons, and Quark Confinement}},
\doihref{http://dx.doi.org/10.1016/0550-3213(83)90064-0}{Nucl. Phys. \textbf{B223} (1983) 433--444}.

\bibitem[Wit86]{Witten:1985bt}
\bysame, \emph{{Topological Tools in Ten-Dimensional Physics}}, \doihref{http://dx.doi.org/10.1142/S0217751X86000034}{Int. J. Mod. Phys. A \textbf{1} (1986) 39}.

\bibitem[Wit98]{Witten:1998cd}
\bysame, \emph{{D-branes and K-theory}}, \doihref{http://dx.doi.org/10.1088/1126-6708/1998/12/019}{JHEP \textbf{12} (1998) 019}, \href{http://arxiv.org/abs/hep-th/9810188}{{\arxivfont arXiv:hep-th/9810188}}.

\bibitem[Wit03]{Witten:2003ya}
\bysame, \emph{{$SL(2,Z)$ Action on Three-Dimensional Conformal Field Theories with Abelian Symmetry}}, {From Fields to Strings: Circumnavigating Theoretical Physics: a Conference in Tribute to Ian Kogan}, 2003, pp.~1173--1200. \href{http://arxiv.org/abs/hep-th/0307041}{{\arxivfont arXiv:hep-th/0307041}}.

\bibitem[Yon22]{Yonekura:2022reu}
K.~Yonekura, \emph{{Heterotic Global Anomalies and Torsion Witten Index}}, \doihref{http://dx.doi.org/10.1007/JHEP10(2022)114}{JHEP \textbf{10} (2022) 114}, \href{http://arxiv.org/abs/2207.13858}{{\arxivfont arXiv:2207.13858 [hep-th]}}.

\bibitem[Zha24]{Zhang:2024oas}
H.~Y. Zhang, \emph{{K-theoretic Global Symmetry in String-constructed QFT and T-duality}}, \href{http://arxiv.org/abs/2404.16097}{{\arxivfont arXiv:2404.16097 [hep-th]}}.

\end{thebibliography}

\end{document}